[1994/12/01]
\documentclass[10pt, reqno]{amsart}
\usepackage{a4wide}
\usepackage[english, activeacute]{babel}
\usepackage{amsmath,amsthm,amsxtra}
\usepackage{epsfig}
\usepackage{amssymb}
\usepackage{latexsym}
\usepackage{amsfonts}
\usepackage{hyperref}
\usepackage{pdftricks}
\usepackage{xparse}
\usepackage{tabularx} 
\usepackage{multicol}

\title{On the variational structure of breather solutions}
\author{Miguel A. Alejo}
\author{Claudio Mu\~noz}
\author{Jos\'e M. Palacios}
\address{Departamento de Matem\'atica, Universidade Federal de Santa Catarina, Brasil}
\email{miguel.alejo@ufsc.br}
\address{CNRS and Departamento de Ingenier\'ia Matem\'atica DIM, Universidad de Chile, Chile}
\email{cmunoz@dim.uchile.cl}
\address{Departamento de Ingenier\'ia Matem\'atica DIM, Universidad de Chile, Chile}
\email{jpalacios@dim.uchile.cl}
\date{\today}
\thanks{}

\chardef\bslash=`\\ 

\newtheorem{thm}{Theorem}[section]
\newtheorem{cor}[thm]{Corollary}
\newtheorem{lem}[thm]{Lemma}
\newtheorem{prop}[thm]{Proposition}

\newtheorem{defn}[thm]{Definition}

\theoremstyle{remark}
\newtheorem{rem}{Remark}[section]

\numberwithin{equation}{section}

\newcommand{\R}{\mathbb{R}}

\newcommand{\Z}{\mathbb{Z}}
\newcommand{\T}{\mathbb{T}}

\newcommand{\la}{\lambda}

\newcommand{\al}{\alpha}
\newcommand{\bt}{\beta}
\newcommand{\ga}{\gamma}

\newcommand{\sech}{\operatorname{sech}}

\newcommand{\sne}{\operatorname{sn}}
\newcommand{\cd}{\operatorname{cd}}
\newcommand{\sd}{\operatorname{sd}}
\newcommand{\cne}{\operatorname{cn}}
\newcommand{\dne}{\operatorname{dn}}

\newcommand{\ba}{\left( \begin{array}{c}}
\newcommand{\ea}{\end{array}\right)}

 \providecommand{\abs}[1]{\lvert#1 \rvert}

\newcommand{\be}{\begin{equation}}
\newcommand{\ee}{\end{equation}}
\newcommand{\bp}{\begin{proof}}
\newcommand{\ep}{\end{proof}}
\newcommand{\bel}{\begin{equation}\label}
\newcommand{\eeq}{\end{equation}}
\newcommand{\bea}{\begin{eqnarray}}
\newcommand{\eea}{\end{eqnarray}}
\newcommand{\bee}{\begin{eqnarray*}}
\newcommand{\eee}{\end{eqnarray*}}
\newcommand{\ben}{\begin{enumerate}}
\newcommand{\een}{\end{enumerate}}
\newcommand{\nonu}{\nonumber}
\newcommand{\bs}{\bigskip}
\newcommand{\ms}{\medskip}

\newcommand{\sn}{\operatorname{sn}}
\newcommand{\nd}{\operatorname{nd}}
\newcommand{\dn}{\operatorname{dn}}
\newcommand{\con}{\operatorname{cn}}

\newcommand{\eval}[2][\right]{\relax
  \ifx#1\right\relax \left.\fi#2#1\rvert}

\let\abs=\envert

\begin{document}
\begin{abstract}
In this paper we give a systematic and simple account that put in evidence that many breather solutions of integrable equations satisfy suitable variational elliptic equations, which also implies that the stability problem reduces in some sense to  $(i)$ the study of the spectrum of explicit linear systems (\emph{spectral stability}), and $(ii)$ the understanding of how bad directions (if any) can be controlled using low regularity conservation laws.  We exemplify this idea in the case of the modified Korteweg-de Vries (mKdV), Gardner, and the more involved sine-Gordon (SG) equation. Then we perform numerical simulations that confirm, at the level of the spectral problem, our previous rigorous results in \cite{AM,AM1}, where we showed that mKdV breathers are $H^2$ and $H^1$ stable, respectively. In a second step, we also discuss the Gardner case, a relevant modification of the KdV and mKdV equations, recovering similar results. Then we discuss the Sine-Gordon case, where the spectral study of a fourth-order 
linear matrix system is the key element to show stability. Using numerical methods, we confirm that all spectral assumptions leading to the $H^2\times H^1$ stability of SG breathers are numerically satisfied, even in the ultra-relativistic, singular regime. In a second part, we study the periodic mKdV case, where a periodic breather is known from the work of Kevrekidis et al. \cite{KKS2}. We rigorously show that these breathers satisfy a suitable elliptic equation, and we also show numerical spectral stability. However, we also identify the source of nonlinear instability in the case described in \cite{KKS2}, and we conjecture that, even if spectral stability is satisfied, nonlinear stability/instability depends only on the sign of a suitable discriminant function, a condition that is trivially satisfied in the case of non-periodic breathers. Finally, we present a new class of breather solution for mKdV, believed to exist from geometric considerations, and which is periodic in time and space, but has nonzero 
mean, unlike standard breathers.
\end{abstract}
\maketitle \markboth{Elliptic equations, breathers, stability} {Miguel A. Alejo, Claudio Mu\~noz and Jos\'e M. Palacios}
\renewcommand{\sectionmark}[1]{}
\tableofcontents

\section{Introduction}

\medskip

The purpose of this paper is twofold: first, to put in evidence that \emph{breathers}, which are nontrivial solutions of integrable dispersive equations, different to solitons and multi-solitons, may satisfy intricate systems of elliptic equations, a fact that  reveals a subtle variational structure underlying the existence and stability of these particular solutions. The validity of these elliptic equations implies, among other things, that breathers can be 
understood as critical points of a suitable Lyapunov functional. However, it is important to note that this functional must be defined in a
\emph{proper} subspace of the so-called \emph{energy space}, namely the space where the standard variational structure of solitons is placed. 
Instead, we must look for functionals described at the leading order by an additional conservation law. Thanks to the integrability of the equation, 
such a quantity is always available. However, the addition of a new conservation law carries together a rigidity structure, in the sense that breathers 
should not exist in nonintegrable 
models, at least breathers with two different, independent component variables. 

\medskip

Since these particular solutions satisfy a well-defined variational characterization, their dynamical stability reduces (in most cases) to the study of the
spectrum of a suitable linear operator defined in a subspace of the energy space, see e.g. our recent papers \cite{AM,AM0}. Except for some simple cases as the already mentioned, this task is far from being simple. This is the second objective of this paper: a numerical understanding of the spectral stability of these linearized operators, since a rigorous description of these systems has escaped to us.

\medskip

As a consequence of our results, the
extension of these stability properties to bigger classes of initial perturbations cannot be obtained via the aforementioned variational structure. 
Indeed, one needs a sort of rigidity or dynamical structure that gives a new kind of stability result for more general Sobolev spaces. As an example of this rigidity property, we refer to the reader to our recent result  \cite{AM1}, where we show that mKdV breathers are $H^1$ stable.

\subsection{Setting of the problem}  In this paper we exemplify the above ideas with a (far from being exhaustive) account of breather solutions of the non-periodic and periodic modified Korteweg-de Vries (mKdV) equation
\be\label{mKdV} 
u_{t} +(u_{xx} + u^3)_x=0, \qquad u(t,x) \in \R, \; t\in\R, \, x \in\R  ~\hbox{ or }~ \T_x:= \R/L\Z; 
\ee
the Gardner equation
\be\label{GE}
w_t +(w_{xx} +\mu w^2+w^3)_x=0, \quad \mu >0,\qquad w(t,x) \in \R, \; (t,x) \in\R^2;
\ee
and the Sine-Gordon (SG) equation
\be\label{SG}
u_{tt} -u_{xx} +\sin u =0, \quad (u,u_t)=(u,u_t)(t,x) \in \R^2, \; (t,x) \in\R^2.
\ee
The above equations  are  well-known \emph{completely integrable} models \cite{Ga,AC,La}, with infinitely many conserved quantities, and a 
suitable Lax-pair formulation. The Inverse Scattering Theory (IST) has been applied by many authors in order to describe the behavior of solutions 
in certain generality, see e.g. \cite{Sch,AC,La,PeGr} and references therein. In particular, the evolution of  \emph{rapidly decaying}\footnote{We emphasize that no soliton-resolution results for subcritical equations are available in standard Sobolev spaces without explicit decay assumptions; in that sense, our results in \cite{AM,AM1} do not need additional decay assumptions.} initial data can be
described by purely algebraic methods. Moreover, solutions for these equations can be obtained explicitly using this theory \cite{AC,La}. Some of them 
are shown to decompose into a very particular set of solutions, usually referred as solitons, breathers and radiation (see e.g. Schuur \cite{Sch} and references therein).

\medskip

Solutions of \eqref{mKdV}, \eqref{GE} and \eqref{SG} are invariant under space and time translations. Indeed, for any $t_0, x_0\in \R$, $u(t-t_0, x-x_0)$
is also a solution. Furthermore, an additional feature of the SG equation is the invariance under Lorentz transformations: given any $v\in (-1,1)$, then
\be\label{Lor}
u(\ga (t-vx),\ga (x-vt)), \quad \ga :=  (1-v^2)^{-1/2},
\ee
is a new solution of (\ref{SG}). 
\ms

The mKdV and  Gardner equations (\ref{mKdV}) and (\ref{GE}) are also relevant due to the existence of solitary wave solutions called \emph{solitons}. A \emph{soliton} is a localized, moving or stationary solution which maintains its form for all time. Similarly, a \emph{multi-soliton} is a (not necessarily) explicit solution describing the interaction of several solitons \cite{HIROTA1}.

\ms

More interesting is the fact that these soliton profiles are often regarded as \emph{global minimizers} of a constrained functional in the $H^1$-topology. For example, mKdV (\ref{mKdV}) has solitons of the form
\be\label{Sol}
u(t,x) = Q_c (x-ct), \quad Q_c(s) := \sqrt{c} Q(\sqrt{c} s), \quad c>0,
\ee
with
\[
Q (s):= \frac{\sqrt{2}}{\cosh (s)} =2\sqrt{2} \partial_s[\arctan(e^{s})].
\]
In the case of the Gardner equation \eqref{GE}, the profile of the soliton solution is slightly more cumbersome, but it is still given explicitly  by the formula
\bel{fexplicita2} w(t,x) := Q_{c,\mu} (x-ct), \qquad Q_{c,\mu} (s) :=\frac{c}{ \displaystyle{ \frac{\mu}{3} +\sqrt{\frac{\mu^2}{9}+\frac{c}{2}}\cosh(\sqrt{c}s) } }.\eeq
By replacing (\ref{Sol}) in (\ref{mKdV}) and \eqref{fexplicita2}  in  \eqref{GE}, one has that $Q_c>0$ and $Q_{c,\mu}>0 $ satisfy respectively the nonlinear elliptic equations
\be\label{eqQc}\begin{array}{ll}
Q_c'' -c\, Q_c +Q_c^3=0, \quad Q_c>0, \quad Q_c \in H^1(\R)~\quad ~\text{(mKdV)},\\\\
Q_{c,\mu}'' -c\, Q_{c,\mu} +\mu Q_{c,\mu}^2+Q_{c,\mu}^3=0, \quad Q_{c,\mu}>0, \quad Q_{c,\mu}\in H^1(\R)~\quad ~\text{(Gardner)}.
\end{array}\ee

These second order, elliptic equations are deeply related to the so-called variational structure of the soliton solution. To be more precise, it is well-known that 
the standard conservation laws in the case of mKdV (\ref{mKdV}) at the $H^1$-level are the \emph{mass}
\be\label{M1}
M[u](t)  :=  \frac 12 \int_\R u^2(t,x)dx = M[u](0),
\ee 
and \emph{energy} 
\be\label{E1}
E[u](t)  :=  \frac 12 \int_\R u_x^2(t,x)dx -\frac 14 \int_\R u^4(t,x)dx = E[u](0).
\ee 
In the case of the Gardner equation \eqref{GE}, the mass has the same definition as in \eqref{M1}, but  the energy is given  by 
\be\label{E2}
E_\mu[w](t)  :=  \frac 12 \int_\R w_x^2 -\frac \mu3\int_\R w^3 - \frac 14 \int_\R w^4= E_\mu[w](0),\ee
which is $H^1$-subcritical. (For mKdV and Gardner, the Cauchy problem is globally well-posed at such a level of regularity or even better, 
see e.g. Kenig-Ponce-Vega \cite{KPV}.)  

\ms

Using these conserved quantities, the variational structure can be characterized as follows: there exists a suitable \emph{Lyapunov functional}, \emph{invariant in time} and such 
that the soliton $Q_c$ or $Q_{c,\mu}$ is a corresponding \emph{extremal point}. Moreover, it is a global minimizer under fixed mass. For the mKdV case, this functional 
is given by \cite{Benj}
\be\label{H0mk}
\mathcal{H}[u](t) = E[u](t) + c \, M[u](t),
\ee
where $c>0$ is the scaling of the solitary wave, and $E[u]$, $M[u]$ are given in (\ref{M1})-(\ref{E1}). Indeed, a simple computation shows that for any $z(t)\in H^1(\R)$ small,
\be\label{Expa1mk}
\mathcal{H}[Q_c+z](t)  =  \mathcal{H}[Q_c] + \int_\R z(Q_c''-cQ_c +Q_c^3) +  O(\|z(t)\|_{H^1}^2).
\ee
The zero order term  above is independent of time, while the first order term in $z$ is zero from (\ref{eqQc}), proving the critical character of $Q_c$. 

\medskip

On the other hand, standard conservation laws for the SG equation (\ref{SG}) at the $H^1\times L^2$-level are the \emph{energy}
\be\label{E}
E[u,u_t](t)  :=   \frac 12 \int_\R (u_x^2 +u_t^2)(t,x)dx  +\int_\R (1-\cos u(t,x))dx  =  E[u,u_t](0),
\ee 
and the \emph{momentum} 
\be\label{P}
P[u,u_t](t) :=  \frac 12 \int_\R  u_t(t,x) u_x(t,x)dx = P[u,u_t](0).
\ee 
It is well known that any suitable well-posedness theory must deal simultaneously with the pair $(u,u_t)$ and not only $u$. Since the nonlinear term $\sin u$ is uniformly bounded independently of the size of $u$, one can find a satisfactory $H^1\times L^2$ global well-posedness theory, see e.g.  Bourgain \cite{B}.

\medskip

If one studies  perturbations of solitons in (\ref{mKdV}) and more general equations,  the concepts of \emph{orbital}, and \emph{asymptotic stability} emerge naturally. In particular, since energy and mass are conserved quantities, it is natural to expect that solitons are stable in a suitable energy space. Indeed, $H^1$-stability of mKdV and more general solitons and multi-solitons has been considered e.g. in Benjamin \cite{Benj}, Bona-Souganidis-Strauss \cite{BSS}, Weinstein \cite{We2}, Maddocks-Sachs \cite{MS}, Martel-Merle-Tsai \cite{MMT}, Martel-Merle \cite{MMcol2} and M. \cite{Mu}. $L^2$-stability of KdV solitons has been proved by Merle-Vega \cite{MV}. Moreover, asymptotic stability properties for gKdV equations have been studied by Pego-Weinstein \cite{PW} and Martel-Merle \cite{MMarma,MMnon}, among many other authors. 

\medskip

On the other hand, the SG equation (\ref{SG}) has soliton solutions usually referred as \emph{kinks}. Indeed, given $v\in (0,1)$, $x_0\in \R$ and $\ga$ defined as in \eqref{Lor},  SG (\ref{SG}) has kinks of the form
\be\label{SolSG}
u(t,x) = \varphi( \ga (x- v t -x_0)), \quad \varphi(s) := 4\arctan e^{s}.
\ee
It is not difficult to see that these solutions can be associated to a well known functional defined in the $H^1\times L^2$ topology.  

\medskip

Kinks solutions of \eqref{SG} are orbitally stable in the natural energy space $H^1\times L^2$. Indeed, the energy (\ref{E}) is a conserved quantity and kinks can be viewed as relative minimizers of a suitable energy functional. For the proofs of this result for the SG case and more general equations, we refer to the works  by Henry-Perez-Wreszinski \cite{HPW}, Grillakis-Shatah-Strauss \cite{GSS}, Soffer-Weinstein \cite{SW}, Cuccagna \cite{Cuc} and the recent result by Kowalczyk, Martel and the second author \cite{KMM}.

\subsection{Modified KdV breathers and their stability} In addition to the above mentioned special solutions of \eqref{mKdV}, \eqref{GE} and \eqref{SG}, there exists another large family of 
explicit and oscillatory solutions, known in the physical and mathematical literature as the \emph{breather} solution, and which is a 
periodic in time, spatially localized real function. Although there is no universal definition for a breather, we will adopt the following convention, that will match the mKdV, Gardner and SG cases.

\begin{defn}[Aperiodic breather]\label{DEF_B}
We say that $B=B(t,x)$ is a breather solution for a particular one-dimensional dispersive equation if there are $T>0$ and $L=L(T)\in \R$ such that, for all $t\in \R$ and $x\in \R$, one has 
\be\label{B_def}
B(t+T,x) = B(t,x-L),
\ee
and moreover, the infimum among times $T>0$ such that property \eqref{B_def} is satisfied for such a time $T$ is uniformly positive. 
\end{defn}

\begin{rem}
Note that the last condition ensures that solitons (and multisolitons) are not breathers, since e.g. $Q_c(x-c (t+T)) = Q_c(x-L -ct)$ for
$L:=cT$ but $T$ can be any real-valued time.\footnote{In the case of NLS equations and their solitons, 
Definition \ref{DEF_B} include them because of the $U(1)$ invariance.}
\end{rem}

For the mKdV equation, the breather was discovered by Wadati \cite{W1}; and for the Gardner 
equation, by \cite{PeGr,Ale0}, using the IST. These solutions have become a canonical example  of  complexity in nonlinear 
integrable systems \cite{La,AC}. Moreover, their  surprising mixed behavior, combining oscillatory and soliton character, has focused 
the attention of many researchers since thirty years ago \cite{Sch,BMW,D}. From the physical point of view,  breather solutions seem 
to be relevant to localization-type phenomena in optics, condensed matter physics and biological processes \cite{Au}. They also play 
an important role in the modeling of freak and rogue waves events on surface gravity 
waves and also of internal waves in the stratified ocean, in Josephson junctions and even in nonlinear optics. See \cite{DyTr,DyTr1,DyTr2,DyTr3,DyTr4} for a 
representative set of these examples.

\medskip

\begin{figure}[!h]
\begin{center}
\includegraphics[width=10cm, height=5.5cm]{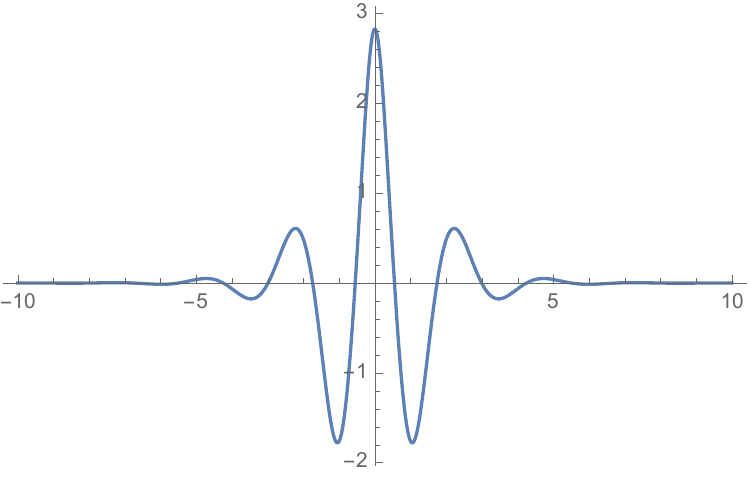}
\caption{mKdV breather at time $t=0$ with $\bt=1$, $\al=2.5$.}\label{mKdV_0}
\end{center}
\end{figure}

The following is the standard definition of the real breather of mKdV \eqref{mKdV} (see \cite{W1,La} and references therein):
\begin{defn}[See Fig. \ref{mKdV_0}]\label{mKdVB} Let $\al, \bt >0$ and $x_1,x_2\in \R$. The breather solution of mKdV \eqref{mKdV} is explicitly given by 
\bea\label{breather}
 B_{\al, \bt}(t,x)   :=  2\sqrt{2} \partial_x \Bigg[ \arctan \Big( \frac{\bt}{\al}\frac{\sin(\al  y_1)}{\cosh(\bt y_2)}\Big) \Bigg] ,
\eea
with 
\be\label{y1y2}
y_1 = x+ \delta t + x_1, \quad y_2 = x+ \ga t + x_2, \quad \delta := \al^2 -3\bt^2, \quad  \ga := 3\al^2 -\bt^2.
\ee
\end{defn}

Note that  breather solutions of the mKdV are periodic in time, but not in space. Additionally, every mKdV breather satisfies Definition \ref{DEF_B} with $T=\frac{2\pi}{\al(\ga-\delta)}>0$ and $L=-\ga T$. A simple but very important remark is that $\delta \neq \ga$, 
for all values of $\al$ and $\beta$ different from zero. This means that the variables  $x+\delta t$ and $x+\ga t$ are {\bf always independent}, a property that characterizes breather solutions, which is not satisfied by standard solitons.  On the other hand, $-\ga$ will be for us the \emph{velocity} of the breather solution, since it corresponds to the velocity of the \emph{carried hump} in the breather profile. Note additionally that breathers have to be considered as \emph{bound states}, since they do not decouple into simple solitons as time evolves.

\medskip

From a mathematical point of view,  breather solutions arise in different contexts. In a geometrical setting, mKdV breathers appear in the
evolution of closed planar curves playing the role of  smooth localized deformations traveling along the closed curve \cite{Ale}. Moreover, 
it is interesting to point out that mKdV  breather solutions have also been considered by Kenig, Ponce and Vega in their proof of the
non-uniform continuity of the mKdV flow in the Sobolev spaces $H^s$, $s<\frac 14$ \cite{KPV2}. On the other hand, they should play an
important role in the  \emph{soliton-resolution} conjecture for the mKdV equation, according to the analysis developed by Schuur in \cite{Sch}. Indeed, according to the IST theory, 
any suitable\footnote{Here suitable means ``with sufficient decay at infinity'' such that IST works properly.} solution of mKdV should decompose, as time goes to infinity, into solitons, breathers and radiation.

\medskip

The question that naturally arises is the following: is there any analogous elliptic equation satisfied by breather solutions, as solitons in \eqref{eqQc}? 
For the mKdV breather the answer is positive. Indeed, any mKdV breather \eqref{breather} $B= B_{\al,\bt}$ satisfies a nonlinear stationary equation:

\begin{thm}[\cite{AM}]\label{GBmkdv} Let $B= B_{\al,\bt}$ be any mKdV breather \eqref{breather} with scalings $\al,\bt>0$. Then, for any  fixed time $t\in \R$, the breather $B$ satisfies the nonlinear stationary equation  
\be\label{H01}
 B_{(4x)} -2(\bt^2 -\al^2) (B_{xx} + B^3)  +(\al^2 +\bt^2)^2 B + 5 BB_x^2 + 5B^2 B_{xx} + \frac 32 B^5 =0.
\ee
\end{thm}

This identity  can be seen as the \emph{nonlinear elliptic equation} satisfied by any mKdV breather profile, and therefore it is 
independent in time and translation parameters $x_1,x_2\in \R$. One can compare with the soliton profile $Q_c(x- ct -x_0)$, which 
satisfies the standard elliptic equation (\ref{eqQc}), obtained as the first variation of the $H^1$ functional (\ref{H0mk}).

\medskip

As a corollary of the previous identity, one can prove that mKdV breathers are critical points of the functional
\be\label{H00}
\mathcal{H}[u](t) := F[u](t) + 2(\bt^2-\al^2)E[u](t) +(\al^2+\bt^2)^2 M[u](t),
\ee
where $F$ is the additional mKdV conservation law:
\be\label{F1mk}
F[u]    :=    \frac 12 \int_\R u_{xx}^2 -\frac 52 \int_\R u^2u_x^2  + \frac 14 \int_\R u^6,
\ee
well-defined for $H^2$ initial data. 

\ms

We should also remark that the original seed of these ideas is certainty not new and it was introduced in a seminal paper by Lax \cite{LAX1},
in the particular case of the two-soliton solution of the Korteweg-de Vries equation. This method has been generalized to several equations 
with soliton solutions \cite{MS,HPZ,Kap,NL}. However, no previous result was available in the case of breathers, apparently because of their
dynamics, which do not resemble any type of simple, decoupled 2-soliton solution. Compared with the previous results, proofs in \cite{AM} are more involved, 
since there is no mass splitting as $t\to +\infty$. Further results where this technique has been successfully applied to the understanding of stability of different soliton solutions are the works \cite{PS,GP,JP}, see also references therein.

\medskip

The underlying question is then the study of the corresponding \emph{stability} of these solutions.  Numerical computations
(see Gorria-Alejo-Vega \cite{AGV}) show that mKdV breathers are \emph{numerically} stable. However, a rigorous proof of orbital stability
for breathers for initial data in Sobolev spaces was missing. In \cite{AM} (see also \cite{AM0}), the authors gave a positive answer 
to the question of breathers stability. The main result is stated in a simple version as follows:

\begin{thm}[\cite{AM}]\label{MT0}
mKdV breathers are $H^2$-orbitally stable.
\end{thm}

The proof of this result uses the fact that $\mathcal{H}$ defined in \eqref{H00} controls the dynamics of perturbations of breathers 
as is done in \eqref{Expa1mk}. Indeed, thanks to \eqref{H01}, the stability question is reduced to the study of \emph{spectral
properties of a suitable linearized operator}, and the control of instability directions by using low regularity conservation laws. However, the big surprise in the proof of \cite{AM} is the fact that, despite having two free scaling parameters $\al$ and $\bt$, mKdV breathers have {\bf only one negative eigenvalue}. The reader can consult \cite{AM} for a more detailed proof, or the present paper for a conditional proof in the SG case.

\medskip

Note that the previous result was improved in \cite{AM1} for the case of $H^1$ perturbations, but our proof was not variational, and it is in some sense more rigid, because it uses a B\"acklund transformation, appearing only in integrable models. In \cite{AM1} it is also proved that mKdV breathers moving to the right in space are asymptotically stable in the energy space $H^1$, which is probably the first result of this type for mKdV breathers.

\begin{rem}
We emphasize that proving stability of breathers in the associated energy space $H^1$ is in some sense unnatural, because the variational structure for breathers is placed at the  $H^2$ level of regularity. This being said, it is possible to show stability in $H^1$, but the standard conserved quantities (energy and mass) are, as far as we know, nearly useless.
\end{rem}

\subsection{Main results} The purpose of this paper is to show that the previous ideas, and in particular Theorem \ref{MT0}, can be generalized 
to several interesting integrable models with breather solutions. The most important of these models is the Sine-Gordon \eqref{SG} scalar field equation, for which the classical \emph{standing} breather is the following
\[
B(t,x;\bt) : = 4 \arctan \Big( \frac{\bt}{\al}\frac{\cos(\al t)}{\cosh(\bt x)}\Big), \quad \al^2 +\bt^2 =1.
\]
In Definition \ref{SGB} it is also presented a general formula of the SG breather, involving all symmetries of the equation. As far as we know, there is no rigorous proof of stability for this solution, at least in any standard Sobolev space. Moreover, it is believed that SG breathers play an important role in the so-called asymptotic stability problem for the kink solution, see \cite{SW}.

\medskip

Along these pages we will prove that SG
breathers satisfy the same rigidity properties as mKdV breathers (see Sections \ref{Sect:3} and \ref{Sect:4}); but before, in Section \ref{Sect:2}, we will describe a numerical method allowing us to understand 
the spectral properties of the cumbersome linearized operators appearing in the stability analysis of SG breathers. Spectral results of this type are now standard in the literature, but not less important, because they are very useful supporting a rigorous proof. For some previous results dealing with the numerical description of the discrete spectrum of matrix linear operators around (NLS) solitons see \cite{CGNT}, and for the study of the spectral gap problem, see \cite{DS} and references therein. For computing the spectra associated to some lumps in a two dimensional setting, see the recent work \cite{AT}. Our numerical method will be applied to the case of mKdV breathers in Section \ref{Sect:2} (a case study where we know that rigorous results do hold), the more involved Gardner breathers, and more importantly to SG breathers in Section \ref{Sect:4}. Additionally, mimicking Theorem \ref{GBmkdv}, we will show the following variational characterization of SG breathers:

\begin{thm}\label{Thm1p4}
Any pair of SG breather $(B,B_t)$ satisfy the set of matrix-valued, nonlinear equations
\[
\begin{aligned}
& B_{txx} + \frac 18 B_t^3 +\frac 38 B_x^2 B_t -\frac 14 B_t \cos B -a B_t -\frac b2 B_x =0 ,\\
& B_{(4x)} +\frac 38 B_x^2 B_{xx} +\frac 34 B_t B_{tx} B_x +\frac 38 B_t^2 B_{xx} +\frac 58 B_x^2\sin B  -\frac 54 B_{xx}\cos B   \\
& \qquad +\frac 14 \sin B\cos B -\frac 18 B_t^2 \sin B   -a(B_{xx} -\sin B) -\frac b2  B_{tx}=0, \qquad 
\end{aligned}
\]
for a set of constants $a,b$ depending only on some particular parameters of the breather.
\end{thm}

For a complete and detailed statement of the previous result, see Theorem \ref{GB}. As the main consequence of this result, it is possible to construct the associated linearized operator around a breather, and try to compute its spectra. However, a rigorous description of this operator has escaped to us. In the following, we will need two spectral assumptions that are standard in the literature:

\begin{itemize}
\item {\bf Assumption 1:} The kernel of a linearized operator around a breather is nondegenerate and it satisfies the gap condition; and

\medskip

\item {\bf Assumption 2:} There is a unique simple negative eigenvalue associated to this linear operator.
\end{itemize}

(See p. \pageref{A12} for a precise description.)  With a slight abuse in the next definition, we will say that breathers satisfying Assumptions 1 and 2 are 
\emph{spectrally stable}\footnote{We should also restrict ourselves to subcritical nonlinearities.}, since they satisfy the same spectral properties than standard solitons. With this definition on hand, in this paper we prove the following conditional result:

\begin{thm}\label{SG_stab_easy}
Under spectral Assumptions 1 and 2, SG breathers are orbitally stable for small $H^2\times H^1$ perturbations. 
\end{thm}

These two spectral conditions are numerically tested, showing agreement and  compelling evidence for their validity, for any suitable region of parameters, including the cases of low and high velocity breathers, see Figs. \ref{SG_beta} and \ref{SG_betaFixvMov1}. In order to clarify the absence of a rigorous proof for the spectral properties, the main difficulty in proving these two assumptions rests in the fourth order,  coupled, matrix-valued character of the linearized operator around a SG breather solution, which formally leads to consider the understanding of an eight-order linear scalar operator, instead of only dealing with a fourth order operator as in mKdV. A rigorous proof of these facts is an interesting open problem.
 
\medskip

Although by using some IST techniques one could possibly obtain a better resolution in Theorem \ref{SG_stab_easy} (but under additional decay assumptions), working in a variational framework has some particular advantages. First of all, the methods and ideas are ``stable'' under variations of the equation: they also say something about nonintegrable models close to the integrable one, just by comparing their respective variational formulations. One example of this property is the method of proof in \cite{BMW}, based on the B\"acklund transformation, which is hard to extrapolate and make it rigorous for the case of nonintegrable equations.

\medskip

Next, in Section \ref{Sect:5} we will study periodic (in space) breathers, or KKSH breathers \cite{KKS2}. By periodic breather we mean the following:
\begin{defn}[Periodic breather]\label{DEF_B_per}
We say that $B=B(t,x)$ is a (nontrivial) periodic breather solution for a particular one-dimensional dispersive equation if it is an aperiodic breather (see Definition \ref{DEF_B}) that is also periodic in space.
\end{defn}
In the standard mKdV case, these breathers are mainly of the form
\[
\begin{aligned}
B & = B(t,x; \al, \bt, k, m) :=2\sqrt{2} \partial_x \Bigg[ \arctan \Big( \frac{\bt}{\al}\frac{\sn(\al (x+ \delta t),k)}
{\nd(\bt (x+\ga t),m)}\Big) \Bigg], \\
 \delta & := \al^2(1+k) +3\bt^2(m-2), \quad  \ga := 3\al^2(1+k) +\bt^2(m-2),
\end{aligned}
\]
where $\sn(\cdot)$ and $\nd(\cdot)$ are the standard Jacobi elliptic functions, see \eqref{Bper} for a more detailed description and general definition. These breathers can be written using only two parametric variables, say $\bt$ and $k$, and have a characteristic period $L=L(\bt,k)$, with $L\to +\infty$ as $k\to 0$. We will make use of a combined theoretical-numerical approach to conclude why periodic breathers may not be stable in some particular regimes. The spectral analysis of this breather will be performed in Section \ref{Sect:6}, but it perfectly matches the previous SG result: KKHS breathers are indeed spectrally stable for a big set of \emph{valid} parameters. However, in order to obtain  nonlinear stability, an additional assumption will be needed:

\begin{itemize}
\item {\bf Assumption 3:} The following sign condition is satisfied: if $M_\#[B]$ is the mass of the breather solution in terms of $\bt$ and $k$, and $a_1, a_2$ are ``variational'' parameters given in \eqref{a1}-\eqref{a2}, then
\be\label{Weinstein22}
  \frac{\partial_k a_1 \partial_\bt M_\#[B] -\partial_\bt a_1 \partial_k M_\#[B]}{ \partial_{k}a_1 \partial_\bt a_2 -\partial_{k}a_2\partial_{\bt}a_1} >0.
\ee
\end{itemize}

\noindent
(See eqn. \eqref{Cond3} for more details.) This requirement is a sort of generalization of the Weinstein's sign condition for soliton solutions, but it is different from the former because it also considers a certain variation of energy (and not only mass), which has been written above in terms of the mass only. Additionally, this sign condition can be evaluated for non-periodic breathers such as mKdV and SG, and it is trivially satisfied, see e.g. \eqref{Key} for the corresponding computation in the SG case. 

\medskip

Assumption 3 is needed for ensuring that one can ``replace'' the first eigenfunction appearing from Assumption 2 above, which is an instability direction, by the breather itself. The advantage of this replacement comes from the fact that a perturbative dynamics lying in the $L^2$ orthogonal space to the breather allows to control variations of the scalings of the breather, and moreover, their dynamics has formally \emph{quadratic variation} in time, meaning that if the error in our initial data is of order $\eta$, then any bad variation of the scalings will be of order at most $\eta^2$, and therefore negligible for the stability result. 
At the rigorous level, the conservation of mass allows to control the orthogonal direction  to the breather itself in terms of quadratic terms only. 
The procedure to replace the first eigenfunction by the breather is standard, but needs to ensure that a certain denominator is never zero, see e.g. \eqref{Positivity}. Under Assumption 3, such a denominator is never zero. Recall that a 
similar result has been proved in \cite[eqn. (4.24)]{AM}, see \cite{Dyuck,Raphael,Mu3} for more examples in the case of soliton solutions. Adding this new assumption, our method of proof works as in the SG case and we are able to state the following 

\begin{thm}\label{T1p8}
Under spectral Assumptions 1, 2 and 3 above, KKSH breathers are stable for $L$-periodic $H^2$-perturbations. 
\end{thm} 

In Fig. \ref{KKHS_d} we  describe numerically the meaning of Assumption 3. In particular, it is inferred that for $k$ small enough (depending only on $\bt$), Assumption 3 is satisfied. Additionally, note that the condition $k\to 0^+$ is equivalent to take the spatial period $L\to +\infty$, and formally recovering the standard mKdV aperiodic breather.
\footnote{See eqn. \eqref{Largo} for this fact.}

\medskip

Moreover, we believe that the lack or invalidity of Assumption 3 is precisely the source of instability in KKHS breathers, in the sense that if this assumption is not satisfied, then they should be unstable, as it happens in the soliton case in the case of supercritical nonlinearities (see \cite[Lemma 1.6]{Mu3} for example), and just in agreement with the numerical computations performed by Kevrekidis et al \cite{KKS2}, for which \eqref{Weinstein22} is not satisfied.

\medskip

Finally, Section \ref{Sect:7} deals with a new class of mKdV periodic breathers, which have the nonstandard property of being of nonzero mean. 

\begin{thm}\label{T1p9}
Given any parameter $\mu>0$, there exists a periodic in space breather solution $B_\mu$, which is solution of \eqref{mKdV} and satisfies the decomposition $B_\mu =\mu + \widetilde B_{\mu}$, where $\widetilde B_\mu$ has zero mean. 
\end{thm} 

For an explicit formula for $B_\mu$, the reader can consult Definition \ref{NEW_BREATHER}. As it is explained in the last comments of this paper, it also satisfies a proper fourth order elliptic equation.

\medskip

This breather has been conjectured to exist by several works on curvature motion of closed curves on the plane, see e.g. \cite{Ale,Kevre,Khare} and references therein. However, its description does not follow the ideas from \cite{KKS2}, because KKHS breathers are zero mean solutions. Instead we use, in a slightly different way,  the method of proof employed in our work \cite{AM1}, which links zero mean breathers with the corresponding zero solution of the equation. In order to find a breather with a nonzero mean, we use as starting point the nonzero constant solution $\mu$, and then apply twice a suitable B\"acklund transformation, as is done in \cite{AM1}. Since the mean of a modified KdV solution is a conserved quantity, this property will be also preserved by the B\"acklund transformation, leading to the desired solution with the  property sought. Concerning this new breather, in this work we only study its simplest properties, and describe its main differences with KKSH breathers, leaving its deep 
understanding, by length reasons, 
to a forthcoming publication. For the moment, we only advance that these breathers satisfy a suitable elliptic equation, as any other breather in this paper. Additionally, we conjecture that this breather should be as stable as the constant solution $u=\mu$ is for the mKdV periodic dynamics.

\medskip

We also mention that, in addition to Theorems \ref{SG_stab_easy}, \ref{T1p8} and \ref{T1p9}, the more involved problem of asymptotic stability for breathers 
could been also considered, as far as a good and rigorous understanding of the associated spectral problem is at hand. 
Usually, these spectral properties are usually harder to establish than the ones involved in the stability problem (because the convergence problem requires
the use of weighted functions, which destroy most of breather's algebraic properties). In addition, breathers can have zero, positive or negative velocity, 
which means that they do not necessarily decouple from radiation. However, it is worth to mention that if the velocity of a periodic mKdV breather is positive, then there is local strong asymptotic stability in the energy space, see \cite{AM}.

\subsection{Organization of this paper} This paper is organized as follows. In Section \ref{Sect:2} we describe, implement and use standard numerical methods to understand and recover the mKdV and the Gardner spectral problem, and consequently the stability of its corresponding breathers. In Section \ref{Sect:3} we prove that SG breathers satisfy an elliptic system of differential equations, revealing its variational character in the proper space $H^2\times H^1$. Also, in Section \ref{Sect:3}, Theorem \ref{MT}, we prove a conditional stability result for SG breathers, but later (see Section \ref{Sect:4}), the required assumptions will be showed to hold numerically. Section \ref{Sect:5}, and in particular, Theorem \ref{GBp} are devoted to the proof that periodic KKHS breathers satisfy suitable elliptic equations, and we find its variational structure.  In Section \ref{Sect:6}, after some numerical tests, we conjecture (see p. \pageref{Conj1}) that KKHS breathers have a dual stability/instability regime. 
Moreover, assuming the validity of some numerical computations, we show that KKHS are stable in a particular set of parameters. Finally, in Section \ref{Sect:7} we present a new kind of periodic mKdV breather whose main property is the fact that it has nonzero mean. 

\ms

{\bf Acknowledgments.} M.A. would like  to express his gratitude with professors Luis Vega, Panos Kevrekidis and Avinash Khare 
for several enlightening  discussions and  valuable comments. He also would like to thank to the Departamento de Ingenier\'ia Matem\'atica (U. Chile) 
for the support while doing this work. C.M. was partially funded by ERC Blowdisol (France), Fondecyt no. 1150202 Chile, Fondo Basal CMM (U. Chile), 
and Millennium Nucleus Center for Analysis of PDE NC130017. He also would like to thank to the Laboratoire de Math\'ematiques d'{}Orsay 
for his kind hospitality during past years, and where part of this work was completed. J.M.P. was partially funded by Fondecyt no. 1150202 Chile.

%

\section{Numerical description of the spectral problem for the mKdV and Gardner breathers}\label{Sect:2}

\medskip

\subsection{Mathematical background} 
The purpose of this section is to explain how one can recover the discrete spectrum of a self-adjoint operator by using some standard numerical schemes. For the purposes of this paper, we will mainly consider breather solutions, which are exponentially decreasing in space, and therefore well-suited for non-exact numerical schemes. As a case study,
let us consider the mKdV breather defined in \eqref{mKdVB}, for which its spectral properties are rigorously well-known, and breathers are stable (Theorem \ref{MT0}), see \cite{AM}.  Since mKdV is invariant under suitable scaling symmetries (see \eqref{Sol}), we can always assume that $\bt=1$. Similarly, thanks to the invariance by shifts in space, we can also assume $x_2=0$. Therefore, for spectral considerations, we are finally reduced  to the study of the breather profile
\[
B_0(x)   :=  2\sqrt{2} \partial_x \Big[ \arctan \Big( \frac{1}{\al}\frac{\sin(\al (x+x_1))}{\cosh x }\Big) \Big].
\]
Thanks to the periodicity of the sine function, we can also assume $x_1\in [0, \frac{2\pi}{\al}]$. This breather profile is solution to \eqref{H01} with $\bt=1$, and from standard linear arguments \cite{AM}, the associated selfadjoint linear operator 
\be\label{LmKdV}
\begin{aligned}
\mathcal L[z] & := z_{(4x)} -2(1-\al^2) z_{xx} +(1+\al^2)^2 z \\
& \quad +5B_0^2 z_{xx} +10B_0(B_0)_x z_x +\Big[5(B_0)_x^2 +10B_0(B_0)_{xx} +\frac{15}2 B_0^4 -6(1-\al^2)B_0^2 \Big]z,
\end{aligned}
\ee
has an expected two dimensional kernel. More difficult is to prove that $\mathcal L$ has, regardless the values of $\al \in (0,\infty)$ and $x_1\in [0, \frac{2\pi}{\al}]$, an always negative, unique eigenvalue $\la_1(\al,x_1)<0$, which is uniformly in $x_1$ away from zero \cite[Proposition 4.8 and Corollary 4.9]{AM}.  Additionally, the spectrum of $\mathcal L$ is isolated and the continuum spectrum is always positive and away from zero, depending on $\al$ and $\beta$. We would like to recover numerically all the properties before mentioned.

\subsection{Numerical Scheme for mKdV}
In order to obtain a numerical approximation of the eigenvalues of $\mathcal L$, we follow the approach based in the approximate Galerkin method for the Hilbert space $L^2(\R)$. First, we consider the subspace $\mathbb V_N$ generated by a finite dimensional subset of an orthonormal basis of $L^2(\R)$:
\[
\mathbb V_N :=\hbox{span} \{ f_0,f_1,f_2,\ldots, f_N\}\subset L^2(\R).
\]
For practical and computational purposes, the best candidate for an orthonormal basis is the one generated by the Hermite polynomials:
\[
f_n (x) : = p_n(x) e^{-x^2/2}, \quad x\in \R, \quad n\geq 0, \qquad \int_\R f_n^2(x)dx=1.
\]
Here $p_n(x)$ are the Hermite polynomials \cite{Wiki} suitably normalized such that the $f_n$ have unit norm.  

\medskip
\noindent
We will approximate the spectrum of $\mathcal L$ by the spectrum of the finite dimensional operator 
\[
\mathcal L_N:= \mathbb P_N \mathcal L \mathbb P_N,
\]
where $\mathbb P_N$ is the projection operator into the subspace $\mathbb V_N$. It is expected that, as long as $N$ is large enough, 
$\mathcal L_N$  will approach both the discrete and continuum spectrum, in particular the eigenvalues of $\mathcal L$, 
any spectral gap present in the exact operator 
$\mathcal L$, and even the multiplicity of the eigenvalues (seen as two very close approximate eigenvalues).\footnote{Additionally, as far as $N$ is large, the continuous spectrum is also discretely approximated.} 

\medskip
\noindent
The above scheme reduces matters to the understanding the eigenvalues of the matrix  
\[
(\mathcal M_N)_{i,j} := \int_\R f_i \mathcal L f_j, \quad i,j=0,\ldots, N, \quad f_i,f_j \in \mathbb V_N.
\]
Note that this $(N+1)\times (N+1)$ matrix is symmetric, so its eigenvalues are real-valued.  Using some symbolic computations done by Mathematica for some simple cases (periodic mKdV), 
or using Matlab for more demanding and time-consuming cases (aperiodic mKdV), we compute $\mathcal M_N$ and its eigenvalues.
We emphasize that, similarly to the work in \cite{CGNT}, each set of eigenvalues computation took in principle 20-30 minutes, but after some fine code reworking and improvement of the algorithm, we were able to reduce the time of computations to a more realistic value. 
We finish spending approximately 5 minutes when computing a set of eigenvalues for $N=150$ test functions. 
 
\medskip

For the operator \eqref{LmKdV}, it was important to understand the first four eigenvalues, ordered by size. 
We recovered the two-dimensional kernel, along with the unique negative eigenvalue. Additionally, at each computation performed, 
we recovered the spectral gap between the kernel and the bottom of the positive spectrum.\footnote{We compared our results with some standard linear operators that have continuous spectrum $[0,+\infty)$ and resonances at the origin, 
recovering the absence of a spectral gap by having plenty of numerical eigenvalues near zero as $N$ was larger and  larger.}   As it was revealed in the numerical computations, for $\al$ large the code becomes unstable, 
and a larger $N$ is needed to recover the expected results.    

\medskip

First of all, Fig. \ref{Table1} represents our findings for the case of a breather with parameters $\bt=1$, $\al=0.5$, and different values for the
shift $x_1$. In Fig. \ref{mKdV_x1}, we plot the behavior of the
only negative eigenvalue with respect to the shift $x_1$. In order to ensure the best possible approximations, we have chosen $N=160$ Hermite test 
functions to compute the corresponding eigenvalues. 

\begin{figure}[!h]
\begin{center}
 \begin{tabular}{||c c c c c||}
 \hline
 $x_1$ & 1st eig. & 2nd & 3rd & 4th \\ [0.5ex] 
 \hline\hline
 0.09 & -4.226  & -0.028 &  0.028 & 1.776 \\ 
 \hline
 0.81 & -4.191 & -0.039 &  0.081 & 1.862 \\
 \hline
 1.53 & -3.491 & -0.005 & 0.006 & 1.886\\
 \hline
 2.15 & -2.557 & -0.000 & 0.000 & 1.901 \\
 \hline
 3.14 & -1.507 & 0.007 & 0.004 & 1.915 \\ 
 \hline
\end{tabular}
\end{center}\caption{The first four eigenvalues of $\mathcal L$ for $\bt=1$, $\al=0.5$, and $x_1$ varying. 
All computations were made with 
$N=160$ test functions. The third and fourth columns represent the approximate kernel of $\mathcal L$, and the fourth eigenvalue is immediately 
followed by a big amount of closer and larger eigenvalues, which represent a discrete approximation of the continuous spectrum.}\label{Table1} 

\end{figure}

\begin{figure}[!h]
\begin{center}
\includegraphics[width=11cm, height=6cm]{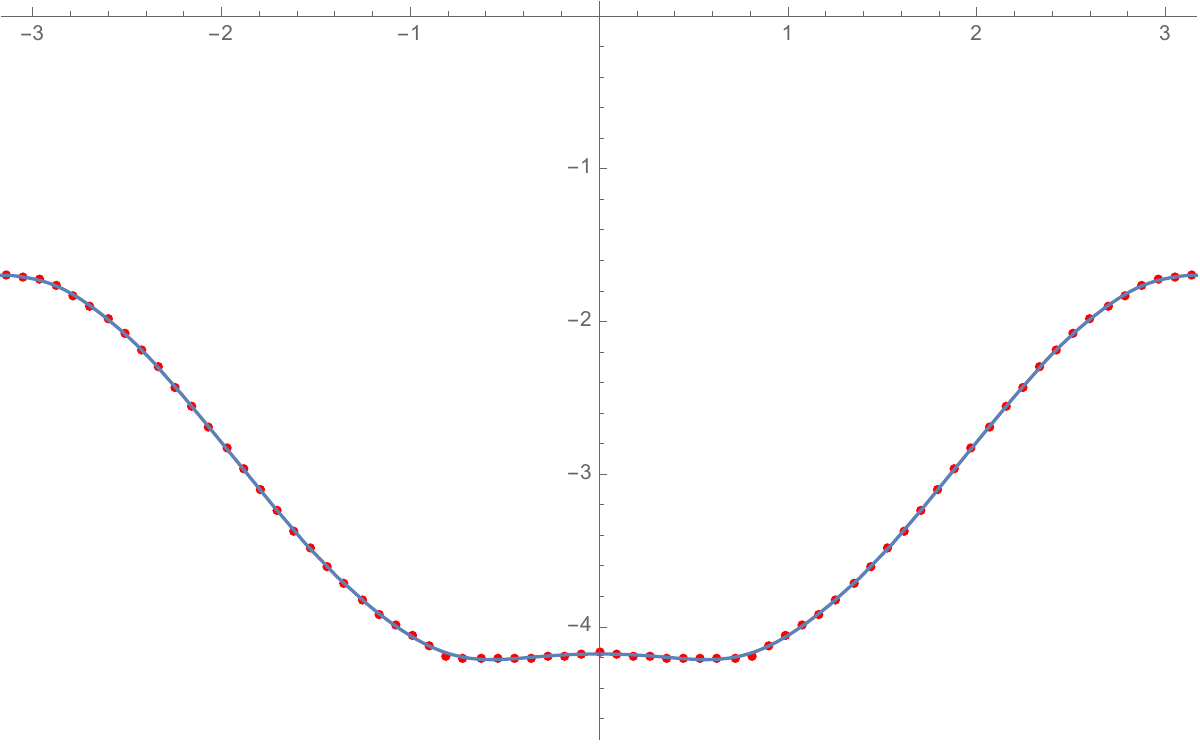}
\caption{Red dots: Variation of the negative eigenvalue of $\mathcal L$ with $\bt=1,$ $\al=0.5$ and $x_1$ varying from -3 to 3. All negative eigenvalues are between -1 and -5. The blue line corresponds to a best polynomial fit of the data. The small upward bump appearing near the center is believed to be error of numerical origin.}\label{mKdV_x1}
\end{center}
\end{figure}

\ms

A second test seeks for the description of the eigenvalues for a higher oscillatory breather, with $\al=1.5$. 
For this case, we obtain the results described in Table \ref{Table12} and Fig. \ref{mKdV_x2}

\begin{figure}[!h]
\begin{center}
 \begin{tabular}{||c c c c c||}
 \hline
 $x_1$ & 1st eig. & 2nd & 3rd & 4th \\ [0.5ex] 
 \hline\hline
0.00 & -22.067  & -0.031 &  -0.000 & 10.043 \\ 
 \hline
 0.99 & -22.571 & -0.000 &  0.000 & 10.086 \\
 \hline
 1.57 & -22.423 &  -0.007 & 0.002 & 10.247\\
 \hline
 2.51 & -22.354 & -0.003 & 0.007 & 10.222 \\
 \hline
 3.14 & -22.571 & -0.000 & 0.000 & 10.084 \\ 
 \hline
\end{tabular}
\end{center}\caption{The first four eigenvalues of $\mathcal L$ for $\bt=1$, $\al=1.5$, and $x_1$ varying. 
All computations were made with $N=164$ test functions. The third and fourth columns represent the approximate kernel of $\mathcal L$.}\label{Table12} 

\end{figure}

\begin{figure}[!h]
\begin{center}
\includegraphics[width=12cm, height=6cm]{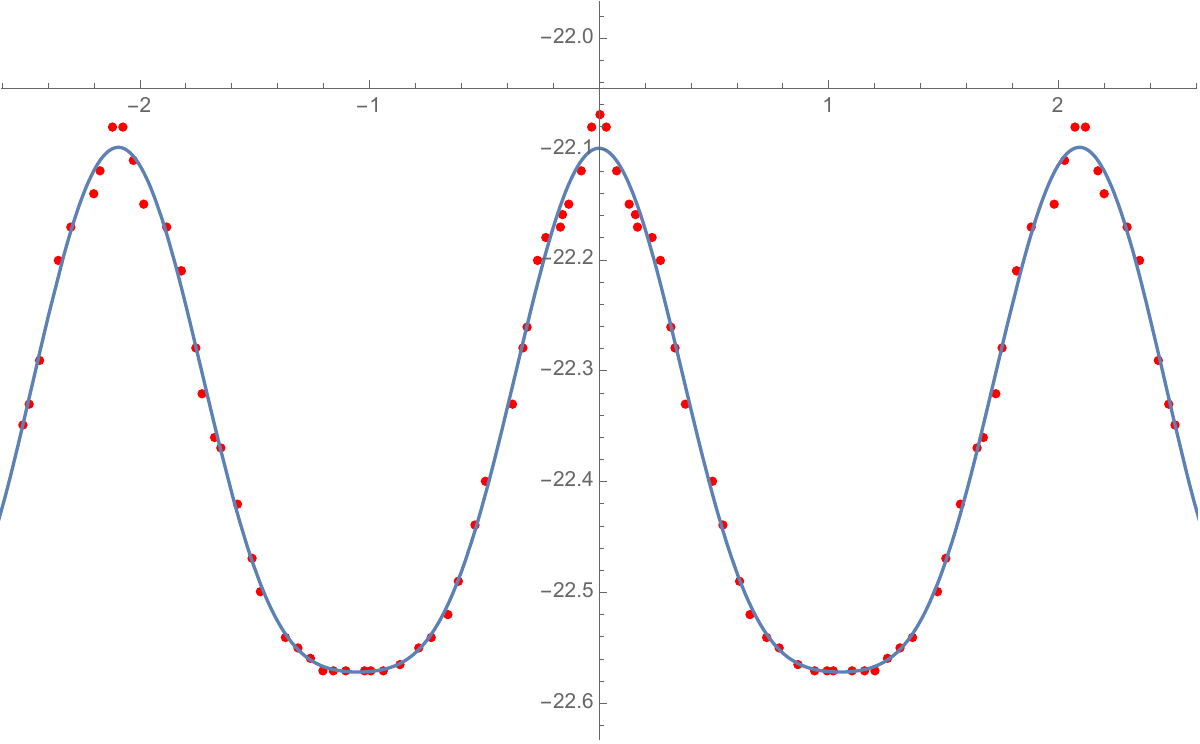}
\caption{Red dots: Variation of the negative eigenvalue of $\mathcal L$ with $\bt=1,$ $\al=1.5$ and $x_1$ varying from -2.6 to 2.6. All data is below -22.05 and above -22.6.
The blue line corresponds to a suitable best polynomial fit of the data, and distance of numerical negative eigenvalues (red points) to this curve are of relative error $\sim 10^{-2}$.}\label{mKdV_x2}
\end{center}
\end{figure}

\ms

For $\al=2.5$, computations become more difficult to perform, due to the oscillatory behavior of the breather solution (see Fig. \ref{mKdV_0}), and $N\geq 160$ is probably needed; an example is the case where $x_1=0.05$ and $x_1=2$, where we performed computations with $N=166$. Our results for the first four eigenvalues are, respectively,
\[
  \{ -71.66, ~ -0.043, ~  0.003, ~28.090 \}, \qquad   \{ -72.335, ~ -0.050, ~  0.055, ~27.926 \}.
\]
We conclude from numerical simulations that $x_1=0$ represents a critical point for the
negative eigenvalue $\la_1(\al,x_1)$, either a maximum or a minimum, depending on the size of the oscillatory scaling $\al$. Additionally, when computing $\la_1$ numerically, some additional, unexpected positive error appears at $x_1=0$ (see e.g. Fig. \ref{mKdV_x2}), that we have not yet fully understood.

\ms

\begin{figure}[!h]
\begin{center}
\includegraphics[width=11.5cm, height=6cm]{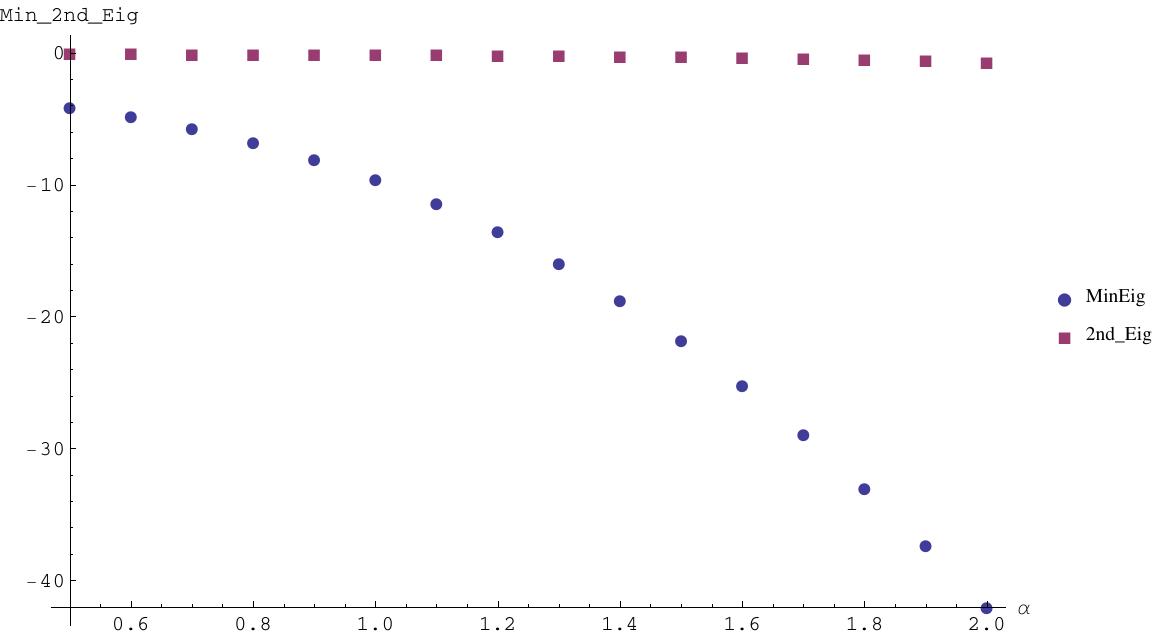}
\caption{Two first minimal eigenvalues of the linearized mKdV operator around the mKdV breather \eqref{mKdVB} with $x_1=0$ and $\beta=1$ while
 $\al$ is varying from $0.5$ to $2.0$. {\color{black} A simple log plot reveals that the absolute value of the minimal eigenvalue depends approximately on the second power of} $\al$.}\label{mKdV_alfmov}
\end{center}
\end{figure}

Finally, we look for the behavior of the minimal eigenvalue with respect to the oscillatory parameter $\al.$ We fix $x_1=0$ and $\beta=1$, 
and we plot the resulting minimum eigenvalue and the second  eigenvalue as a function of $\al.$ We get the results showed in Figure \ref{mKdV_alfmov}.
Note that the larger $\alpha$ is, the less accurate is the second eigenvalue (believed to be zero), but up to $\al=2$, the relative error stays bounded below $10^{-1}$.

\medskip

We can conclude, from the previous numerical analysis, that the rigorous description obtained in \cite{AM} is certainly recovered in a satisfactory way. The problem now is to apply these methods to other more complicated models and breather solutions for which the associated linearized spectral problem  is far from being rigorously characterized.

\subsection{Application to the Gardner case}

For the Gardner equation \eqref{GE}, the breather solution $B_{\al,\bt,\mu}$  can be obtained by using different methods 
(e.g. Inverse Scattering, Hirota method, etc), and it is characterized by the introduction of the parameter $\mu$ that 
controls the quadratic nonlinearity. 
\begin{defn}\label{BreatherGE} Let $\al, \bt,\mu \in \R\backslash\{0\}$ such that $\Delta=\al^2+\bt^2-\frac{2\mu^2}{9}>0$, 
and $x_1,x_2\in \R$. The real breather solution of the Gardner equation \eqref{GE} is given explicitly by the formula
\be\label{GEBre}B_{\al, \bt,\mu}(t,x;x_1,x_2)   
 :=   2\sqrt{2}\partial_x\Bigg[\arctan\Big(\frac{G_{\al, \bt,\mu}(t,x)}
{F_{\al, \bt,\mu}(t,x)}\Big)\Bigg],
\ee
with $y_1$ and $y_2$ defined in \eqref{y1y2}, and
\[
\begin{aligned}
G_{\al, \bt,\mu}(t,x) & :=   \frac{\bt\sqrt{\al^2+\bt^2}}{\al\sqrt{\Delta}}\sin(\al y_1) -\frac{\sqrt2 \mu\bt[\cosh(\bt y_2)+\sinh(\bt y_2)]}{3\Delta},\\
F_{\al, \bt,\mu}(t,x) & :=   \cosh(\bt y_2)-\frac{\sqrt2\mu\bt[\al\cos(\al y_1)-\bt\sin(\al y_1)]}{3\al\sqrt{\al^2+\bt^2}\sqrt{\Delta}},
\end{aligned}
\]
\end{defn}
Gardner breather solutions have also been used  \cite{Ale1} to prove the ill-posedness for the Gardner equation \eqref{GE} in the line, 
showing that solutions cannot depend in an uniformly continuous form on their initial data in the Sobolev spaces $H^s(\R)$  for $s < 1/4$.

\begin{rem}
 Note  that we can take the limit when $\al\rightarrow0$ in \eqref{BreatherGE}, obtaining the \emph{double pole} solution for the Gardner equation \eqref{GE}, 
 which it is a natural generalization of the well-known double pole solution of mKdV. See \cite{AM} for more details.
\end{rem}

For the Gardner case, we have the following result, that we state without proof, but that after some tedious computations, can be  obtained following the proofs in \cite{AM,AM0}:

\begin{thm}\label{GBGE} Let $B_\mu\equiv B_{\al,\bt,\mu}$ be any Gardner breather \eqref{BreatherGE}. Then, for any  fixed $t\in \R$, $B_\mu$ 
satisfies the nonlinear stationary equation  
\bea\label{EcBGE}
& G_\mu[B_\mu]:=& B_{\mu,4x}  -   2(\beta^2 -\alpha^2) (B_{\mu,xx} +\mu B_\mu^2 + B_\mu^3) +  
(\alpha^2 + \beta^2)^2 B_\mu + 5 B_\mu B_{\mu,x}^2 + 5B_\mu^2 B_{\mu,xx}
\nonu\\
&&+ \frac 32 B_\mu^5  + \frac{5}{3}\mu B_{\mu,x}^2  + \frac{10}{3}\mu B_\mu B_{\mu,xx} + \frac{10}{9}\mu^2 B_\mu^3 + \frac 52 \mu B_\mu^4  =0.
\eea
In particular, $B_\mu$ is a critical point of the functional
\bea\label{LyapunovGE}
\mathcal H_\mu[w](t) := F_\mu[w](t) + 2(\bt^2-\al^2)E_\mu[w](t) + (\al^2 +\bt^2)^2 M[w](t),
\eea
where $E_\mu$ and $M$ are defined in \eqref{M1} and \eqref{E2}, and
\be\label{F1ga}
\begin{aligned}
 F_\mu[w](t)  &  :=   \frac 12 \int_\R w_{xx}^2dx -\frac 53\mu\int_\R ww^2_xdx + \frac{5}{18}\mu^2\int_\R w^4dx \\
& \quad   - \frac{5}{2} \int_\R w^2w_x^2dx+ \frac{\mu}{2} \int_\R w^5+ \frac{1}{4} \int_\R w^6dx
\end{aligned}
\ee
is the third conserved quantity for the Gardner equation.
\end{thm}

Note that \eqref{EcBGE} and \eqref{BreatherGE} reduces to \eqref{H01} and \eqref{mKdVB} (up to translations) when $\mu=0$.
Note additionally that the functionals $\mathcal{H}$ and $\mathcal H_\mu$ for the mKdV and Gardner equations are surprisingly the same. 

\medskip

As a consequence of Theorem \ref{GBGE}, and following our ideas from \cite{AM}, it is possible to prove that Gardner breathers are $H^2$-stable:

\begin{thm}[$H^2$-stability of Gardner breathers]\label{T1gardner} Let $\al, \bt,\mu \in \R\backslash\{0\}$ such that $\Delta=\al^2+\bt^2-\frac{2\mu^2}{9}>0$.
There exist positive parameters $\eta_0, A_0$, depending on $\al,\beta$ and $\mu$, such that the following holds.  Consider $u_0 \in H^2(\R)$,
and assume that there exists $\eta \in (0,\eta_0)$ such that 
\be\label{In}
\|  u_0 - B_{\mu}(t=0;0,0) \|_{H^2(\R)} \leq \eta.
\ee
Then there exist $x_1(t), x_2(t)\in \R$ such that the solution $u(t)$ of the Cauchy problem for the Gardner equation 
\eqref{GE}, with initial data $u_0$, satisfies
\be\label{Fn1}
\sup_{t\in \R}\big\| u(t) - B_{\mu}(t; x_1(t),x_2(t)) \big\|_{H^2(\R)}\leq A_0 \eta.
\ee
\end{thm}
Precise estimates on the derivatives of the parameters $x_1(t)$ and $x_2(t)$ are easy to obtain using modulation theory; we left the details to the interested reader. Essential for the proof of Theorem \ref{T1gardner} is the study of the associated linear operator appearing from Theorem \ref{GBGE} and particularly, equation \eqref{EcBGE}.

\begin{defn}
Let $z\in H^4(\R)$, and $B_{\mu}$ be any Gardner breather. We define  $\mathcal L_\mu$ as the 
linearized operator associated to $B_\mu$, as follows: 
\be\label{LGE}
\begin{aligned}
\mathcal L_\mu[z] & := z_{(4x)} -2(\beta^2-\al^2) z_{xx} +(\al^2+\beta^2)^2 z \\
& \quad +5B_\mu^2 z_{xx} +10B_\mu B_{\mu,x} z_x +(5B_{\mu,x}^2 +10B_\mu B_{\mu,xx} +\frac{15}{2} B_\mu^4 -6(\beta^2-\al^2)B_\mu^2)z\\
& \quad + \frac{10}{3}\mu B_{\mu}z_{xx} + \frac{10}{3}\mu B_{\mu,x}z_{x} + \mu(10 B_\mu^3 - 4(\beta^2-\al^2)B_\mu + \frac{10}{3}B_{\mu,xx}
+ \frac{10}{3}\mu B_\mu^2)z.\\
\end{aligned}
\ee
\end{defn}

 \begin{figure}[!h]
\begin{center}
\includegraphics[width=14cm, height=5.5cm]{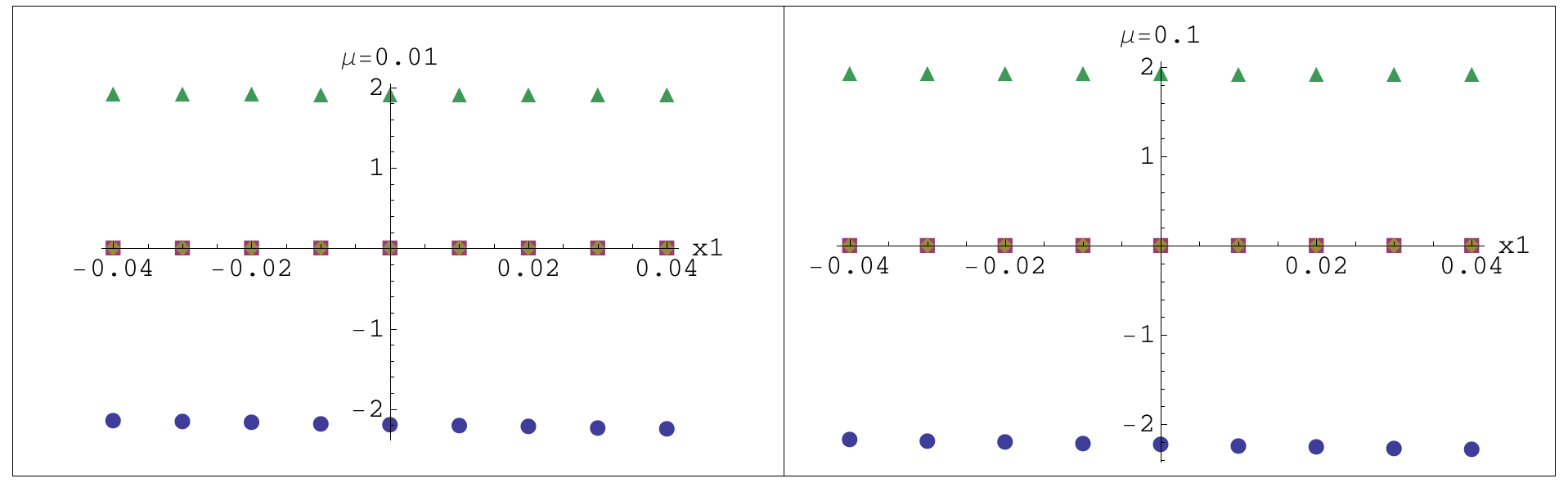}
\caption{The graph of the negative eigenvalue, the double zero kernel and the fourth eigenvalues for the 
linearized operator $\mathcal L_\mu$ around
a Gardner breather, for $\bt=1,~\al=0.5$ and two differents  $\mu=0.01$ and $\mu=0.1$, 
while $x_1$ varying between -0.4 and 0.4. Computations were made with $N=50$ test functions.}\label{muTable}
\end{center}
\end{figure}
 
\medskip

{\small
\begin{figure}[!htb]
\begin{center}
 \begin{tabular}{||c c c c c |c| c c c c||}
 \hline& & $\mu=0.01$  &  & &&  &~~~\hspace{0.9cm}$\mu=0.1$ & & \\ [0.15ex]
 \hline
 $x1$ & 1st eig. & 2nd & 3rd & 4th && 1st eig. & 2nd & 3rd & 4th \\ [0.15ex] 
 \hline
 -0.04 &  -2.244  & $0.0019$ &  0.0032 &  1.905&&-2.283  & $0.0015$ &  0.0027 &  1.919\\ 
 \hline
 -0.03 &  -2.231  & $0.0018$ &  0.0031 &  1.906&& -2.269  & $0.0015$ &  0.0027 &  1.920 \\
 \hline
 -0.02 &  -2.218 & $ 0.0017$ & 0.0030 & 1.907&&-2.255  & $ 0.0014$ & 0.0026 & 1.921\\
 \hline
 -0.01 &  -2.205 & $ 0.0017 $ &  0.0029 & 1.908&& -2.242 & $ 0.0014 $ &  0.0025 & 1.922 \\
 \hline
 0.0 &  -2.192 & $0.0016$ &  0.0028 & 1.909&&-2.228 & $0.0013$ &  0.0024 & 1.923\\ 
 \hline
 0.01 &  -2.179 & $0.0016$ & 0.0027 &  1.910&&-2.214 & $0.0013$ & 0.0023 &  1.924 \\ 
 \hline
 0.02 &  -2.167 & $0.0015$ & 0.0026 & 1.911&& -2.201 & $0.0013$ & 0.0022 & 1.925 \\ 
 \hline
 0.03 &  -2.154  & $0.0015$ & 0.0025 & 1.912&&-2.188  & $0.0012$ & 0.0021 & 1.926\\ 
 \hline
 0.04 &  -2.141 & $0.0014$ &  0.0024 & 1.913&&-2.174 & $0.0012$ &  0.0021 & 1.927\\ 
 \hline%
\end{tabular}
\end{center}\caption{The first four eigenvalues for the linearized operator $\mathcal L_\mu$ around
a Gardner breather for $\bt=1,~ \al = 0.5,~ \mu = 0.01$ and $\mu=0.1$ and $x_1$ varying, as corresponding to Fig. \ref{muTable}. 
All computations were made with $N=50$ test functions. The third and fourth columns (left) and seventh and eigth columns (right) 
represent respectively the approximate kernel of $\mathcal L_\mu$.}\label{TableGar1} 
\end{figure}
}

\medskip

We will apply our previous numerical tests to the understanding of the operator $\mathcal L_\mu$ (Figs. \ref{muTable} and \ref{TableGar1}). First of all, 
we consider two different cases to be studied depending on the values of the parameter $\mu$, with  fixed scalings $\al,\bt$, 
while varying the shift $x_1$.  Firstly, we study the discrete spectra of the linearized operator $\mathcal L_\mu$ 
when parameter $\mu$ is small ($\approx 0.01$) compared to the fixed scalings $\al,\bt$. After that we study the intermediate range
for parameter $\mu\approx 0.1$. Note that the value of $\mu$ as it was indicated in 
\eqref{BreatherGE} is bounded above by the value $3\sqrt{\frac{\al^2+\bt^2}{2}}$ and henceforth this will be the explicit limit value for big $\mu$. 

\bigskip

\section{The Sine-Gordon breather}\label{Sect:3}

\subsection{Preliminaries} One of the most important examples of breather solution is the one obtained from the SG equation \eqref{SG}. For this case, we have the following definition.

\begin{figure}[!h]
\begin{center}
\includegraphics[width=10cm, height=5.5cm]{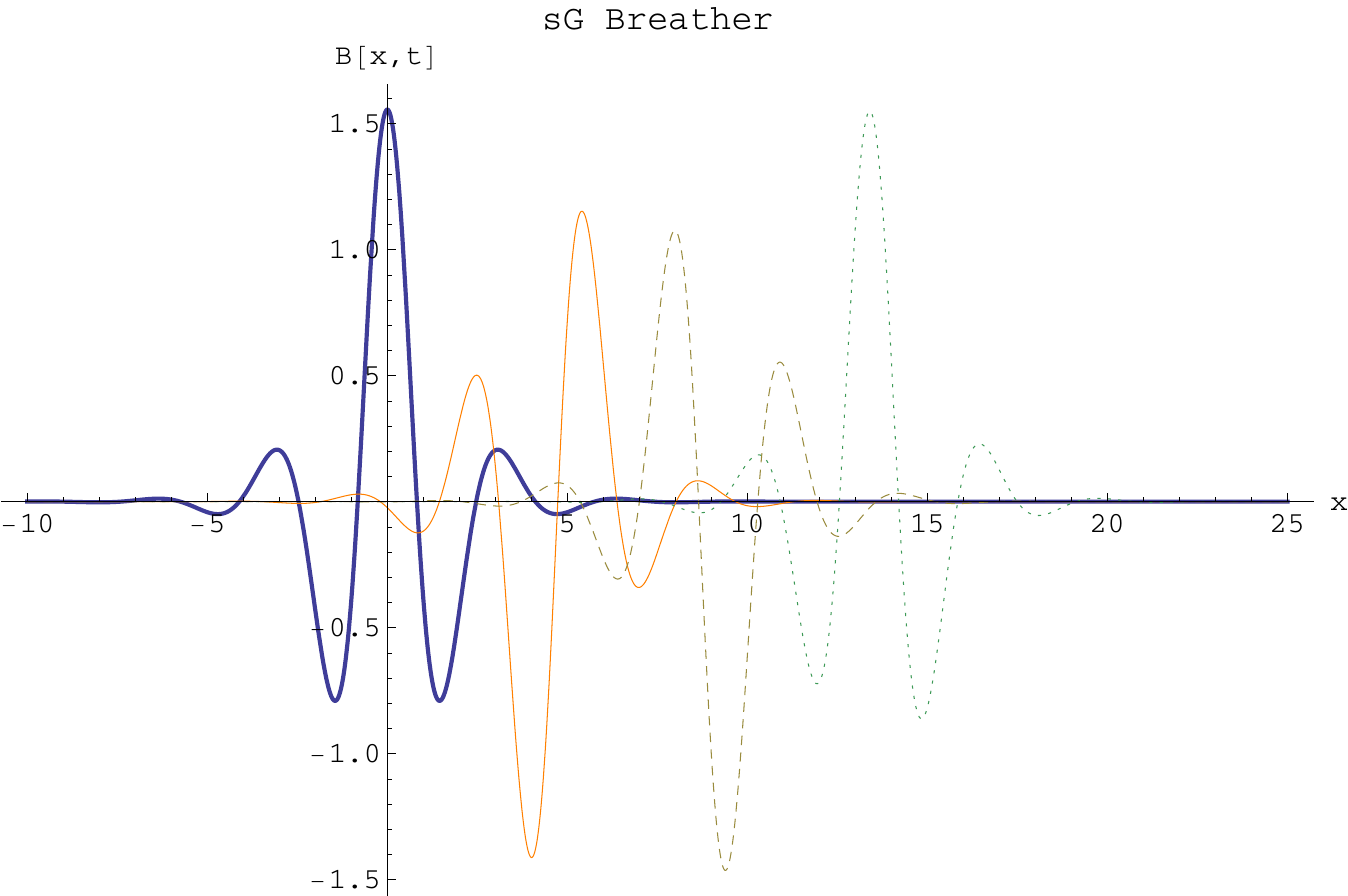}
\caption{Evolution of a Sine-Gordon breather moving rightward in space as time evolves (blue color to green color).}\label{SG_fig}
\end{center}
\end{figure}

\begin{defn}[\cite{La}, see also Fig. \ref{SG_fig}]\label{SGB} Let $v\in (-1,1)$, $\ga:=(1-v^2)^{-1/2}$, $\bt \in (0,\ga)$ and $x_1,x_2\in \R$. 
Any breather solution
of Sine-Gordon \eqref{SG} is given by the expression 
\be\label{BB}
B :=  B_{\bt,v}(t,x;x_1,x_2)   :=  4 \arctan \Big( \frac{\bt}{\al}\frac{\cos(\al y_1)}{\cosh(\bt y_2)}\Big) ,
\ee
with 
\be\label{deltagamma}
y_1 := t -vx + x_1, \quad y_2 := x-vt +x_2,\quad \al := \sqrt{\ga^2 -\bt^2}.
\ee
The parameters $\bt$ and $v$ correspond to the scaling and velocity of the breather, and the case $v=0$ represents a standing SG breather. 
\end{defn}

Note that the SG breather satisfies Definition \ref{DEF_B} with $T=\frac{2\pi}{\al(1-v^2)}>0$ and $L=vT \in \R$. Also, the previous definition takes into account the velocity $v$ of the breather via a Lorentz boost, therefore it is slightly different to the one written in \cite{La}. However, after redefinition of the parameters, it is not difficult to check that they represent the same solution. Additionally, note that in the SG case the two parameters $\al$ and $\bt$ are not independent, unlike the mKdV case. In order to make sense for a suitable Cauchy theory, our previous definition requires additionally a description of the time derivative
of a breather solution. Since we are going to work with several time-dependent parameters, it is certainly necessary to give a precise 
definition of this second nonlinear mode.

\begin{defn} For any $x_1,x_2 \in \R$ fixed,  we define the time derivative of $B$, denoted by $B_t = (B_{\bt,v})_t$, as follows
\be\label{Bt}
B_t (t,x; x_1,x_2) : = -4\al\bt \Bigg[ \frac{ \al  \sin (\al y_1)\cosh (\bt y_2) -\bt v \cos (\al y_1)\sinh (\bt y_2)}
{\al^2 \cosh^2 (\bt y_2) +\bt^2 \cos^2(\al y_1)} \Bigg].
\ee
\end{defn}

We introduce now  some useful notation. Recall that $v\in (-1,1)$, $\ga =(1-v^2)^{-1/2}\geq1$, and $\bt \in (0,\ga)$. Define the parameters 
\be\label{ab}
a:= -\frac 14 + \bt^2 +v^2 (2\ga^4 - \ga^2 +\bt^2), \qquad b:=4v (\ga^4 -\bt^2). 
\ee
Note that $a+\frac 14 >0$ and $b \in \R$. The reader may observe that whenever $v=0$ (the static breather), we have the simplified expressions $a + \frac 14= \bt^2 \in (0,1)$,  and $b=0$.  

\subsection{Variational characterization} As it was announced in the introduction of this work (see Theorem \ref{Thm1p4}), in this paper we will prove the following generalization of \eqref{H01} for the SG case.

\begin{thm}\label{GB} Let $(B, B_t)$ be any SG breather of parameters $v\in (-1,1)$, $\bt \in (0,\ga)$, and $a,b$ as in (\ref{ab}).
Then, for any  fixed $t\in \R$, $(B,B_t)$ satisfy the nonlinear equations
\be\label{EcB2}
B_{txx} + \frac 18 B_t^3 +\frac 38 B_x^2 B_t -\frac 14 B_t \cos B -a B_t -\frac b2 B_x =0 ,
\ee
and
\bea\label{EcB}
B_{(4x)} +\frac 38 B_x^2 B_{xx} +\frac 34 B_t B_{tx} B_x +\frac 38 B_t^2 B_{xx} +\frac 58 B_x^2\sin B  -\frac 54 B_{xx}\cos B  & &  \nonu \\
+\frac 14 \sin B\cos B -\frac 18 B_t^2 \sin B   -a(B_{xx} -\sin B) -\frac b2  B_{tx}=0. \qquad 
\eea
\end{thm}

\begin{proof}[Proof of \eqref{EcB}, assuming \eqref{EcB2}]
Taking time derivative in  equation (\ref{EcB2}), and replacing $B_{tt}$ using (\ref{SG}), we have
\[
\begin{aligned}
0 & = (B_{tt})_{xx} +\frac 38B_t^2 B_{tt} +\frac 34 B_t B_x B_{tx} + \frac 38 B_x^2 B_{tt}  
 \quad -\frac 14 B_{tt} \cos B +\frac 14 B_t^2 \sin B -a B_{tt} -\frac b2 B_{tx}\\
& =  (B_{xx}-\sin B)_{xx} +\frac 38B_t^2 (B_{xx}-\sin B) +\frac 34 B_t B_x B_{tx} + \frac 38 B_x^2 (B_{xx}-\sin B) \\
& \quad  -\frac 14 (B_{xx}-\sin B) \cos B +\frac 14 B_t^2 \sin B -a (B_{xx}-\sin B) -\frac b2 B_{tx}\\
& = B_{(4x)} -(B_x \cos B)_{x} +\frac 38B_t^2B_{xx} -\frac 38 B_t^2\sin B  +\frac 34 B_t B_x B_{tx} + \frac 38 B_x^2 B_{xx} -\frac 38B_x^2 \sin B \\
& \quad  -\frac 14 B_{xx} \cos B + \frac 14 \sin B \cos B +\frac 14 B_t^2 \sin B -a (B_{xx}-\sin B) -\frac b2 B_{tx}\\
& = B_{(4x)} +\frac 38 B_x^2 B_{xx} +\frac 34 B_t B_{tx} B_x +\frac 38 B_t^2 B_{xx} +\frac 58 B_x^2\sin B  -\frac 54 B_{xx}\cos B    \\
& \quad  +\frac 14 \sin B\cos B -\frac 18 B_t^2 \sin B   -a(B_{xx} -\sin B) -\frac b2  B_{tx} = \hbox{ l.h.s. of }  \eqref{EcB}.
\end{aligned}
\]

\medskip

\noindent
\emph{Proof of \eqref{EcB2}}. We use the specific structure of the breather. From \eqref{Bt} and \eqref{BB} we have
\[
B_t = -4\al\bt \frac{ \al  \sin (\al y_1)\cosh (\bt y_2) -\bt v \cos (\al y_1)\sinh (\bt y_2)}{\al^2 \cosh^2 (\bt y_2) +\bt^2 \cos^2(\al y_1)},
\]
and
\be\label{Bx}
B_x= 4\al\bt\frac{  v\al \sin (\al y_1)\cosh (\bt y_2)  -\bt  \cos (\al y_1)\sinh (\bt y_2)}{\al^2 \cosh^2 (\bt y_2) +\bt^2 \cos^2(\al y_1)}.
\ee
From here we have
\bee
\frac 38B_x^2 B_t & =&  \frac{-24\al^3\bt^3}{(\al^2 \cosh^2 (\bt y_2) +\bt^2 \cos^2(\al y_1))^3}  [ v\al \cosh (\bt y_2) \sin (\al y_1) -\bt  
\sinh (\bt y_2)\cos (\al y_1)]^2 \times \\
& &  \qquad \times[ \al \cosh (\bt y_2) \sin (\al y_1) -\bt v \sinh (\bt y_2)\cos (\al y_1)] \\
& =&  \frac{-24\al^3\bt^3 \, h_1(t,x)}{(\al^2 \cosh^2 (\bt y_2) +\bt^2 \cos^2(\al y_1))^3},
\eee
with
\bee
h_1(t,x) & :=& \al^3 v^2 \cosh^3(\bt y_2) \sin^3(\al y_1) -\bt^3 v \sinh^3(\bt y_2) \cos^3(\al y_1) \\
& & - v(2+v^2)\al^2\bt \sinh(\bt y_2) \cosh^2(\bt y_2) \sin^2(\al y_1) \cos(\al y_1) \\
& & +(1+2v^2)\al \bt^2 \sinh^2(\bt y_2) \cosh(\bt y_2) \sin(\al y_1) \cos^2(\al y_1).  
\eee
Similarly,
\[
\frac 18 B_t^3  =  \frac{-8\al^3\bt^3\, h_2(t,x)}{(\al^2 \cosh^2 (\bt y_2) +\bt^2 \cos^2(\al y_1))^3},
\]
with 
\bee
h_2(t,x) & :=&   \al^3 \cosh^3 (\bt y_2) \sin^3 (\al y_1)   -3\al^2\bt v \cosh^2 (\bt y_2) \sin^2 (\al y_1)\sinh (\bt y_2)\cos (\al y_1) \\
& &    + 3   \al \bt^2 v^2 \cosh (\bt y_2) \sin (\al y_1)\sinh^2 (\bt y_2)\cos^2 (\al y_1) -\bt^3 v^3 \sinh^3 (\bt y_2)\cos^3 (\al y_1).
\eee
We compute now the term $B_{txx}$. From (\ref{Bt}) we have
\[
B_{tx} = \frac{ -4\al\bt \, h_3(t,x)}{(\al^2 \cosh^2 (\bt y_2) +\bt^2 \cos^2(\al y_1))^2},
\]
where
\bea\label{h3}
h_3(t,x) & :=& [ \al \bt (1-v^2) \sin(\al y_1)\sinh(\bt y_2) -v\ga^2 \cos(\al y_1) \cosh(\bt y_2)  ]\times \nonu\\
& &  \qquad \times [ \al^2 \cosh^2(\bt y_2) + \bt^2 \cos^2(\al y_1)]  \nonu \\
& & -2 \al \bt [\al \sin(\al y_1) \cosh(\bt y_2) - \bt v \cos(\al y_1) \sinh(\bt y_2)]\times  \nonu\\
& & \qquad \times [\al \sinh(\bt y_2)\cosh(\bt y_2) + v\bt \sin(\al y_1)\cos(\al y_2)].
\eea
Similarly, after a long computation,
\[
B_{txx} = \frac{-4\al\bt \, h_4(t,x) }{(\al^2 \cosh^2 (\bt y_2) +\bt^2 \cos^2(\al y_1))^3},
\]
where
\bea\label{h4}
h_4(t,x) & :=&     [ \al^2 \cosh^2(\bt y_2) + \bt^2 \cos^2(\al y_1)] (h_3)_x  \nonu\\
& & - 4 \al \bt [\al \sinh(\bt y_2)\cosh(\bt y_2) + v\bt \sin(\al y_1)\cos(\al y_2)] h_3.
\eea
On the other hand, using the well-known formulae
\[
\cos (4\theta) = 1-8\sin^2 \theta + 8\sin^4 \theta, \quad \sin^2 \theta =\frac{\tan^2 \theta}{1+\tan^2\theta},
\]
we have
\[
\cos (4\arctan s) = \frac{ (s^2 -2s-1)(s^2 +2s -1) }{(1+s^2)^2} = \frac{s^4 -6s^2 +1}{(1+s^2)^2}, \quad s\in \R.
\]
Therefore, from (\ref{BB}) we obtain
\be\label{coB}
\cos B = \frac{\bt^4 \cos^4(\al y_1) -6\al^2\bt^2 \cosh^2(\bt y_2)\cos^2(\al y_1) +\al^4 \cosh^4(\bt y_2)}{(\al^2 \cosh^2 (\bt y_2) +\bt^2 \cos^2(\al y_1))^2},
\ee
and thus
\[
\begin{aligned}
-\frac 14 B_t \cos B &= \frac{\al\bt }{(\al^2 \cosh^2 (\bt y_2) +\bt^2 \cos^2(\al y_1))^3}  [ \al \cosh (\bt y_2) \sin (\al y_1) -\bt v \sinh (\bt y_2)\cos (\al y_1)] \times \\
&   \times[\bt^4 \cos^4(\al y_1) -6\al^2\bt^2 \cosh^2(\bt y_2)\cos^2(\al y_1) +\al^4 \cosh^4(\bt y_2)] \\
& =  \frac{\al\bt \, h_5(t,x) }{(\al^2 \cosh^2 (\bt y_2) +\bt^2 \cos^2(\al y_1))^3},
\end{aligned}
\]
where 
\bee
h_5(t,x) & :=&  \al \bt^4 \sin(\al y_1) \cos^4(\al y_1) \cosh(\bt y_2)   -\al^4 \bt v \cos(\al y_1) \sinh(\bt y_2) \cosh^4(\bt y_2)  \\
& & +\al^5 \sin(\al y_1) \cosh^5(\bt y_2) -\bt^5 v \cos^5(\al y_1) \sinh(\bt y_2)  \\
& &   + 6\al^2 \bt^3 v \cos^3(\al y_1) \sinh(\bt y_2) \cosh^2(\bt y_2)  -6 \al^3 \bt^2 \sin (\al y_1) \cos^2(\al y_1) \cosh^3(\bt y_2).  
\eee
Collecting the above identities, we get
\bea\label{Concl1}
  B_{tx}+ \frac 18 B_t^3 +\frac 38 B_x^2 B_t -\frac 14 B_t \cos B   = \al \bt \frac{[-4h_4 - 8\al^2 \bt^2 h_2 -24\al^2 \bt^2 h_1 + h_5] }
  {(\al^2 \cosh^2 (\bt y_2) +\bt^2 \cos^2(\al y_1))^3}.
\eea
In the following lines, we simplify the numerator in the previous expression. In order to carry out this computation, the key point will 
be the denominator, denoted by $g$:
\be\label{g}
g(t,x):= \al^2 \cosh^2 (\bt y_2) +\bt^2 \cos^2(\al y_1).
\ee
First of all, note from (\ref{h3}) that $h_3$ obeys the unique decomposition
\[
h_3 =: h_{31}g - h_{32}g_x, \quad g_x =2\al \bt [\al \sinh(\bt y_2)\cosh(\bt y_2) + v\bt \sin (\al y_1) \cos(\al y_1)].
\] 
Therefore, from (\ref{h4}),
\bea
h_4 & =&  g[h_{31}g - h_{32}g_x]_x  - 2g_x [h_{31}g - h_{32} g_x] \nonu\\
& =& g^2 (h_{31})_x + g[h_{31} g_x - (h_{32}g_x)_x - 2h_{31}g_x ]  + 2h_{32} g_x^2 \nonu \\
& =& g^2 (h_{31})_x - (h_{31}g_x +   (h_{32}g_x)_x) g + 2h_{32} g_x^2.   \label{h42}
\eea
On the other hand,
\[
\begin{aligned}
  -8\al^2 \bt^2 ( h_2 + 3h_1)  & =  -8\al^5 \bt^2 (1+3v^2)  \sin^3 (\al y_1) \cosh^3 (\bt y_2)   + 8\al^2 \bt^5 v (3 + v^2) \cos^3 (\al y_1) \sinh^3 (\bt y_2) \\
&   \quad  + 24 \al^4 \bt^3 v (3 +v^2) \sin^2 (\al y_1) \cos (\al y_1) \sinh (\bt y_2) \cosh^2 (\bt y_2)  \\
&  \quad - 24  \al^3  \bt^4  (3v^2+1 )  \sin (\al y_1) \cos^2 (\al y_1)\sinh^2 (\bt y_2)\cosh (\bt y_2). 
\end{aligned}
\]
Therefore, after some simplifications,
\bea\label{A1}
& &  h_5 -8\al^2 \bt^2 ( h_2 + 3h_1)  - 8h_{32} g_x^2   = \nonu\\
& & =  g \Big[  \al^3(1-32\bt^2) \sin (\al y_1) \cosh^3(\bt y_2)   -\al^2 \bt v (1-32\bt^2) \cos(\al y_1) \sinh(\bt y_2) \cosh^2 (\bt y_2) \nonu \\
& & \qquad - \bt^3 v (1+32\al^2v^2) \cos^3(\al y_1) \sinh(\bt y_2) +  \al \bt^2  (1+32\al^2v^2) \sin(\al y_1)\cos^2(\al y_1) \cosh(\bt y_2) \nonu\\
& & \qquad +24\al^3 \bt^2 (1-v^2) \sin(\al y_1) \cosh(\bt y_2)  -24\al^2 \bt^3v (1-v^2) \cos(\al y_1) \sinh(\bt y_2) \Big].
\eea
On the other hand, we consider the second term in (\ref{h42}). We have
\bea\label{A2}
& &  4(h_{31}g_x +   (h_{32}g_x)_x) g  = \nonu\\
& & =  8\al\bt  g \Big[Ê 2\al^2 \bt (2-v^2) \sin (\al y_1) \cosh^3(\bt y_2) -3\al^2 \bt (1-v^2)\sin(\al y_1) \cosh(\bt y_2) \nonu\\
& & \quad \quad  -2\al v(\bt^2 +\ga^2) \cos(\al y_1) \sinh(\bt y_2) \cosh^2(\bt y_2)   +3\al \bt^2 v (1-v^2) \cos(\al y_1) \sinh(\bt y_2)  \nonu\\
& & \qquad \quad +2\al \bt^2 v(2v^2-1)\cos^3(\al y_1)\sinh(\bt y_2)  -2\bt v^2 (\al^2 +\ga^2) \sin(\al y_1) \cos^2(\al y_1) \cosh(\bt y_2) \Big]. \nonu\\
& & 
\eea
Collecting the last two estimates, we obtain
\bea\label{A3}
(\ref{A1}) + (\ref{A2}) =  g^2 [ \al(1-16\bt^2 v^2) \sin(\al y_1) \cosh(\bt y_2)   -\bt v (1+16\al^2) \cos(\al y_1) \sinh (\bt y_2)].
\eea
We finally add the first term in (\ref{h42}), such that 
\bee
-4 g^2 (h_{31})_x +  (\ref{A3}) & =&   \al \bt g^2 [ \al(1 + 4\al^2 v^2 -8 \bt^2 v^2 -4 \bt^2) \sin(\al y_1) \cosh(\bt y_2) \\
& & \quad - v\bt (1+ 8\al^2 -4\bt^2 +4\al^2 v^2)\cos(\al y_1) \sinh(\bt y_2)] .
\eee
Comparing with (\ref{Bt}) and (\ref{Bx}), in order to obtain (\ref{EcB2}) we have
\[
(\ref{Concl1}) = a B_t + \frac b2 B_x,
\]
provided $a$ and $b$ are chosen as in (\ref{ab}). The proof is complete.
\end{proof}

\begin{rem}
The structure of equations \eqref{EcB2}-\eqref{EcB} is inherent to solutions of \eqref{SG}. For example, the static kink solution 
$(\varphi(x), 0)$ defined in \eqref{SolSG} is also a solution of the same set of equations, no matter what are $a$ and $b$.  In some sense, the proof of  \eqref{EcB2} is independent of the breather itself, but the proof of \eqref{EcB} explicitly involves the structure of the breather. 
\end{rem}

Although the parameters $x_1,x_2$ are chosen independent of time, a simple argument ensures that the previous lemma still hold under time dependent, translation parameters $x_1(t)$ and $x_2(t)$. 

\begin{cor}\label{Eqtt}
Let $(B^0,B^0_t)$ be any SG breather as in (\ref{BB}), and $x_1(t), x_2(t)\in \R$ two continuous functions, defined for all $t$ in a given interval $I$. Consider the modified breather
\[
( B,B_t) (t,x):=(B^0,B^0_t)(t,x; x_1(t),x_2(t)), \qquad (\hbox{cf. \eqref{BB}}). 
\]
Then $(B,B_t)$ satisfy (\ref{EcB2}) and (\ref{EcB}), for all $t$ in the considered interval $I$.
\end{cor}

\begin{proof}
A direct consequence of the invariance of equations (\ref{EcB2}) and (\ref{EcB}) under translations in the variables $x_1$ and $x_2$.
\end{proof}

Let $(u_0,u_1)\in H^2(\R)\times H^1(\R)$ and $(u(t),u_t(t)) \in H^2(\R)\times H^1(\R)$ be the corresponding local in time solution of
the Cauchy problem associated to (\ref{SG}), with initial condition $(u(0), u_t(0))=(u_0,u_1)$, to be specified later (cf. e.g. Bourgain \cite{B}).  Let us consider the second order conserved quantity for \eqref{SG}: 
\bea\label{F1}
F[u,u_t](t) \!  &  := & \! \frac 12 \int_\R (u_{xx}^2 + u_{tx}^2)(t,x)dx -\frac 1{32}\int_\R (u_t^4 + u_x^4)(t,x)dx 
-\frac 3{16}\int_\R u_t^2(t,x)u_x^2(t,x)dx \nonu \\ 
& &  +\frac 5{8}\int_\R u_x^2(t,x) \cos u(t,x)dx  +\frac 18 \int_\R (\sin^2 u(t,x) + u_t^2(t,x) \cos u(t,x))dx.
\eea
For this functional we have the following simple result.

\begin{lem}\label{dF10} Given $(u,u_t)(t)$, local $H^2\times H^1$-solution of (\ref{SG}) with initial data $(u_0,u_1)$ in the same  vector space, the functional $F[u,u_t](t)$ is a conserved quantity.  In particular, $(u,u_t)$ is a global-in-time $H^2\times H^1$-solution.
\end{lem}
At this point we make use of the integrability of the equation, since a quantity as $F$ it is not present in a general, 
non-integrable scalar field equation. The verification of Lemma \ref{dF10} is a tedious but direct computation.

\medskip

Now we present some important consequences of Theorem \ref{GB} and Lemma \ref{dF10}. These results imply that there is a suitable Lyapunov functional that describes the local $H^2\times H^1$ dynamics of SG breathers. Indeed, define the functional $\mathcal H$ as the following linear combination of $F$, the energy $E$ in \eqref{E} and the momentum $P$ from \eqref{P}:
\be\label{H1}
\mathcal{H}[u,u_t] = F[u,u_t] + a E[u,u_t] +  b P[u,u_t],
\ee
where $a$ and $b$ are the constants previously introduced in \eqref{ab}.  Clearly the functional $\mathcal H$ is conserved along the flow. We have then the following direct result, proven in Appendix \ref{B}.

\begin{cor}\label{corSG}
SG breathers are critical points of $\mathcal H$.  
\end{cor}

\subsection{A rigorous statement for Theorem \ref{SG_stab_easy}} The main consequence of previous results is the possibility of proving stability, independent of how complicated breather dynamics can be. After a suitable understanding of the associated linearized problem associated to \eqref{EcB2}-\eqref{EcB}, we are able to prove the following conditional result:

\begin{thm}\label{MT}
Under Assumptions 1 and 2 in page \pageref{A12}, SG breathers are stable under small $H^2\times H^1$ perturbations. More precisely, there are $\eta_0>0$ and $K_0>0$, only depending on $\bt$, such that if $0<\eta<\eta_0$ and
\be\label{Initial_Condition}
\| (u_0,u_1) - (B_{\bt,v}, (B_{\bt,v})_t)(t=0,\cdot ; 0, 0)    \|_{H^2\times H^1} <\eta,
\ee
then there are real-valued parameters $x_1(t)$ and $x_2(t)$ for which the global $H^2\times H^1$-solution $(u,u_t)(t)$ of \eqref{SG} with initial data $(u_0,u_1)$ satisfies
\[
\sup_{t\in \R} \| (u,u_t)(t) - (B_{\bt,v}, (B_{\bt,v})_t)(t,\cdot ; x_1(t), x_2(t))    \|_{H^2\times H^1} <K_0\eta,
\]
with similar estimates for the derivatives of the shift parameters $x_1,~x_2$.
\end{thm}

A mathematical proof of Assumptions 1 and 2 has escaped to us, even in the case $v=0$,\footnote{Although we have a proof for the case where $v=x_1=x_2=0$ and $\bt=\frac12$, however this particular case is not sufficient to describe correctly a general dynamics, where nonzero shifts are always present.} mainly due to the matrix, highly coupled, fourth-order character of the linearized system associated to SG breathers. However, as explained in the Introduction, in this paper we present compelling numerical evidence revealing that Assumptions 1 and 2 do hold for any breather solution, regardless its size, initial velocity or position, see Figs. \ref{SG_beta}, \ref{SG_betaFixvMov1} and \ref{TableBetaFixVmov}. 

\ms

The proof of Theorem \ref{MT} follows the ideas of our proof in \cite{AM}, with some interesting changes due to the matrix-valued character of the solution. The proof is decomposed in several steps, and it is finally done in Section \ref{8}.

\begin{rem}
We find that Theorem \ref{MT} is in fact a surprise, since from Grillakis-Shatah-Strauss \cite{GSS},  it was expected that real-valued, 
non topological solutions to scalar field equations were certainly unstable. However, the fact that breathers have no sign, do not 
satisfy a simple second order ODE, and the more important fact that the equation is completely integrable reveal deep obstructions 
to a more general behavior of solutions.  We emphasize that our results above are independent of the velocity $v$: standing and highly 
relativistic breathers should be both stable.
\end{rem}

\begin{rem}
Looking for previous contributions to the stability problem we have found the work by Ercolani, Forest and McLaughlin \cite{EFM}, where a
sketch of the stability proof is presented. To be more precise, their argument states that, by using the inverse scattering theory applied 
to a small perturbation of the breather, the corresponding solution to \eqref{SG} must remain uniformly close in time to a modulated breather
solution. However, it is important to stress that any rigorous argument involving the inverse scattering method requires a \emph{nontrivial amount of decay}
on the initial data, an assumption that is not needed in our case. 
\end{rem}

The purpose of the next Section is to give the main ingredients for the Proof of Theorem \ref{MT}. In a first subsection, we prove some energy and momentum identities,  allowing
us to show that, at least formally, there is only one negative direction (see Definition \ref{Neg_dir} for a rigorous explanation 
of this term), 
and moreover, it is possible to control such a bad direction by using only the breather $(B,B_t)$ as replacement.

\section{Mathematical Analysis of the SG breather}\label{Sect:4}

\ms

\subsection{Energy identities} In the next lines, we summarize some well known identities for the energy and momentum of a fixed breather. For the sake of completeness, we give full proofs of these results. Recall the definition of breather $(B,B_t)$ in \eqref{BB}-\eqref{Bt}.

\begin{lem} 
Let $(B,B_t)$ be any SG breather, with parameters for $\bt>0$, $v\in (-1,1)$, and $x_1,x_2\in \R$. Let $P[B,B_t]$ be the momentum of a breather defined in (\ref{P}). Then
\be\label{Mom_B}
P[B,B_t](t)= -8\beta v.
\ee
\end{lem}

\begin{proof}
From (\ref{Bt}) and (\ref{Bx}), we have
\[
P[B,B_t] = -8\al^2\bt^2 \int_\R \frac{h_p(t,x)}{(\al^2 \cosh^2(\bt y_2) + \bt^2 \cos^2 (\al y_1))^2},
\]
where $h_p(t,x)$ is given by the expression
\bee
h_p(t,x)& :=& \al^2 v  \sin^2(\al y_1)\cosh^2(\bt y_2)   -\al\bt (1+v^2) \sin(\al y_1)\cos(\al y_1) \sinh(\bt y_2)\cosh(\bt y_2) \\
& &  +\bt^2 v  \cos^2(\al y_1)\sinh^2(\bt y_2) .
\eee
Now the purpose is to use double angle formulae to avoid the squares. More precisely, it is well known that
\be\label{double1}
\cos^2(\al x) =\frac 12 (1+\cos (2\al x)), \quad \sin^2(\al x) =\frac 12 (1-\cos (2\al x)),
\ee
and
\be\label{double2}
\cosh^2 (\bt x) =\frac 12 (1+\cosh(2\bt x)), \quad \sinh^2(\bt x) =\frac 12 (\cosh (2\bt x)-1).
\ee
We replace these identities in the previous expression above. We obtain
\[
P[B,B_t] = -8\al^2\bt^2 \int_\R \frac{h_p(t,x)}{\Big(\frac{\al^2 + \bt^2}{2} + \frac{\al^2}{2}\cosh(2\bt y_2) + \frac{\bt^2}{2}\cos(2\al y_1)\Big)^2},
\]
where now $h_p(t,x)$ is rewritten  as
\bee
h_p(t,x)& :=& \frac v4 (\al^2 -\bt^2)  + \frac v4 (\al^2 + \bt^2) \cosh(2\bt y_2) - \frac v4(\al^2 + \bt^2) \cos(2\al y_1)\\
& & -\frac v4(\al^2 - \bt^2) \cos(2\al y_1)\cosh(2\bt y_2)   -\frac 14 \al\bt (1+v^2) \sin(2\al y_1)\sinh(2\bt y_2).
\eee
Let $f_p(t,x) := \frac 1{4\al\bt} [\bt \sin(2\al y_1) + v\al \sinh(2\bt y_2)]$. Then
\[
(f_p)_x(t,x) =  \frac 12v [\cosh(2\bt y_2) -\cos(2\al y_1) ].
\]
Recall the definition of $g$ given in (\ref{g}) but rewriten in terms of the double angle formulas \eqref{double1},\eqref{double2}
\be\label{defG}
g(t,x) := \frac12 (\al^2 + \bt^2) + \frac{\al^2}{2}\cosh(2\bt y_2) + \frac{\bt^2}{2}\cos(2\al y_1).
\ee
Then $(f_p)_x(t,x) g(t,x) - g_x(t,x) f_p(t,x) = h_p(t,x).$ From this identity we finally obtain
\[
P[B,B_t] = -8\al^2 \bt^2 \int_\R \Big( \frac{f_p}{g}\Big)_x(t,x)dx = -8\bt v.
\]
\end{proof}
We compute now the energy of a breather. See Lamb \cite{La} for a similar result in the case where $v=0$. 

\begin{lem}\label{ME} Let $(B,B_t)$ be any SG breather of parameters for $\bt>0$, $v\in (-1,1)$, and $x_1,x_2\in \R$. Then
\be\label{MassB}
E[B,B_t](t) =16\bt .
\ee
Moreover, this result does not change if we replace $x_1$ and $x_2$ by time dependent parameters.\footnote{But now $(B,B_t)$ ceases being an exact solution of SG.}
\end{lem}
\begin{proof}
This is a classical result. From (\ref{Bt}) and (\ref{Bx}), we have
\bee
\frac 12 (B_t^2 + B_x^2) & =&  \frac{8\al^2 \bt^2}{(\al^2 \cosh^2 (\bt y_2) +\bt^2 \cos^2(\al y_1))^2} \times \\
& & \quad \times \Big[  (1+v^2)\al^2 \cosh^2 (\bt y_2) \sin^2 (\al y_1)  + (1+v^2)\bt^2  \sinh^2 (\bt y_2)\cos^2 (\al y_1)  \\
&& \qquad  -4\al\bt v \sin(\al y_1) \cos(\al y_1) \sinh(\bt y_2) \cosh(\bt y_2) \Big].
\eee
On the other hand, from (\ref{coB}),
\[
1-\cos B = \frac{ 8\al^2\bt^2 \cosh^2(\bt y_2)\cos^2(\al y_1) }{(\al^2 \cosh^2 (\bt y_2) +\bt^2 \cos^2(\al y_1))^2}.
\]
Therefore using double angle formulas, we have
\[
\frac 12 (B_t^2 + B_x^2) + (1-\cos B) = \frac{8\al^2 \bt^2 \tilde h(t,x)}{(\frac{\al^2 + \bt^2}{2} + \frac{\al^2}{2}\cosh(2\bt y_2) + \frac{\bt^2}{2}\cos(2\al y_1))^2},
\]
where
\bee
\tilde h(t,x) &:=&  \frac14 ((1+v^2)(\al^2-\bt^2)+1) + \frac 14 ((1+v^2)(\al^2+\bt^2)+1) \cosh (2\bt y_2) \\
&&  -\al\bt v \sin(2\al y_1)\sinh(2\bt y_2) - \frac 14 ((1+v^2)(\al^2+\bt^2)-1) \cos(2\al y_1)  \\
& &  + \frac14((1+v^2)(\bt^2-\al^2)+1) \cosh(2\bt y_2)\cos (2\al y_1)  .
\eee
Now with the relation among $\al,\bt$ \eqref{deltagamma} and $g$ given by \eqref{defG}, a direct computation shows that 
\[
\tilde h = g (h_e)_x - h_e g_x, 
\]
where $h_e(t,x):= \frac 1{2\al\bt}[\al\sinh (2\bt y_2) + \bt v\sin (2\al y_1)]. $ From this identity we obtain
\[
E[B,B_t] = 8\al^2 \bt^2 \int_\R \Big( \frac{h_e}{g}\Big)_x = 16\bt.
\]
\end{proof}

\begin{rem}
The reader may compare \eqref{MassB} with a similar result for the mass of the mKdV breather, see \cite[eqn. (2.4)]{AM}, where $M[B] =4\bt$. In that sense, both results reveal that the mass (or energy) does not depend on the oscillatory parameter, but only on the main scaling $\bt$.  This property seems inherent to aperiodic breathers.
\end{rem}

\subsection{Linear operators and stability tests} An essential consequence of the last two identities are the following stability conditions, which will be useful when dealing with coercivity estimates in Subsection \ref{4}.

\begin{cor}\label{WCcor} Let $(B,B_t)$ be any SG breather of the form \eqref{BB}-\eqref{Bt}. For $t\in \R$ fixed, let
\be\label{LAB}
 \Lambda B := \partial_\bt B, \quad  \Lambda B_t :=  \partial_\bt B_t.
\ee
Then $(\Lambda B, \Lambda B_t)$ are Schwartz in the space variable, and the following  identities are satisfied
\be\label{LAB2}
\partial_\bt E[ B,B_t]   = 16>0, \quad \partial_\bt P[B, B_t] = -8v,
\ee
independently of time.
\end{cor}

\begin{proof}
By simple inspection one can see that  for each time $t$ one has that $\Lambda B$ and $\Lambda B_t$  are well-defined Schwartz functions. The proof of  (\ref{LAB2})  is  consequence of (\ref{MassB}) and \eqref{Mom_B}.
\end{proof}

Now, we introduce the following two directions associated to spatial translations. Let $(B,B_t)$ as defined in  \eqref{BB}-\eqref{Bt}, with main parameters $\bt>0$ and $v\in (0,1)$. We define 
\be\label{B12}
\begin{cases}
 B_1(t ; x_1,x_2) := \partial_{x_1} B(t ; x_1, x_2), \\
 B_2(t ; x_1,x_2): =\partial_{x_2} B(t ;  x_1, x_2).
 \end{cases}
\ee
Similarly, we introduce the terms involving time derivatives
\be\label{B12t}
\begin{cases}
 (B_t)_1(t ; x_1,x_2) := \partial_{x_1} B_t(t ; x_1, x_2)\\
  (B_t)_2(t ; x_1,x_2): =\partial_{x_2} B_t(t ;  x_1, x_2).
\end{cases}
\ee
It is clear that, for all $t\in \R$, $\bt>0$ and $x_1,x_2\in \R$, both $B_1$ and $B_2$ are real valued, Schwartz functions.\footnote{
Additionally, we can express $B_t$ and $B_x$ in terms of $B_1$ and $B_2$: we have
\[
B_t = B_1 -v B_2, \qquad B_x = -v B_1 + B_2.
\]
}
 More explicitly,
\be\label{B1}
B_1= \frac{ -4\al^2\bt  \sin (\al y_1)\cosh (\bt y_2)}{\al^2 \cosh^2 (\bt y_2) +\bt^2 \cos^2(\al y_1)},\qquad B_2 = \frac{-4\al\bt^2  \cos (\al y_1)\sinh (\bt y_2)}{\al^2 \cosh^2 (\bt y_2) +\bt^2 \cos^2(\al y_1)}.
\ee


Now we introduce the main linear operator associated to the breather $(B,B_t).$ As it will be clear below, this matrix operator will be of fourth order in one component ($\mathcal L_1$), and of second order in the second one $(\mathcal L_2)$. Two nontrivial nondiagonal terms  $\mathcal B_1$ and $\mathcal B_2$ couple both previous components in a nontrivial fashion, which is never zero in general, even in the case $v=0$.

\begin{defn}
Let  $\mathcal L$ be matrix linear operator
\be\label{L}
\mathcal L[z,w] := \left( \begin{array}{cc} \mathcal L_1  & \mathcal B_1 \\  \mathcal B_2 & \mathcal L_2 \end{array}\right)      \left(\begin{array}{c}  z \\ w  \end{array}\right),
\ee
where
\be\label{L1}
\begin{aligned}
\mathcal L_1[z] & :=  z_{(4x)} - \Big[ a-  \frac 38 (B_x^2 + B_t^2) +\frac 54 \cos B \Big]z_{xx}   + \Big[ \frac 34B_x B_{xx} +\frac 34 B_t B_{tx} + \frac 54 \sin B B_x \Big]z_x   \\
&   \quad + \Big[ a\cos B  + \frac 58B_x^2 \cos B +\frac 54 B_{xx}\sin B    +\frac 14(\cos^2 B-\sin^2 B) -\frac 18 B_t^2 \cos B \Big]z,
\end{aligned}
\ee
\be\label{L2}
\mathcal L_2[w] := -w_{xx} + \frac 14\Big[4a + \cos B -\frac 32(B_t^2 +B_x^2)Ê\Big] w,
\ee
\be\label{B1a}
\mathcal B_1[w] := \frac 14\Big[  3 B_{tx}B_x  + 3 B_tB_{xx} {\color{black}-} B_t \sin B \Big] w + \frac 14\Big[ 3B_tB_x -2b\Big]w_x, 
\ee
and finally
\be\label{B2a}
 \mathcal B_2[z] := \frac12 (b- \frac 32 B_t B_x )z_x -\frac 14 B_t \sin B z.
\ee
In order to ensure a correct definition for $\mathcal L$, we will set $\mathcal L$ on the Hilbert space $L^2(\R)^2$, 
with dense domain $H^4(\R) \times H^2(\R)$.

\end{defn}

A simple consequence of Theorem \ref{GB} is the following partial kernel description.

\begin{cor}\label{kernel} Let $a,b$ be defined as in (\ref{ab}), and $B_1,B_2$ given in \eqref{B1}.
We have
\[
\mathcal L[B_1,(B_t)_1] =\mathcal L[B_2,(B_t)_2] = \ba 0 \\ 0 \ea.
\]
Moreover, $(B_1,(B_t)_1)^T $ and $(B_2,(B_t)_2)^T $ are linearly independent in $\R$.
\end{cor}
 
\medskip

Now we consider the natural directions associated to the scaling parameters. First of all, we define the quadratic form associated to $\mathcal L$, namely
\be\label{Q}
\mathcal Q[z,w]:= \frac 12 \int_\R (z,w)\mathcal L[z,w].
\ee
A more detailed version of $\mathcal Q$ is the following expression, obtained after integration by parts:
\be\label{QQ3}
\begin{aligned}
\mathcal Q[z,w] & =   \int_\R \{ z\mathcal L_{1}[z]  + z \mathcal B_1[w] +  w \mathcal B_2[z] + w\mathcal L_{2}[w]\} \\
& =    \int_\R z_{xx}^2 +\int_\R  w_x^2  + \int_\R \Big[a -  \frac 38 (B_x^2  + B_t^2)  +\frac 54 \cos B  \Big] z_x^2 \\
& \quad    + \int_\R \Big[ \frac 58 B_x^2 \cos B  +{\color{black}\frac 54} B_{xx}\sin B   +\frac 14 (\cos^2 B-\sin^2 B) -\frac 18 B_t^2 \cos B + a\cos B   \Big] z^2 \\
& \quad + \int_\R \Big[(a +\frac 14) \cos B -\frac 38( B_t^2+B_x^2) +\frac 34 B_{tx}B_x  +\frac34B_tB_{xx} -\frac 14B_t \sin B \Big] w^2  \\
&   \quad +\int_\R \Big[ (b-\frac 32 B_t B_x) z_x w -{\color{black}\frac 12} B_t\sin B zw \Big].
\end{aligned}
\ee

The following concept is standard.

\begin{defn}\label{Neg_dir}
Any nonzero pair $(z,w) \in H^4\times H^2$ is said to be a positive (null, negative) direction for $\mathcal Q$ if we have $\mathcal Q[z,w] >0$ $(= 0,< 0)$.
\end{defn}

Recall the definitions of $\Lambda B$ and $\Lambda B_t$ introduced in (\ref{LAB}). For this direction, one has the result below.

\begin{cor}\label{Scaling} The following are satisfied:
\ben
\item
Let $(B,B_t) $ be any SG breather. Consider the \emph{scaling} direction $(\Lambda B, \Lambda B_t)$ introduced in (\ref{LAB}). Then $(\Lambda B, \Lambda B_t)$ is a negative direction for $\mathcal Q$. Moreover,
\be\label{Neg}
\mathcal Q[\Lambda B, \Lambda B_t] =-32(1+3v^2)\bt<0.
\ee
\item The quadratic functional $\mathcal Q$ is bounded below, namely
\be\label{Bdd_below}
\mathcal Q[z,w] \geq -c_{\bt,v}\|(z,w)\|_{H^2(\R)\times H^1(\R)}^2,
\ee
for some nonnegative constant $c_{\bt,v}$. Moreover, $\mathcal L$ has \emph{at least} one negative eigenvalue, 
and therefore a minimal one, which is simple.
\een
\end{cor}

\begin{proof}
The proof of \eqref{Bdd_below} is standard. On the other hand, from (\ref{EcB}), we get after derivation with respect to $\beta$,
\be\label{Eq1}
\mathcal L_1 [\Lambda B ] + \mathcal B_2[\Lambda B_t] =  a'(\bt)  ( B_{xx} - \sin B) + \frac 12 b'(\bt) B_{tx},
\ee
and from (\ref{EcB2}),
\be\label{Eq2}
\mathcal L_2[\Lambda B_t] + \mathcal B_1[\Lambda B] = -a'(\bt) B_t  - \frac 12 b'(\bt) B_x.
\ee
Integrating against $\Lambda B$ and $\Lambda B_t$ respectively, we get from (\ref{LAB2}),
\[
\begin{aligned}
\mathcal Q[\Lambda B, \Lambda B_t]   & =  - \frac 12 a' (\bt) \partial_\bt \int_\R (B_x^2 + B_t^2 + 1-\cos B)   -\frac 12 b'(\bt)\partial_\bt \int_\R B_t B_x\\
& =   -a'(\bt) \partial_\bt E[B,B_t]  -b'(\bt) \partial_\bt P[B,B_t] \\
&= -32(1+v^2)\bt  -64v^2\bt  \\
& = -32(1+3v^2)\bt<0,
\end{aligned}
\]
where we have used (\ref{ab}) to obtain $a'(\bt) = 2(1+v^2)\bt $ and $b'(\bt)=-8v\bt$. This last identity proves (\ref{Neg}).  
\end{proof}

From the previous result we conclude the following useful identity:
\begin{cor}
Let $(B_0,\tilde B_0)$ denote the following direction
\[
(B_0,\tilde B_0) := -\frac 1{2\bt}(\Lambda B,\Lambda B_t). 
\]
Then
\be\label{AtA}
\mathcal L[B_0,\tilde B_0] = (1+v^2)\left( \begin{array}{c} \sin B-B_{xx} \\ B_t \end{array} \right) +2v  \left( \begin{array}{c} B_{tx} \\ -B_x \end{array} \right) ,
\ee
and
\be\label{Key}
- \int_\R (B_0,\tilde B_0) \cdot \mathcal L[B_0,\tilde B_0] = -\frac 1{4\bt^2}\mathcal Q[\Lambda B, \Lambda B_t]  = \frac 8\bt (1+3v^2)>0.
\ee
\end{cor}
\begin{proof}
A direct consequence of (\ref{Eq1}) and (\ref{Eq2}).
\end{proof}

\begin{rem}
The previous result states that $(B_0,\tilde B_0)$ is a very good candidate to replace the first eigenfunction of $\mathcal L$ coming from Assumption 2 below. 
This idea, originally coming from Weinstein \cite{Weinstein}, has been recently used in several works (see \cite{AM} for example) where no knowledge of the ground state is at hand. 
\end{rem}

\subsection{Spectral analysis}\label{4}

Let $z=z(x)$ and $w=w(x)$ be two functions, to be specified below, and let $(B,B_t)(t,x;x_1,x_2)$ be any breather solution, with parameters $x_1,x_2$ possibly depending on time.  
In this section we describe part of the spectrum of the operator $\mathcal L$ defined in (\ref{L})-(\ref{L2}).  We start with the following result. 
 
\begin{lem} $\mathcal L$ is a linear, unbounded operator in $L^2(\R)^2$, with dense domain $H^4(\R)\times H^2(\R)$. Moreover, $\mathcal L$ is self-adjoint. 
\end{lem}
It is a surprising fact that $\mathcal L$ is actually self-adjoint, due to the non constant terms appearing in the definition of $\mathcal L$.

\begin{proof}
We prove that $\mathcal L$ is symmetric. Let $(f,g)\in H^4(\R)\times H^2(\R)$. Then, integrating by parts, one has
\bee
\int_\R (f,g) \mathcal L[z,w] & =& \int_\R (f,g) \left( \begin{array}{cc} \mathcal L_1  & \mathcal B_1 \\  \mathcal B_2 & \mathcal L_2 \end{array}\right)      \left(\begin{array}{c}  z \\ w  \end{array}\right) \\
&=& \int_\R \{ f\mathcal L_{1}[z]  + f \mathcal B_1[w] +  g \mathcal B_2[z] + g\mathcal L_{2}[w]\}\\
& =& \int_\R \{ z\mathcal L_{1}[f]  + z \mathcal B_1[g] +  w \mathcal B_2[f] + w\mathcal L_{2}[g]\}  \\
&=& \int_\R (z,w) \left( \begin{array}{cc} \mathcal L_1  & \mathcal B_1 \\  \mathcal B_2 & \mathcal L_2 \end{array}\right)      \left(\begin{array}{c}  f \\ g  \end{array}\right) = \int_\R (z,w) \mathcal L[f,g].
\eee
\end{proof}
A consequence of the above result is the fact that the spectrum of $\mathcal L$ is real valued. Furthermore, $\mathcal L$ is a compact perturbation of a constant coefficients operator.
Now we look at the kernel of $\mathcal L$. Let us fix $t\in\R$, $v\in (-1,1)$ and $\bt \in (0,\ga)$. It is not difficult to check that any breather $(B,B_t)$ as in \eqref{BB}-\eqref{Bt} can be seen as a function independent of time, by considering 
\be\label{BBnew}
\begin{aligned}
(B,B_t)(t,x;x_1,x_2) &= (B,B_t)(0,x; \tilde x_1, \tilde x_2), \quad  \tilde x_1 :=t+x_1 , ~\tilde x_2 := -vt +x_2, \\
&=: (B,B_t)(x;\tilde x_1,\tilde x_2).
\end{aligned}
\ee
Moreover, it turns out that if\footnote{Of course, we also have a dependence on $\bt$ and $v$, but it is not needed at this moment.} 
\[
(z(x;\tilde x_1,\tilde x_2),w(x;\tilde x_1,\tilde x_2),\la (\tilde x_1,\tilde x_2)) \in H^4\times H^2 \times\R
\]
satisfy the eigenvalue-eigenfunction problem $\mathcal L[z,w]=\la (z,w)^T$ with $(B,B_t)$ as in \eqref{BBnew}, then the new function
\[
(\tilde z,\tilde w)(x; \tilde x_1,\tilde x_2) := (z, w)(x-\tilde x_2; \tilde x_1,\tilde x_2)
\]
will satisfy the same spectral equation with $(B,B_t)$ replaced by $(B,B_t)(x; \tilde x_1- \tilde x_2,0)$. After redefining $\tilde x_1$, we can assume without loss of generality that $(B,B_t)$ is only depending on $x$, $\bt$, $v$ and $\tilde x_1$, 
and therefore $(\tilde z,\tilde w)$ will also depend on the same variables. We will have $\la=\la(\bt,v, \tilde x_1)$ only.  Finally, since the breather was periodic in the variable $x_1$, with period $2\pi /\al$, we can also assume
\[
\tilde x_1 \in \Big[0, \frac{2\pi}\al \Big], \quad \al = \sqrt{\ga^2-\bt^2}.
\]
In conclusion, with no loss of generality we set through this section  $(B,B_t) = (B,B_t)(0,x;x_1,0)$, with $x_1$ lying in the interval $[0,2\pi/\al]$ and each breather-like potential periodic on that variable. The same property applies for the functions $(B_1, (B_t)_1)$ and $(B_2, (B_t)_2)$.

\ms

This being said, in this paper we will need the following two assumptions:

\ms 

\begin{itemize}\label{A12}
\item[]~{\bf Assumption 1:} (\emph{Nondegeneracy of the kernel})
 For each $v\in (-1,1)$, $x_1\in \R$ and $\bt\in (0,\ga)$, $\ker \mathcal L$ is spanned by the two elements $(B_1, (B_t)_1)^T$ and $(B_2, (B_t)_2)^T$; and there is a (uniform in $x_1$) gap between the kernel and the bottom of the positive spectrum. 
\end{itemize}

\ms
\begin{itemize}
\item[]~{\bf Assumption 2:} (\emph{Unique, simple negative eigenvalue})
For each $\bt \in(0,\ga)$, $v\in (-1,1)$ and $x_1 \in [0,2\pi \al^{-1}]$, the operator  $\mathcal L$ has a unique simple, negative eigenvalue $\la_{1}=\la_{1}(\bt,v,x_1)<0$ associated to the unit $L^2\times L^2$-norm  eigenfunction $(\tilde B,\hat B)^T$. Moreover, there is $\la_{1}^0<0$ depending on $\bt$ and $v$ only, such that  $\la_{1} \leq \la_{1}^0$ for all $x_1$.
\end{itemize}

\ms 
In Subsection \ref{S4.5} we perform numerical computations that reveals that both conditions are naturally correct for every set of parameters that we numerically tested. The main consequence of the two preceding assumptions is the following direct coercivity property (recall the  quadratic form $\mathcal Q$, associated to $\mathcal L$, and defined in (\ref{Q})): 

\begin{lem}
Let $(z,w)\in H^2(\R)\times H^1(\R)$, and $(B,B_t)$ be any SG breather, and let $(\tilde B,\hat B)^T$ be the first eigenfunction from Assumption 2. If $(z,w)$ are such that they satisfy the orthogonality conditions
\be\label{ortho1}
\int_\R (z,w)(B_1,(B_1)_t)^T =\int_\R (z,w)(B_2,(B_2)_t)^T=0,  
\ee
then there is $\mu_0=\mu_0(\bt,v)>0$ such that
\be\label{Coerr}
\mathcal Q[z,w] \geq \mu_0 \|(z,w)\|_{H^2\times H^1}^2 - \frac 1{\mu_0} \abs{\int_\R (z,w)(\tilde B,\hat B)^T}^2.
\ee
\end{lem}

%
%

Following a similar strategy as in \cite{AM}, we must use a different orthogonality condition in order to ensure a good control on the scaling parameter $\bt$, in such a form that we can run a stability argument without using modulations, which are very difficult to estimate for breather solutions\footnote{In other words, we control the variations of the scaling parameter $\bt$ at least at the second order in the error term $(z,w)$.} Indeed, consider the new direction
\be\label{AtAdef}
\ba A \\ \tilde A \ea := (1+v^2)\left( \begin{array}{c} \sin B -B_{xx} \\ B_t \end{array} \right) + 2v \left( \begin{array}{c} B_{tx} \\ -B_x \end{array} \right)\in H^2 \times H^1.
\ee
Note that from (\ref{AtA}), we have
\[
\mathcal L[B_0,\tilde B_0] = \ba A \\ \tilde A \ea.
\]
We must have in mind that  $(A,\tilde A)^T$ is a sort of \emph{generalized negative direction} (see \eqref{Key}), that will replace $(\tilde B, \hat B)$ in \eqref{Coerr}. The advantage of taking this new set of orthogonality conditions comes from the fact that  $(A,\tilde A)^T$ are {\bf naturally associated to the energy and momentum conservation laws}, which are only $H^1\times L^2$ based, and therefore, low-regularity conserved quantities compared with the $H^2\times H^1$ global dynamics. Indeed, will prove the following

\begin{prop}\label{Coer}
Let $(z,w)\in H^2\times H^1$ satisfying the orthogonality conditions (\ref{ortho1}), and $(A,\tilde A)$ the direction defined in \eqref{AtAdef}. Then there is $\nu_0=\nu_0(\bt,v)>0$ such that
\[
\mathcal Q[z,w] \geq \nu_0 \|(z,w)\|_{H^2\times H^1}^2 - \frac 1{\nu_0} \abs{\int_\R (z,w)(A,\tilde A)^T}^2.
\]
\end{prop}

\begin{proof}
We follow a similar strategy as stated in \cite{AM}. It is enough to prove that, under the orthogonality conditions (\ref{ortho1}), and the additional constraint
\[
\int_\R \ba z \\ w \ea \cdot \ba A \\ \tilde A\ea=0,
\]
we have
\[
\mathcal Q[z,w] \geq \nu_0 \|(z,w)\|_{H^2\times H^1}^2.
\]
We write 
\[
(z,w) = (\tilde z, \tilde w) + \delta_0 (\tilde B,\hat B), \quad (B_0,\tilde B_0) = (b_0, \tilde b_0) + \ga_0 (\tilde B,\hat B).  
\]
Note that 
\[
\int_\R (B_0,\tilde B_0)(B_1,(B_t)_1)^T =\int_\R (B_0,\tilde B_0)(B_2,(B_t)_2)^T = 0,
\]
and we can assume
\[
\int_\R (\tilde z, \tilde w) (\tilde B,\hat B)^T = \int_\R (b_0, \tilde b_0) (\tilde B,\hat B)^T =0.
\]
Therefore,
\be\label{AAA1}
\begin{aligned}
\mathcal Q[z,w]  & = \mathcal Q[\tilde z +\delta_0 \tilde B, \tilde w + \delta_0 \hat B]   \\
& = \mathcal Q[\tilde z, \tilde w]    +\delta_0^2 \mathcal Q[ \tilde B, \hat B]  = \mathcal Q[\tilde z, \tilde w] - \delta_0^2\la_0^2.
\end{aligned}
\ee
Now we must replace $\delta_0$ and $\la_0$ by suitable expressions. First of all,
\[
0=\int_\R (z,w)(A,\tilde A)^T= \int_\R (z,w) \mathcal L[B_0,\tilde B_0] =  \int_\R (B_0,\tilde B_0)\mathcal L[z,w]. 
\]
Since $\mathcal L[z,w] = \mathcal L[\tilde z,\tilde w] -\delta_0 \la_0^2(\tilde B,\hat B)^T$ and $(B_0,\tilde B_0) = (b_0,\tilde b_0)+ \ga_0 (\tilde B,\hat B) $, we obtain
\be\label{AAA2}
 \int_\R (b_0,\tilde b_0)\mathcal L[\tilde z, \tilde w] =\ga_0\delta_0 \la_0^2. 
\ee
This last expression involves $\ga_0$, $\delta_0$ and $\la_0$. We want now an expression for $\ga_0$. We have
\bee
\int_\R (A,\tilde A)(B_0,\tilde B_0)^T &=& \int_\R (B_0,\tilde B_0)\mathcal L[B_0,\tilde B_0] \\
&=& \int_\R \{ (b_0, \tilde b_0) + \ga_0 (\tilde B,\hat B)\} \{ \mathcal L[b_0,\tilde b_0] -\ga_0\la_0^2 (\tilde B,\hat B)^T \}\\
& =& \mathcal Q[b_0,\tilde b_0] -\ga_0^2\la_0^2.
\eee
Note in addition that from (\ref{AtA}) and (\ref{Key}),
\[
\int_\R (A,\tilde A)(B_0,\tilde B_0)^T = \int_\R (B_0,\tilde B_0) \mathcal L[B_0,\tilde B_0]   = -\frac 8\bt (1+3v^2).
\] 
Therefore,
\[
\ga_0^2\la_0^2 =  \mathcal Q[b_0,\tilde b_0] + \frac 8\bt (1+3v^2).
\]
Replacing this last expression in \eqref{AAA1}, and using \eqref{AAA2}, we get
\be\label{Positivity}
\begin{aligned}
\mathcal Q[z,w] & = \mathcal Q[\tilde z, \tilde w]  - \frac{ (\ga_0\delta_0 \la_0^2)^2}{\ga_0^2\la_0^2} \\
& = \mathcal Q[\tilde z, \tilde w] - \frac{\displaystyle{ \Big( \int_\R (b_0,\tilde b_0)\mathcal L[\tilde z, \tilde w] \Big)^2 }}{ \displaystyle{\mathcal Q[b_0,\tilde b_0] + \frac 8\bt (1+3v^2)}}.
\end{aligned}
\ee
Note that the denominator above is {\bf always positive}.  If now $(\tilde z,\tilde w) =\la (b_0,\tilde b_0)$, for some $\la\neq 0$, we have
\[
 \Big( \int_\R (b_0,\tilde b_0)\mathcal L[\tilde z, \tilde w] \Big)^2 =   \Big( \int_\R (\tilde z, \tilde w)\mathcal L[\tilde z, \tilde w] \Big) \Big( \int_\R (b_0,\tilde b_0)\mathcal L[b_0,\tilde b_0] \Big) = \mathcal Q[\tilde z,\tilde w] \mathcal Q[b_0,\tilde b_0],
\] 
and therefore,
\[
 \frac{\displaystyle{ \Big( \int_\R (b_0,\tilde b_0)\mathcal L[\tilde z, \tilde w] \Big)^2 }}{ \displaystyle{\mathcal Q[b_0,\tilde b_0] + \frac 8\bt (1+3v^2)}} =  \frac{\displaystyle{  \mathcal Q[\tilde z,\tilde w] \mathcal Q[b_0,\tilde b_0] }}{ \displaystyle{\mathcal Q[b_0,\tilde b_0] + \frac 8\bt (1+3v^2)}} = \rho  \mathcal Q[\tilde z,\tilde w] ,
\]
with $\rho \in (0,1)$ independent of $(z,w)$. We have then,
\[
\mathcal Q[z,w]  =  (1-\rho) \mathcal Q[\tilde z, \tilde w] . 
\]
Now if $(\tilde z,\tilde w) $ lies in the orthogonal vector space spanned by $(b_0,\tilde b_0)$,  $(\tilde B,\hat B)$, and the kernel $(B_1,(B_1)_t)$, $(B_2,(B_2)_t)$, we have that $\mathcal Q[\tilde z,\tilde w]$ defines an internal product, for which the Cauchy-Schwarz's inequality holds:
\[
\Big( \int_\R (b_0,\tilde b_0)\mathcal L[\tilde z, \tilde w] \Big)^2  < \mathcal Q[\tilde z,\tilde w] \mathcal Q[b_0,\tilde b_0],
\]  
(note that there is no equality since $(b_0,\tilde b_0)$ and $(\tilde z, \tilde w)$ are not orthogonal). We conclude now that 
\[
 \frac{\displaystyle{ \Big( \int_\R (b_0,\tilde b_0)\mathcal L[\tilde z, \tilde w] \Big)^2 }}{ \displaystyle{\mathcal Q[b_0,\tilde b_0] + \frac 8\bt (1+3v^2)}}  < \frac{\displaystyle{ \mathcal Q[\tilde z,\tilde w] \mathcal Q[b_0,\tilde b_0] }}{ \displaystyle{\mathcal Q[b_0,\tilde b_0] + \frac 8\bt (1+3v^2)}}  =\rho \mathcal Q[\tilde z,\tilde w],
\]
with $\rho\in(0,1)$. We conclude that
\[
\mathcal Q[z,w]  >  (1-\rho) \mathcal Q[\tilde z, \tilde w]  \geq 0. 
\]
Therefore, in \eqref{AAA1} we get 
\[
\mathcal Q[\tilde z, \tilde w]  \geq \delta_0^2 \la_0^2.
\]
Finally, we have
\[
\begin{aligned}
\mathcal Q[z,w] &  >  (1-\rho) \mathcal Q[\tilde z, \tilde w] \\
& =  \frac12 (1-\rho) \mathcal Q[\tilde z, \tilde w] + \frac12  (1-\rho) \mathcal Q[\tilde z, \tilde w] \\
& \geq  \frac12 (1-\rho) \mu_0 \| (\tilde z,\tilde w)\|_{H^1\times L^2}^2   +   \frac12  (1-\rho) \delta_0^2 \la_0^2 \\
& \gtrsim  \| (\tilde z,\tilde w)\|_{H^2\times H^1}^2  + \| (\tilde B,\hat B)\|_{H^2\times H^1}^2  \gtrsim \|(z,w)\|_{H^2\times H^1}^2.
\end{aligned}
\]
\end{proof}

\subsection{Proof of the Main Theorem}\label{8}

In this subsection we prove Theorem \ref{MT}.  This proof follows similar lines as in \cite{AM}. Assume $(u_0,u_1)$ satisfy the hypothesis \eqref{Initial_Condition}, for some $\eta<\eta_0$ small. Let $(u,u_t)$ be the associated $H^2\times H^1$ solution to \eqref{SG} with initial data $(u_0,u_1)$. Given $K^*>1$ to be chosen later, we denote $T^*=T^*(K^*)>0$ as the maximal time for which, for all time $T\in (0,T^*]$,
\[
\sup_{t\in [0,T]}\|(u,u_t)(t) -(B,B_t)_{\bt,v}(0, \cdot ; \tilde x_1(t), \tilde x_2(t)) \|_{H^2\times H^1} <K^* \eta,
\]
is satisfied for some choice of modulation parameters $\tilde x_1(t)$ and $\tilde x_2(t)$, not necessarily unique. For the sake of simplicity, and if no confusion arises, we denote
\[
(B,B_t):=(B,B_t)_{\bt,v}.
\]
If we assume that $T^*(K^*)$ is finite, we can choose (assuming $\eta_0$ smaller if necessary), using the Implicit Function Theorem, special parameters  $x_1(t)$ and $x_2(t)$ such that 
\[
\int_\R (z,w)(B_1,(B_1)_t)^T =\int_\R (z,w)(B_1,(B_1)_t)^T=0,  
\]
where
\be\label{z}
z(t,x):= u(t,x) - B(t,x; x_1(t),x_2(t)),
\ee
 \be\label{w}
 w(t,x) := u_t(t,x) - B_t(t,x; x_1(t),x_2(t)),
\ee
and $B_1$, $B_2$ are given in (\ref{B12})-(\ref{B12t}). These conditions are well-defined since the matrix with coefficients
\[
\left(
\begin{array}{cc}
\displaystyle{ \int_\R \big[ B_1^2 +((B_1)_t)^2 \big] }&\displaystyle{ \int_\R B_1B_2 + (B_1)_t (B_2)_t} \\
\displaystyle{ \int_\R B_1B_2 + (B_1)_t (B_2)_t  } &\displaystyle{ \int_\R B_2^2 +((B_2)_t)^2}
\end{array} 
\right)
\]
has nonzero determinant everywhere (cf. Corollary \ref{kernel}).  Moreover, thanks to \eqref{Initial_Condition}, we have
\[
\|(z,w)(0)\|_{H^1\times L^2} \lesssim \eta,
\]
with constant independent of $K^*$.

\medskip

Consider now the decomposition \eqref{z}-\eqref{w}, with $(B,B_t)$ depending on the modulation parameters $x_1(t)$ and $x_2(t)$. A simple argument reveals that  $\mathcal H[B,B_t](t)$ is still independent of time (see \cite{AM} for a similar proof). Therefore we can apply Corollary \ref{Eqtt} and Lemma \ref{EE0} at times $t=0$ and $t>0$ fixed to obtain
\[
\mathcal Q[z,w](t) \leq  \mathcal Q[z,w](0) + \sup_{t\in [0,T^*]}\mathcal N[z,w](t) \lesssim \eta^2 + (K^*)^3\eta^3,
\]
Now, from Proposition \ref{Coer} we have
\be\label{Est1}
\|(z,w)(t)\|_{H^2\times H^1}^2 \lesssim \eta^2 + (K^*)^3\eta_0^3 +\abs{\int_\R (z,w) (A,\tilde A)^T(t)}^2,
\ee
Recall that $(A,\tilde A)$ was defined in \eqref{AtAdef}. Finally, using the energy and momentum conservation laws \eqref{E}-\eqref{P}, evaluated at two different times, we obtain a good control on the term 
\[
\abs{\int_\R (z,w) (A,\tilde A)^T(t)}^2.
\]
Indeed, we have
\[
\begin{aligned}
&   \frac 12 \int_\R ((B_x+z_x)^2 +(B_t+w)^2)(t)  +\int_\R (1-\cos (B+z))(t) = \\
& \qquad = \frac 12 \int_\R ((B_x+z_x)^2 +(B_t+w)^2)(0)  +\int_\R (1-\cos (B+z))(0),
\end{aligned}
\]
from which 
\[
\begin{aligned}
\int_\R z(  \sin B -B_{xx} ) (t) + \int_\R B_t w(t) & = \int_\R z( \sin B -B_{xx} ) (0) + \int_\R B_t w(0)  \\
& \quad + O\Big(\sup_{t\in [0,T^*]}\|(z,w)(t)\|_{H^1\times L^2}^2 \Big).
\end{aligned}
\]
Similarly,
\[
 \int_\R (B_x+z_x)(B_t + w)(t) = \int_\R (B_x+z_x)(B_t + w)(0), 
\]
so that 
\[
\int_\R ( B_{tx} z-B_x w  )(t) =\int_\R ( B_{tx} z -B_x w  )(0) + O\Big(\sup_{t\in [0,T^*]}\|(z,w)(t)\|_{H^1\times L^2}^2 \Big).
\]
Consequently, reconstructing $(A,\tilde A)$ as it was defined in \eqref{AtAdef},
\[
\begin{aligned}
\abs{\int_\R (z,w)(A,\tilde A)^T(t)} & \lesssim  \abs{\int_\R (z,w)(A,\tilde A)^T(0)} +\sup_{t\in [0,T^*]} \|(z,w)(t)\|_{H^2\times H^1}^2\\
& \lesssim  \eta + (K^*)^2\eta^2.
\end{aligned}
\]
Finally, replacing in (\ref{Est1}), we get
\[
\|(z,w)(t)\|_{H^2\times H^1}^2 \lesssim \eta^2 + (K^*)^3\eta^3.
\]
Taking $K^*$ large enough, and then $\eta(K^*)>0$ small, we have
\[
\|(z,w)(t)\|_{H^2\times H^1}^2\leq \frac 14 (K^*)^2 \eta^2,
\]
so we improve the original estimate on $(z,w)$, which contradicts the finiteness of $T^*$. Therefore, for all $K^*$ large enough, $T^*=+\infty.$

\subsection{Numerical analysis}\label{S4.5} The purpose of this paragraph is to give strong evidence 
of the fact that Assumptions 1 and 2 in page \pageref{A12} do hold. We have run a similar numerical scheme as in 
the case of the mKdV and Gardner breathers, but now it is important to mention that the resulting linear 
system is no longer a scalar valued one; instead, we have to deal with a vector valued linear 
system, which formally implies that we need twice as much test functions as usual, compared to the mKdV case.

\medskip

\begin{figure}[!h]
\begin{center}
\includegraphics[width=15cm, height=10cm]{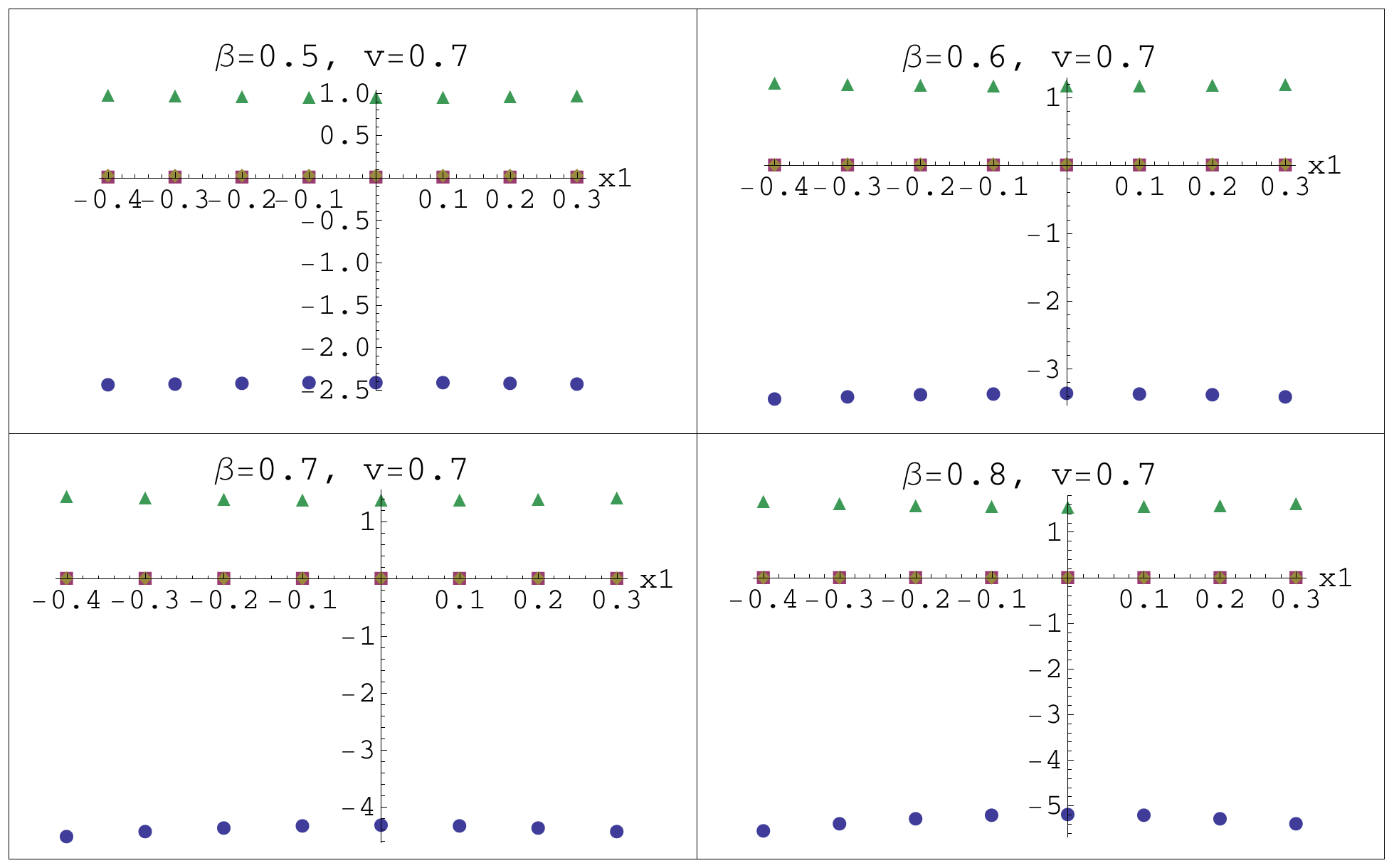}
\caption{The graph of the four minimal eigenvalues 
(from the first negative eigenvalue(blue circle), to the  double zero kernel (purple box and brown diamond) and the first positive eigenvalue (green triangle))
of the SG breather with $\bt=0.5,0.6,0.7$ and $0.8$, 
with $x_1$ varying between -0.4 and 0.3. 
The velocity of the relativistic breather is taken as $v=0.7$. Note that two numerical eigenvalues are placed near zero, see also Fig. \ref{TableBetaFixVmov}.}\label{SG_beta}
\end{center}
\end{figure}

\begin{figure}[!h]
\begin{center}
\includegraphics[width=12cm, height=6cm]{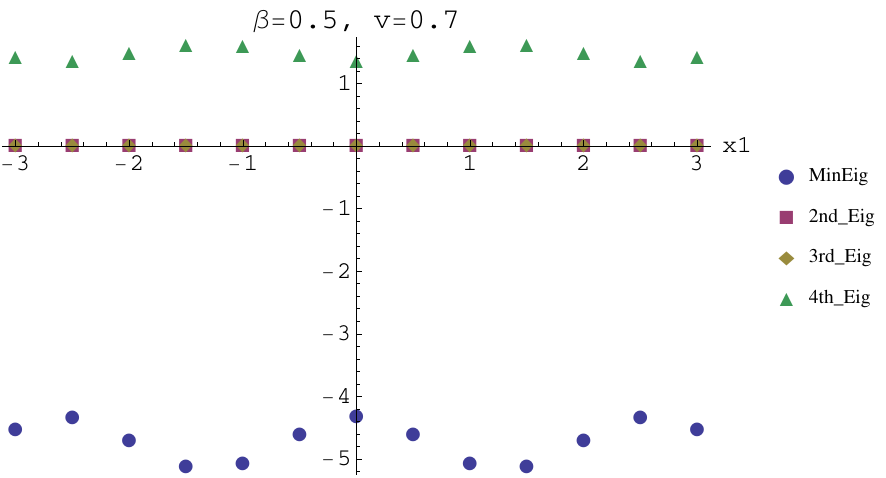}
\caption{The graph of the four minimal eigenvalues of the SG breather case, with $\bt=0.5,~v=0.7$ and with $x_1$ varying between -3.0 and 3.0.}\label{sG_Bigx1}
\end{center}
\end{figure}

{\bf First test.} Recall that we can always assume $x_2=0$ and $x_1\in [0, 2\pi/\al]$. We have run a primary test with $v=0.7$,
different values of $\bt$, and $x_1$ varying between -0.4 and 0.3. It seems that the SG case is numerically
simpler than the mKdV one, because of the fact that the SG breather requires one derivative less for its definition. 
Therefore, with $N=25$ (that is to say, 50 test functions), we have obtained the results summarized in Fig. \ref{SG_beta}, 
in agreement with Assumptions 1 and 2 up to two significative digits. We also check the expected oscillatory behavior of the minimal 
eigenvalue of the linearized operator $\mathcal{L}$ around a SG breather for variations of the shift $x_1$. We show this result in Fig. \ref{sG_Bigx1} 
where in that example, we consider  $\beta=0.5$ and $v=0.7$ and shift parameter $x_1$ varying from $-3.0$ to $3.0$.

\ms

\begin{figure}[!htb]
\begin{center}
\includegraphics[width=12.5cm, height=6cm]{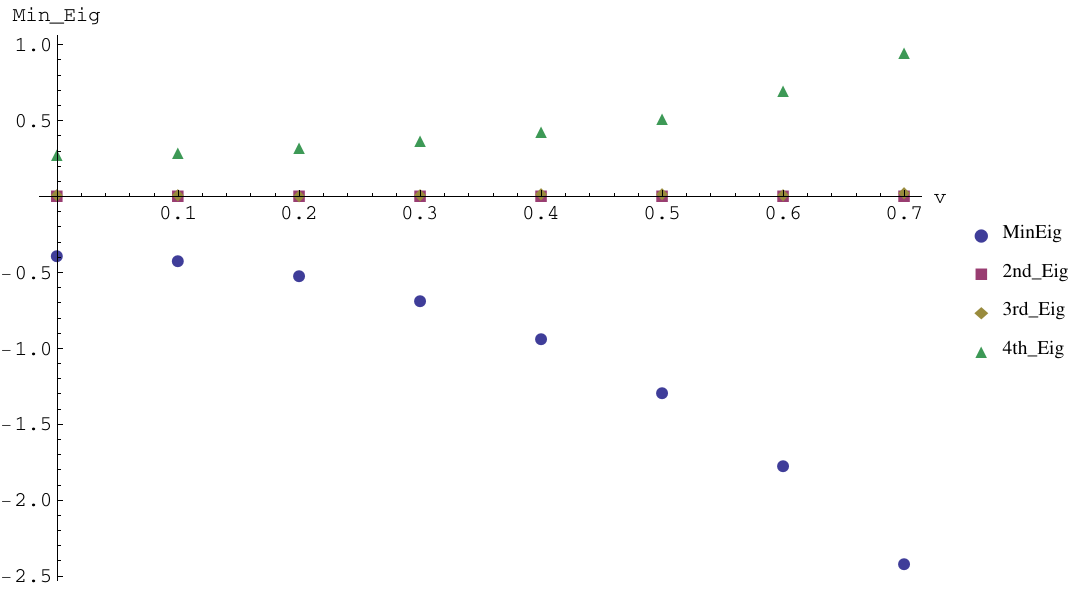}
\caption{The graph of the four minimal eigenvalues of the SG breather with $\bt=0.5$ and the velocity $v$ of the relativistic breather
varying between $0.0$ to $0.7$. The shifts $x_1=0.1,~ x_2=0$ and we take $N=50$ eigenfunctions (see also Figs. \ref{SG_Log} and \ref{TableBetaFixVmov}, the table below left).}\label{SG_betaFixvMov1}
\end{center}
\end{figure}

{\bf Second test.} When $\beta$ is fixed (e.g.  $\beta=0.5$),
and we move the velocity $v$ of the relativistic SG breather from $0.0$ to $0.7$, we obtain the  
results written in Fig. \ref{TableBetaFixVmov} (table of eigenvalues).  The corresponding graphics are shown in Fig. \ref{SG_betaFixvMov1}.


\begin{figure}[!htb]
\begin{center}
\includegraphics[width=10cm, height=5.5cm]{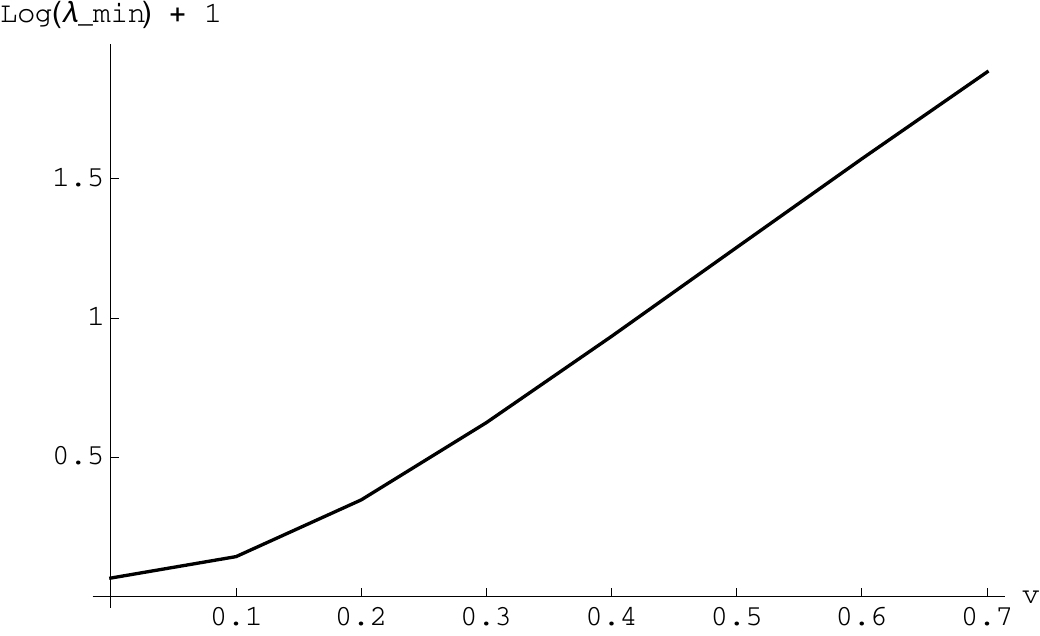}
\caption{The graph of the logarithm of the \emph{absolute} value of the minimal eigenvalue, 
plus one, for the SG breather with $\bt=0.5$ and the velocity $v$ of the relativistic breather varying between 
$0.0$ to $0.7$. The shifts are $x_1=0.1,~ x_2=0$. The continuous curve is the set of straight lines that join each pair of points, 
and the approximate slope among two adjacent points varies from  3.1 to 3.2.}\label{SG_Log}
\end{center}
\end{figure}


{\small
\begin{figure}[htc]
\begin{center}
\begin{tabular}{||c c c c c |c|c c c c c||}
 \hline
 $v$ & 1st eig. & 2nd & 3rd & 4th &&  $x_1$ & 1st eig. &  2nd &  3rd & 4th \\ [0.1ex]
 \hline\hline
 0.0 & -0.3932  & 0.0020 & 0.0124 & 0.2783 && -0.4 & -5.555  & 0.0004 & 0.0011 & 1.658 \\ [0.1ex] 
 \hline
 0.1 & -0.4246 & 0.0023 & 0.0107 & 0.2883&& -0.3 & -5.404  & 0.0003 & 0.0012 & 1.612 \\ [0.1ex]
 \hline
 0.2 & -0.5206 & 0.0040 & 0.0069 & 0.3191&& -0.2 & -5.290 & 0.0002 & 0.0013 & 1.576\\ [0.1ex]
 \hline
 0.3 & -0.6871 & 0.0037  & 0.0085& 0.3698&& -0.1 & -5.218 & 0.0001 & 0.0014 & 1.553 \\ [0.1ex]
 \hline
 0.4 & -0.9367  & 0.0023  &  0.0162&  0.4292&& 0.0 & -5.194   & 0.0001 & 0.0014 & 1.545\\ [0.1ex] 
 \hline
 0.5 & -1.2892& 0.0026 & 0.0180 & 0.5143&& 0.1 & -5.218 & 0.0001 & 0.0014 & 1.553\\ [0.1ex] 
 \hline
 0.6 & -1.7724 &  0.0067 & 0.0094   & 0.6994&& 0.2 & -5.290 & 0.0002 & 0.0013 & 1.576\\ [0.1ex] 
 \hline
 0.7 & -2.4203  &  0.0036 &  0.0254   & 0.9489&& 0.3 & -5.404 & 0.0003 & 0.0012 & 1.612\\ [0.1ex] 
 \hline
\end{tabular}
\end{center}\caption{The first four eigenvalues of $\mathcal L$: (left), for $\bt=0.5,~x_1=0.1,~ x_2 = 0$, and  the velocity $v$ 
of the relativistic breather varying from $0.0$ to $0.7$,
as corresponding to Fig. \ref{SG_betaFixvMov1}. (right), for $\bt=0.8,~v=0.7,~ x_2 = 0$, and $x_1$ varying from $-0.4$ to $0.3$,
as corresponding to bottom right Fig. \ref{SG_beta}. 
All computations were made with $N=50$. The third and fourth columns (left) and eigth and ninth columns (right) represent the approximate kernel 
of $\mathcal L$ respectively.}\label{TableBetaFixVmov} 

\end{figure}
}
\medskip

Additionally, from the numerical tests we have found the following 

\ben
\item For $v$ fixed, as long as $\beta$ approaches zero, we see that the negative eigenvalue converges to zero, a phenomenon that it is in concordance with the convergence to zero in $L^\infty$ norm of the breather as $\bt \to 0$. 

\ms 
\item As $v$ increases to the speed of light 1, the quality and accuracy of the resulting numerical results strongly decreases, implying that one needs to take $N$ even larger to recover the desired spectral stability.  
\een 
In conclusion, our numerical computations reveal that, in any of parameters regime considered, {\bf there is only one negative eigenvalue, as well as a well-defined two dimensional kernel}, and a spectral gap between the kernel and the continuum spectrum, supporting Assumptions 1 and 2 in page \pageref{A12}.

\bs

\section{Periodic mKdV breathers}\label{Sect:5}

\ms

\subsection{Definitions} We consider now the case of the periodic (in space) mKdV equation. Periodic mKdV breathers (or KKSH breathers), in the sense of Definition \ref{DEF_B_per}, were found by Kevrekidis, Khare,
Saxena and Herring  \cite{KKS1,KKS2} by using elliptic functions and a matching of free parameters. More precisely, we consider the equation 
\eqref{mKdV} where now
\[
u:  \R_t\times \T_x \mapsto \R_x,
\]
is periodic in space, and $\T_x =\T = \R/ L\Z  =(0,L)$ denotes a torus with period $L$, to be fixed later. Given $\al,\bt>0$, $x_1,x_2\in \R$ and
$k,m\in [0,1],$ KKSH breathers are given by the explicit formula \cite{KKS1}
\be\label{Bper} 
B= B(t,x; \al, \bt, k, m, x_1,x_2)   :=  \partial_x \tilde B := 2\sqrt{2} \partial_x \Big[ \arctan \Big( \frac{\bt}{\al}\frac{\sn(\al y_1,k)}
{\nd(\bt y_2,m)}\Big) \Big],
\ee
with $\sn(\cdot ,k)$ and $\nd (\cdot , m)$ the standard Jacobi elliptic functions of elliptic modulus $k$ and $m$, respectively, but now
\be\label{Y1_Y2}
y_1:=x+\delta t + x_1,  \quad y_2:=x+\ga t +x_2,
\ee
and
\be\label{DG}
 \delta := \al^2(1+k) +3\bt^2(m-2), \quad \hbox{ and }\quad \ga := 3\al^2(1+k) +\bt^2(m-2).
 \ee

\begin{figure}[!h]
\begin{center}
\includegraphics[width=11cm, height=5.5cm]{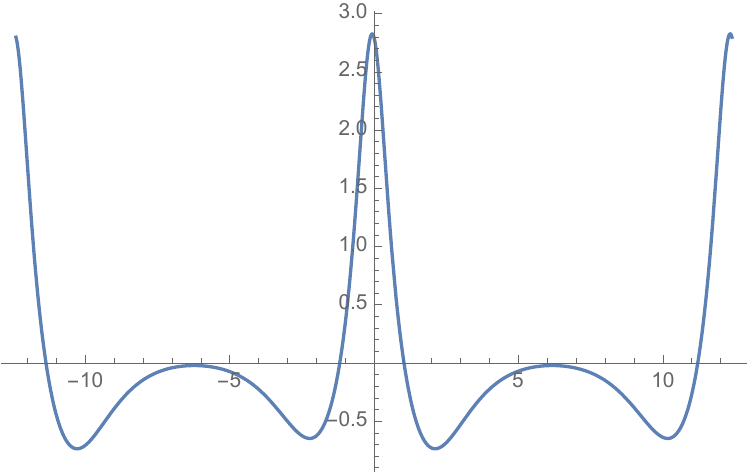}
\caption{The graph of the KKSH breather at time $t=0$, for $\bt=1$, $k=0.001$, $m=0.999$ and $L=12.398$.}\label{One_case}
\end{center}
\end{figure}

See Fig. \ref{One_case} for a description of a KKHS breather solution and Appendix \ref{JEF} for a more detailed account on
the Jacobi elliptic functions $sn$ and $nd$ presented in \eqref{Bper}. Additionally, in order to be a periodic solution of mKdV, 
the parameters $m,k,\al$ and $\bt$ must satisfy the commensurability conditions on
the spatial periods
\be\label{Cond1}
\frac{\bt^4}{\al^4}=\frac{k}{1-m}, \qquad K(k)=\frac{\al}{2\bt} K(m),
\ee
where $K$ denotes the complete elliptic integral of the first kind, defined as \cite{By}
\be\label{Kint}
K(r):=\int_0^{\pi/2}(1-r\sin^2(s))^{-1/2}ds = \int_0^1((1-t^2)(1-rt^2))^{-1/2}dt,
\ee
and which satisfies 
\[
K(0) =\frac \pi2 \quad \hbox{ and }\quad \lim_{k\to 1^-}K(k)=\infty.
\]
Under these assumptions, the spatial period is given by
\be\label{Largo}
L:=\frac{4}{\alpha}K(k) =\frac{2}{\bt} K(m).
\ee
Note that conditions \eqref{Cond1}  formally imply that $B$ has only four independent parameters (e.g. $\bt,~k$ and translations $x_1,~x_2$). 
Additionally, if we assume that 
the ratio $\bt/\al$ stays bounded, we have that $k$ approaches $0$ as $m$ is close to $1$. Using this information, the standard non
periodic mKdV breather \eqref{mKdVB} can be formally recovered as the limit of very large spatial period $L\to +\infty$, obtained e.g. 
if $k\to 0.$  In that sense, we can think of \eqref{Bper} as a nontrivial periodic bifurcation at infinity of the aperiodic mKdV breather.

\ms

\begin{figure}[!h]
\begin{center}
\includegraphics[width=10cm, height=5.5cm]{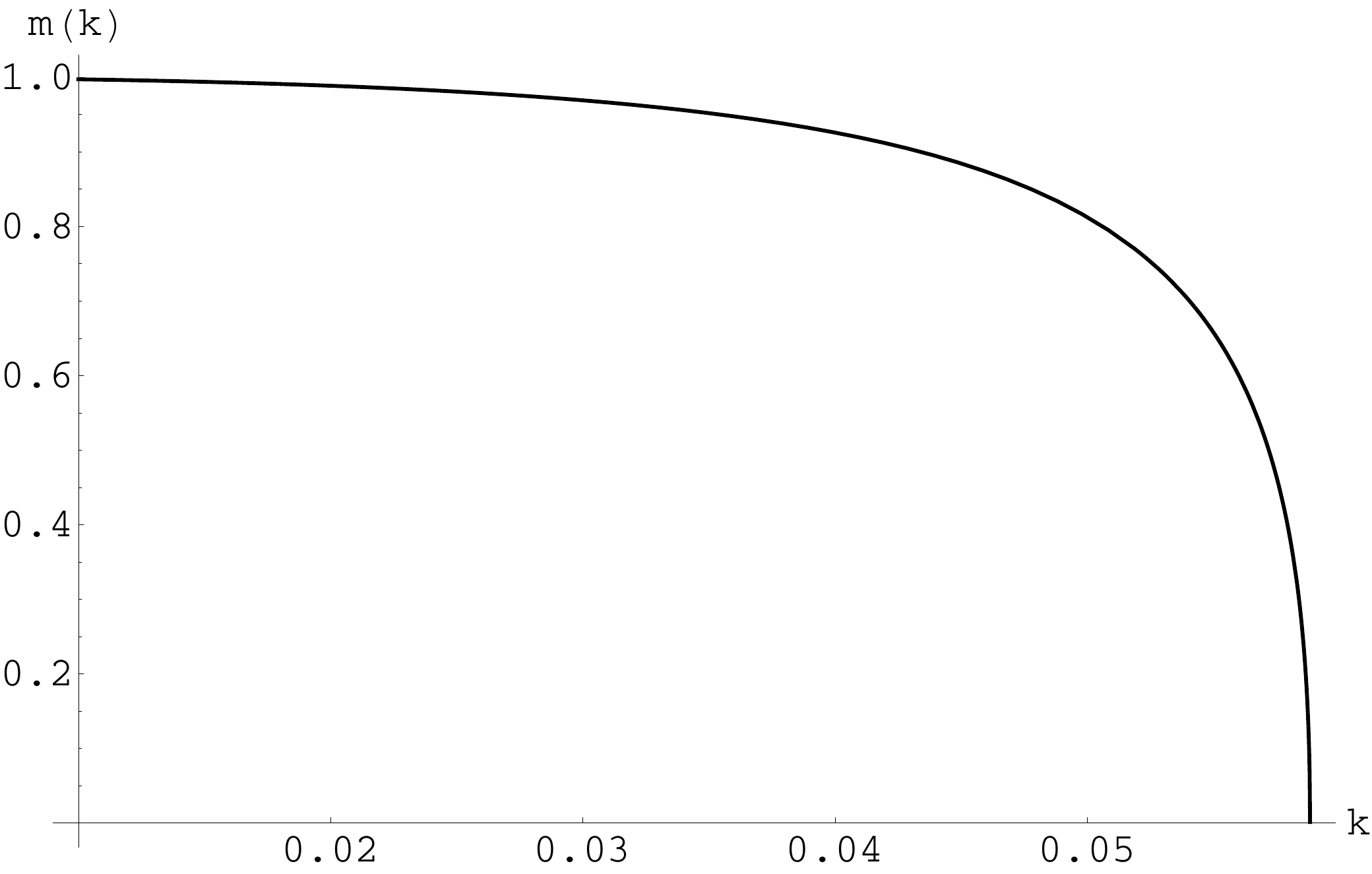}\caption{The graph of $m$ as a function of
$k$ obtained solving conditions \eqref{Cond1}. Although $m$ runs from 0 to 1, 
not all values of $k \in [0,1]$ are allowed, being the limiting value of $k\sim0.05883626$.}\label{km}
\end{center}
\end{figure}

It is important to mention that conditions \eqref{Cond1} impose a particular set of restrictions on $k$ and $m$. 
In terms of $k$, one has $k\in [0,k^*)$, with $k^*\sim 0.058$, while $m$ is decreasing with respect to $k$, with $m(k=0)=1$ and $m(k\sim k^*) \sim 0$. 
Below, Fig. \ref{km} describes the behavior of $k$ and $m$ more clearly.

\medskip

In what follows, we will use the convention that 
\be\label{dependence}
m=m(k), \quad \alpha=\alpha(\bt,k),
\ee
obtained by solving $m=m(k)$ in \eqref{Cond1} numerically, and then solving for $\alpha$ algebraically.

\bigskip

\subsection{Variational characterization} Our first result is the following theorem.

\begin{thm}\label{GBp} Assume \eqref{Cond1}. Let $B$ be any KKSH breather. Define 
\be\label{a1}
a_1 :=2(\bt^2(2-m)-\al^2(1+k)) = -\frac12(\delta+\ga),
\ee
and
\be\label{a2}
a_2:=   \al^4(1+k^2-26k) + 2\al^2\bt^2(2-m)(1+k) + \bt^4m^2 . 
\ee
Then $B$ satisfies the generalized nonlinear elliptic equation  
\be\label{EcBp}
B_{(4x)} + 5 BB_x^2 + 5B^2 B_{xx} + \frac 32 B^5 - a_1(B_{xx}+B^3)  + a_2 B  =0.
\ee
\end{thm}

We emphasize that the first condition in \eqref{Cond1} is essential for the proof of \eqref{EcBp}. However, for the proof we do not need the 
second one. Note additionally that $a_1$ and $a_2$ converge to the corresponding constants for the aperiodic mKdV case when $k\to 0$ and $m\to 1$, 
see \eqref{H01}. 

\ms

Now we introduce the conserved quantities
\be\label{Mass_per}
M_\#[u](t)  :=  \frac 12 \int_\T u^2(t,x)dx,
\ee
\[
E_\#[u](t)  :=  \frac 12 \int_\T u_x^2(t,x)dx -\frac 14 \int_\T u^4(t,x)dx,
\] 
and
\[
F_\#[u](t)    :=    \frac 12 \int_\T u_{xx}^2(t,x)dx -\frac 52 \int_\T u^2u_x^2(t,x)dx  + \frac 14 \int_\T u^6(t,x)dx,
\]
which are preserved in the space
\be\label{H2T}
H^2(\T) := \{ v \in H^2(0,L) \ : \ v(0)=v(L), \; v_x(0) =v_x(L), \; v_{xx}(0) = v_{xx}(L)\},
\ee
if $u(t,\cdot) \in H^2(\T)$ is a solution of \eqref{mKdV}. The proof of this result is straightforward if we work by density
in a space of smooth functions and we note that  $u(t,0) = u(t,L)$ and $u_x(t,0) = u_x(t,L) $  for all time imply
\[
u_t(t,0) = u_t(t,0), \quad  u_{xt}(t,0) = u_{xt}(t,L) ,
\]
and therefore, using \eqref{mKdV}, $u_{xxx}(t,0)=u_{xxx}(t,L).$

\begin{cor}\label{timeD}
KKSH breathers are critical points of the functional
\[
\mathcal H_\#[u] := F_\#[u](t) +a_1E_\#[u](t) + a_2M_\#[u](t),
\]
defined in the space $H^2(\T)$ and preserved along the mKdV periodic flow.
\end{cor}

As in the previous sections, the next step is the study of the nonlinear stability of this solution. However, we must emphasize that the existence of a 
suitable elliptic equation does not imply stability. Even worse, \emph{spectrally stable} breathers may be nonlinearly unstable. Indeed, 
we present below compelling evidence that 
periodic mKdV breathers are \emph{unstable} under a suitable type of periodic perturbations, at least for $L$ not so 
large\footnote{If $L$ is large it seems that the periodic breather is close in a certain topology to the aperiodic breather, 
which satisfies stability properties in a very general open neighborhood.}. Our results do agree with the numerical ones obtained by 
Kevrekidis et al \cite{KKS2,KKS1}, and our numerical computations below. In this case, the periodic character of the solution leads to
nontrivial interactions between adjacent  breathers, which probably play an important role in the instability character of this solution.

\subsection{Proof of Theorem \ref{GBp}}\label{3}

Let $B$ be a periodic KKSH breather. Without loss generality, we can assume $x_1 =x_2 =0$, and after taking time derivative we
assume $t=0$, since \eqref{mKdV} is invariant under space and time translations, as well as \eqref{GBp}. Recall that from \eqref{Bper}
\[
\tilde B_t := \delta \tilde B_1 + \ga \tilde B_2,
\]
where $\tilde B_j := \partial_{x_j}\tilde B$. We also have \cite{AS,By}:
\[
\sn'(s,k) = \con(s,k) \dn(s,k), \quad \Big(\dn(s,k) := \frac1{\nd(s,k)} \Big),  \quad \con' (s,k)=- \sn(s,k)\dn(s,k),
\]
\[
\dn' (s,k)= -k \sn (s,k)\con(s,k),
\]
and $\con(0,k)=\dn(0,k) =1$, $\sn(0,k)=0$.

\ms

We start with some notation. Let
\[
\sn_1 := \sn(\al y_1,k), \quad \con_1 := \con(\al y_1,k) , \quad \dn_1 := \dn(\al y_1,k);
\]
\[
\sn_2 := \sn(\bt y_2,m), \quad \con_2 := \con(\bt y_2,m) , \quad \dn_2 := \dn(\bt y_2,m).
\]
We have
\[
\tilde B = 2\sqrt{2} \arctan \Big(\frac\bt\al \sn_1\dn_2 \Big).
\]
Define
\be\label{hth}
h := \al \con_1\dn_1\dn_2 - \bt m \sn_1\sn_2\con_2, \quad \tilde h:=\al \delta \con_1\dn_1\dn_2 - \bt m\ga \sn_1\sn_2\con_2,
\ee
so that 
\be\label{hx}
h_x = - \Big[ \al^2\sn_1\dn_2(k\con_1^2+\dn_1^2)  + 2\al\bt m\con_1\dn_1\sn_2\con_2 + \bt^2 m \sn_1\dn_2(\con_2^2-\sn_2^2) \Big],
\ee
and
\be\label{ht}
h_t = - \Big[\al^2 \delta \sn_1 \dn_2 (k \con_1^2 + \dn_1^2) + \al\bt m(\delta+\ga) \con_1 \dn_1 \sn_2\con_2  + \bt^2 m\ga \sn_1\dn_2 (\con_2^2 -\sn_2^2) \Big].  
\ee
Similarly
\be\label{gg}
g := \al^2 + \bt^2 \sn_1^2\dn_2^2, \quad g_x = 2\bt^2\sn_1 \dn_2 ( \al\con_1\dn_1 \dn_2 - \bt m\sn_1 \sn_2\con_2),
\ee
and
\be\label{gt}
g_t =2\bt^2 \sn_1 \dn_2 (\al \delta \con_1\dn_1\dn_2 - \bt m \ga \sn_1\sn_2\con_2) =2\bt^2 \sn_1 \dn_2 \tilde h.
\ee
Then
\[
B= \frac{2\sqrt{2}\al\bt h}{g}, \quad B(x=0) =2\sqrt{2} \bt,
\]
and
\[
\tilde B_t = \delta \tilde B_1 + \ga \tilde B_2  =  \frac{2\sqrt{2}\al\bt \tilde h}{g}, \quad \tilde B_t(x=0) = 2\sqrt{2}\bt \delta.
\]
Similarly
\be\label{Bxper}
B_{x} = \frac{2\sqrt{2}\al\bt  \hat h}{g^2}, \quad \hat h:= h_x g- g_xh.
\ee
It is not difficult to check that $h_x(x=0) =g_x(x=0) =0.$
In particular $B_x(x=0)=0$. From this identity we see that 
\[
B_{xx}(x=0) =\frac{2\sqrt{2}\al\bt \hat h_x}{g^2}(x=0) = -2\sqrt{2} \bt [ (2+3m)\bt^2 +(1+k)\al^2 ].
\]
First of all, we have from \eqref{mKdV} and \eqref{DG}
\[
\tilde B_t +B_{xx} + B^3 = (\tilde B_t +B_{xx} + B^3)(x=0) = 0.
\]
This identity can be proved to hold {\bf for any} $t,x_1,x_2\in \R$.  On the other hand, if
\[
\mathcal M:=  \frac 12\int_0^x B^2, \qquad \mathcal M_t =  \int_0^x B B_t,
\]
we have
\[
B \tilde B_t - \mathcal M_t + \frac 12 B_{x}^2 + \frac 14B^4 =(B \tilde B_t +  \frac 12 B_{x}^2 + \frac 14B^4)(x=0) =:  \frac 12c_0,
\]
where for $t=x_1=x_2=0$ we have that  $c_0$ is explicitly given by
\be\label{c0}
c_0 := 16\bt^2 [\al^2 (1+k) + \bt^2(3m-4)]. 
\ee
However, in the general case  $c_0$ may depend on time. 

\ms 

Replacing in \eqref{EcBp} we obtain
\bea
& & B_{(4x)} + 5 BB_x^2 + 5B^2 B_{xx} + \frac 32 B^5 - a_1(B_{xx}+B^3)  + a_2 B   \nonu \\
& &  \qquad = -(B_t + 3B^2B_x)_{x}+ 5 BB_x^2 + 5B^2 B_{xx} + \frac 32 B^5 - a_1(B_{xx}+B^3)  + a_2 B \nonu  \\
& &  \qquad = -B_{tx}  - BB_x^2 + 2 B^2 B_{xx} + \frac 32 B^5 - a_1(B_{xx}+B^3)  + a_2 B  \nonu \\
& &  \qquad = -B_{tx}  - B(c_0 - 2 B\tilde B_t +2\mathcal M_t -\frac 12 B^4) -  2 B^2 ( B^3 + \tilde B_t) + \frac 32 B^5 + a_1\tilde B_t  + a_2 B \nonu  \\
& &  \qquad = -B_{tx}  - 2 B \mathcal M_t   + a_1\tilde B_t  + (a_2-c_0) B .\label{Ecuacion}
\eea
Now we prove that this last quantity is identically zero. We compute $\mathcal M_t.$ Denote
\be\label{F_G}
F:= \frac{1}{2\al} \sn_1 \sn_1' + \frac 1{2\bt } \nd_2 \nd_2' , \quad G:= \bt^2 \sn_1^2 + \al^2 \nd_2^2,
\ee
where, with a slight abuse of notation we denote
\[
\sn_1' := \sn'(\al y_1,k), \quad \con_1' := \con'(\al y_1,k) , \quad \dn_1' := \dn'(\al y_1,k);
\]
and so on.  We claim that 
\be\label{B22}
B^2 = 8\al^2\bt^2(\frac{F}{G})_x-4\al^2\frac{k}{\bt^2}(\bt^2 \sn_1^2 - \al^2 \nd_2^2).
\ee

{\bf Proof of \eqref{B22}}.  From \eqref{Bper} we have
\be\label{Bquad}
B^2 =8\al^2\bt^2\frac{\al^2 \nd_2^2\sn_1'^2 + \bt^2\sn_1^2\nd_2'^2 - 2\al\bt \sn_1 \nd_2 \sn_1 ' \nd_2'}{(\bt^2 \sn_1^2 + \al^2 \nd_2^2)^2}.
\ee
Note that
\bee
F_xG - FG_x & =&  \frac12  (  \sn_1'^2 +  \sn_1\sn_1''  +  \nd_2'^2 +  \nd_2\nd_2''  )(\bt^2 \sn_1^2 + \al^2 \nd_2^2) \\
& &  -( \bt \sn_1 \sn_1' +\al  \nd_2 \nd_2')  (\bt \sn_1\sn_1' + \al \nd_2 \nd_2') .
\eee
Now we use the well-known identities \cite{By}
\be\label{ecu1}
\begin{aligned}
\sn_1'^2 &= 1 - (1+k)\sn_1^2  + k \sn_1^4, \qquad \sn_1'' = -(1+k) \sn_1 +2k \sn_1^3, \\
\nd_2'^2 &= -1+(2-m)\nd_2^2 +(m-1)\nd_2^4, \qquad \nd_2'' =(2-m) \nd_2 +2(m-1)\nd_2^3.
\end{aligned}
\ee
Replacing above we obtain
\bee
F_xG - FG_x & =&  \al^2 \nd_2^2  \sn_1'^2 + \bt^2 \sn_1^2  \nd_2'^2 - 2\al\bt \sn_1 \sn_1'\nd_2 \nd_2' \\
& & + \textcolor{black}{ \frac12 (\bt^2k \sn_1^6 + \al^2  k  \nd_2^2 \sn_1^4 + \bt^2 (m-1) \sn_1^2 \dn_2^4 + \al^2 (m-1) \dn_2^6)} \\
& = & \al^2 \nd_2^2  \sn_1'^2 + \bt^2 \sn_1^2  \nd_2'^2 - 2\al\bt \sn_1 \nd_2 \sn_1' \nd_2'\\
& & +  \frac{k}2 \sn_1^4(\bt^2 \sn_1^2+\al^2 \nd_2^2) + \frac{(m-1)}2 \nd_2^4(\bt^2 \sn_1^2+\al^2\nd_2^2).
\eee
Therefore, from \eqref{Bquad},
\be\label{B2per}
B^2  =  8\al^2 \bt^2 \frac{(F_xG-FG_x)}{G^2} - 4\al^2\bt^2\textcolor{black}{\frac{k\sn_1^4 + (m-1)\nd_2^4}{(\bt^2 \sn_1^2 + \al^2 \nd_2^2)}}. 
\ee
Since $\frac{\bt^4}{\al^4}=\frac{k}{1-m}$, \eqref{B2per} simplifies as follows:
\[
B^2 = 8\al^2\bt^2(\frac{F}{G})_x-4\al^2 \frac{k}{\bt^2}(\bt^2\sn_1^2 - \al^2\nd_2^2),
\]
as desired. From \eqref{B22} we have
\be\label{Mass_x}
\frac 12\int_0^x B^2 =  4\al^2\bt^2 \frac{F}{G} -4\al^2\bt^2 \frac{F}{G}(0)  -2\al^2\frac{k}{\bt^2} \int_0^x (\bt^2\sn_1^2 - \al^2\nd_2^2).
\ee
Since
\[
\frac{F}{G} = \frac{\bt \sn_1 \con_1 \dn_1\dn_2^2 +\al m \sn_2\con_2 \nd_2}{2\al\bt  g} =: \frac{\bar h}{2\al\bt  g} ,
\]
we obtain
\[
\mathcal M_t = 2\al\bt \frac{(g \bar h_t - g_t \bar h)}{g^2} -2\al\bt \Big(\frac{(g \bar h_t - g_t \bar h)}{g^2} (x=0)\Big) \Big|_{t=0}  
-2\al^2\frac{k}{\bt^2}  (\delta \bt^2 \sn_1^2 -  \ga \al^2 \nd_2^2  +\ga \al^2).
\]
The second term above can be computed explicitly. We have
\[
\mathcal M_t = 2\al\bt \frac{(g \bar h_t - g_t \bar h)}{g^2}  -2\bt^2 (\delta+m\ga)  -2\al^2\frac{k}{\bt^2}  (\delta \bt^2 \sn_1^2 - 
\ga \al^2 \nd_2^2  +\ga \al^2).
\]
Let us calculate $B_{tx}$. Using \eqref{Bxper}, we have
\[
B_{tx} =  \frac{2\sqrt{2} \al\bt}{g^3} ( g \hat h_t -2g_t \hat h).
\]
Now we compute the term $-B_{tx} - 2B\mathcal M_t$. We have
\bea\label{Final0}
-B_{tx} - 2B\mathcal M_t  & =&   - \frac{2\sqrt{2} \al\bt}{g^3} \Big[ \hat h_t g -2  g_t \hat h + 4 \al\bt (g \bar h_t - g_t \bar h) h   \nonu\\
& & \qquad  - 4 \al^2\frac{k}{\bt^2} g^2 (\delta \bt^2 \sn_1^2 -  \ga \al^2 \nd_2^2  +\ga \al^2)  h  \Big]  + 4\bt^2 (\delta+m\ga) B.
\eea
We consider the term $\hat h + 2\al\bt h \bar h.$ We have
\be\label{Final1}
\hat h + 2\al\bt h \bar h = h_x g +  h( 2\al\bt  \bar h -g_x).
\ee
The term $2\al\bt  \bar h -g_x$ reads now
\bee
2\al\bt  \bar h -g_x & = & 2\al\bt (\bt \sn_1 \con_1 \dn_1\dn_2^2 +\al m \sn_2\con_2 \nd_2)  \\
& & - 2\bt^2 ( \al \sn_1 \con_1\dn_1 \dn_2^2 -m \bt \sn_1^2 \sn_2\con_2 \dn_2) \\
& = & 2\bt m\sn_2\con_2 (\al^2 \nd_2 + \bt^2 \sn_1^2 \dn_2 ) \ =  2\bt m\sn_2\con_2 \nd_2 g.
\eee
Consequently,  $\eqref{Final1} = g(h_x  + 2\bt m\sn_2\con_2 \nd_2  h)$, and replacing in \eqref{Final0},
\be\label{Final3}
\begin{aligned}
-B_{tx} - 2B\mathcal M_t  & =  - \frac{2\sqrt{2} \al\bt}{g^2} \Big[ \hat h_t  -2  g_t (h_x  + 2\bt m\sn_2\con_2 \nd_2  h) + 4 \al\bt  \bar h_t  h \\
 &  \qquad    - 4 \al^2\frac{k}{\bt^2} g (\delta \bt^2 \sn_1^2 -  \ga \al^2 \nd_2^2  +\ga \al^2)  h  \Big] + 4\bt^2 (\delta+m\ga) B. 
\end{aligned}
\ee
Since $ \hat h_t = h_{tx} g + h_x g_t - g_{xt}h - g_x h_t$, we are left to compute the term
\be\label{Final2}
4\bt h (\al \bar h_t  -\frac1{4\bt} g_{tx} - m \sn_2\con_2 \nd_2 g_t ) -h_x g_t   -g_x h_t.
\ee
Note that 
\[
\al \bar h_t  -\frac1{4\bt} g_{tx} = ( \al \bar h -\frac1{4\bt} g_x)_t.
\]
From \eqref{gg} we have
\[
 \al \bar h -\frac1{4\bt} g_x = m \sn_2 \con_2 \nd_2 g+  \frac12 \bt  \Big[ \al \sn_1 \con_1 \dn_1\dn_2^2  -  \bt m \sn_1^2 \sn_2 \con_2\dn_2   \Big],
\]
and
\[
\begin{aligned}
\eqref{Final2} &=  4\bt m  ( \sn_2 \con_2 \nd_2)_t  g h   +  2\bt^2 h \Big[  \al\sn_1 \con_1 \dn_1\dn_2^2  -\bt m \sn_1^2 \sn_2 \con_2\dn_2 \Big]_t   -(h_x g_t   + g_x h_t) \\
& = 4\bt^2 m \ga   ( \sn_2 \con_2 \nd_2)' g h  + 2\bt^2 h \Big[ \al \sn_1 \con_1 \dn_1\dn_2^2  -\bt m \sn_1^2 \sn_2 \con_2\dn_2 \Big]_t -(h_x g_t   + g_x h_t) .
\end{aligned}
\]
Now we have
\bee
& & (\al \sn_1 \con_1 \dn_1\dn_2^2  -\bt m \sn_1^2 \sn_2 \con_2\dn_2 )_t  =  \al^2 \delta  \dn_2^2 [\dn_1^2(\con_1^2 -\sn_1^2) -k \sn_1^2\con_1^2] \\
&& \qquad \qquad  -2 \al\bt m (\delta+\ga) \sn_1\con_1\dn_1\sn_2\con_2\dn_2 -\bt^2 m \ga \sn_1^2 [  \dn_2^2 (\con_2^2-\sn_2^2) -m \sn_2^2\con_2^2 ],
\eee
and since $h = \al \con_1\dn_1\dn_2 - \bt m \sn_1\sn_2\con_2$,
\[
\begin{aligned}
&   2 \bt^2 h (  \al \sn_1 \con_1 \dn_1\dn_2^2  -\bt m \sn_1^2 \sn_2 \con_2\dn_2 )_t   \\
& =  2\bt^2 \Big[  \al^3 \delta \con_1\dn_1 \dn_2^3[\dn_1^2(\con_1^2 -\sn_1^2) -k\sn_1^2\con_1^2 ]  - 2 \al^2 \bt m(\delta+\ga) \sn_1\con_1^2 \dn_1^2 \sn_2\con_2 \dn_2^2 \\
&    \qquad \quad - \al\bt^2 m \ga \sn_1^2\con_1\dn_1\dn_2 \ [\dn_2^2(\con_2^2-\sn_2^2) -m\sn_2^2\con_2^2]  \\
&  \qquad \quad-  \al^2\bt m \delta \sn_1\sn_2\con_2 \dn_2^2 \ [\dn_1^2(\con_1^2 -\sn_1^2) -k\sn_1^2\con_1^2 ]   \\
&   \qquad \quad+  2\al\bt^2 m^2(\delta+\ga) \sn_1^2 \con_1\dn_1\sn_2^2 \con_2^2 \dn_2 + \bt^3 m^2\ga  \sn_1^3 \sn_2 \con_2  [\dn_2^2(\con_2^2-\sn_2^2) -m\sn_2^2\con_2^2] \Big] .
\end{aligned}
\]
On the other hand, using \eqref{ht}, \eqref{hx}, \eqref{gg} and \eqref{gt},
\bee
& & -(g_t h_x   + g_x h_t)= \\
& & \quad = 2\bt^2 \sn_1\dn_2 \Big[  ( \al \delta\con_1\dn_1\dn_2-  \bt m \ga \sn_1\sn_2\con_2) \times\\
& &  \qquad   \times (\al^2 \sn_1 \dn_2 (k\con_1^2 +\dn_1^2) + 2 \al\bt m \con_1\dn_1\sn_2 \con_2 +\bt^2m \sn_1\dn_2(\con_2^2-\sn_2^2) )\\
& &  \qquad \qquad  \qquad \qquad + \ (\al \con_1\dn_1 \dn_2 -\bt m\sn_1\sn_2\con_2) \times\\
& & \qquad  \times ( \al^2\delta \sn_1 \dn_2 (k\con_1^2 +\dn_1^2) +  \al\bt m (\delta+\ga) \con_1\dn_1\sn_2 \con_2 +\bt^2m \ga \sn_1\dn_2(\con_2^2-\sn_2^2)) \\
& &\quad  = 2\bt^2 \Big[  2\al^3\delta \sn_1^2 \con_1 \dn_1\dn_2^3 (k\con_1^2 +\dn_1^2) +    \al^2\bt m (3\delta+\ga)\sn_1 \con_1^2\dn_1^2\sn_2 \con_2\dn_2^2 \\
& &  \qquad  + \al\bt^2m (\delta+\ga) \sn_1^2 \con_1\dn_1\dn_2^3 (\con_2^2-\sn_2^2)  -\al^2\bt m (\delta+\ga) \sn_1^3 \sn_2\con_2\dn_2^2(k\con_1^2+\dn_1^2)\\
& &\qquad  -\al \bt^2 m^2(\delta +3\ga) \sn_1^2  \con_1 \dn_1 \sn_2^2 \con_2^2 \dn_2  - \ 2\bt^3 m^2\ga \sn_1^3 \sn_2\con_2\dn_2^2 (\con_2^2-\sn_2^2)  \Big].
\eee
Rearranging similar terms, and using the identities (\cite{By})
\[
\sn_1^2 + \con_1^2 =1, \quad k \sn_1^2 + \dn_1^2 =1, \quad m \con_2^2 + 1-m =\dn_2^2, \ldots,
\]
and the fact that $k\al^4 = (1-m)\bt^4$, we get
\bee
& & 2\bt^2 h ( \al \sn_1 \con_1 \dn_1\dn_2^2  -\bt m \sn_1^2 \sn_2 \con_2\dn_2 )_t -(h_x g_t   + g_x h_t)= \\
& &  = 2\bt^2 \Big[  \al^3 \delta \con_1\dn_1\dn_2^3 ( \dn_1^2 + k\sn_1^2 \con_1^2) + \bt^3m^2 \ga \sn_1^3 \sn_2 \con_2(\sn_2^2 \dn_2^2-\con_2^2 )  \\
& & \qquad \quad  -\al^2\bt m \ga \sn_1\sn_2\con_2\dn_2^2 ( \dn_1^2 + k\sn_1^2 \con_1^2  ) 
  - \al\bt^2 m \delta \sn_1^2 \con_1 \dn_1 \dn_2  (  \sn_2^2 \dn_2^2 - \con_2^2   ) \Big] \\  
& & =     2\bt^2 (\al\delta \con_1\dn_1 \dn_2 -\bt m \ga \sn_1 \sn_2 \con_2 )( \al^2 \dn_2^2 (\dn_1^2 + k \sn_1^2 \con_1^2)  + \bt^2 m \sn_1^2 (\con_2^2 -\sn_2^2 \dn_2^2) )\\
& & = 2\bt^2  \tilde h [    \al^2 \dn_2^2 (1-k \sn_1^4 ) + \bt^2 \sn_1^2 \dn_2^2 -(1-m) \bt^2 \sn_1^2 -\bt^2 m \sn_1^2\sn_2^2 \dn_2^2   ] \\
& & =2\bt^2  \tilde h [  \dn_2^2 \{ \al^2  (1-k \sn_1^4 ) + \bt^2 \sn_1^2  -\bt^2 \sn_1^2(1-\dn_2^2) \}  -   \al^4 \bt^{-2} k \sn_1^2   ] \\
& &  = 2\bt^2  \tilde h g ( \dn_2^2   -  \al^2 \bt^{-2}k \sn_1^2  ).
\eee
From \eqref{Final3} we conclude that 
\[
\begin{aligned}
-B_{tx} - 2B\mathcal M_t  & =  - \frac{2\sqrt{2} \al\bt}{g} \Big[ h_{tx} + 2  \tilde h  (  \bt^2 \dn_2^2   -  \al^2 k \sn_1^2  )  + 4\bt^2 m \ga   ( \sn_2 \con_2 \nd_2)'  h  \nonu \\
 &  \qquad   \qquad  - 4 \al^2\frac{k}{\bt^2}  (\delta \bt^2 \sn_1^2 -  \ga \al^2 \nd_2^2  +\ga \al^2)  h  \Big] + 4\bt^2 (\delta+m\ga) B. \label{Final4}
\end{aligned}
\]
Note that (\cite{By}) $ ( \sn_2 \con_2 \nd_2)'  =  \con_2^2-\sn_2^2  + m \sn_2^2\con_2^2 \nd_2^2.$ Consequently,
\[
\begin{aligned}
&  4\bt^2 m \ga   ( \sn_2 \con_2 \nd_2)'  - 4 \al^2\frac{k}{\bt^2}  (\delta \bt^2 \sn_1^2 -  \ga \al^2 \nd_2^2  +\ga \al^2) =  \\
&  \qquad =    4\bt^2 m \ga (\con_2^2-\sn_2^2)  -4\al^2 k \delta \sn_1^2 +  4 \bt^2  \ga \nd_2^2 (   m^2 \sn_2^2\con_2^2 +  k \frac{\al^4}{\bt^4}(1-\dn_2^2) ) \\
&  \qquad =    4\bt^2 m \ga (\con_2^2-\sn_2^2)  -4\al^2 k \delta \sn_1^2 +  4  \bt^2 m  \ga  \sn_2^2 \nd_2^2 ( m \con_2^2 + k\frac{\al^4}{\bt^4} )\\
&  \qquad =    4\bt^2 m \ga (\con_2^2-\sn_2^2)  -4\al^2 k \delta \sn_1^2 +  4  \bt^2 m \ga   \sn_2^2  =   4\bt^2 m \ga \con_2^2   -4\al^2 k \delta \sn_1^2 .  
\end{aligned}
\]
Finally we compute $h_{tx}$. From \eqref{ht} we have
\bee
h_{tx} & =&  -\Big[  \al^2(k\con_1^2 +\dn_1^2) ( \al\delta \con_1\dn_1\dn_2 -\bt m(2\delta+\ga) \sn_1\sn_2 \con_2 ) \\
& &  \qquad + \bt^2 m(\con_2^2 -\sn_2^2) (\al(\delta+2\ga) \con_1\dn_1\dn_2 -\bt m\ga \sn_1\sn_2 \con_2 ) \\
& & \qquad - 4\al^2 k  \delta  \sn_1^2 (\al \con_1\dn_1\dn_2)   - 4 \bt^2  \ga  ( \bt m \sn_1\sn_2 \con_2) \dn_2^2  \Big].
\eee
Using the last two identities and some standard simplifications, \eqref{Final4} becomes
\[
\begin{aligned}
-B_{tx} - 2B\mathcal M_t  & = - \frac{2\sqrt{2} \al\bt}{g} \Big[  [-\al^2 (1+k) \delta  + \bt^2((2-m)\delta + 2m\ga)] (\al \con_1\dn_1\dn_2)  \\
&  \qquad \qquad   + [ \al^2 (1+k) (2\delta+\ga) +\bt^2 (2-3m)\ga ] ( \bt m \sn_1\sn_2 \con_2)\Big]  + 4\bt^2 (\delta+m\ga) B\\
&  =   -a_1 \tilde B_t - \tilde a_2 B,
\end{aligned}
\] 
where $a_1$ is defined in \eqref{a1}, and
\[
\tilde a_2 =\al^4 (1+k)^2 - 2\al^2\bt^2(1+k)(m+6) + \bt^4(2-m)(18-m).  
\]
Comparing with  \eqref{Ecuacion} we have ($c_0$ is given by \eqref{c0})
\[
a_2 = \tilde a_2 + c_0 = \al^4(1+k)^2 + 2\al^2\bt^2 (1+k)(2-m) +\bt^4(m^2+28m-28).
\]
Finally we use that $k\al^4 = \bt^4(1-m)$ to get
\[
a_2 =\al^4(1+k^2 -26 k) + 2\al^2\bt^2 (1+k)(2-m) +\bt^4m^2,
\]
as in \eqref{a2}. The proof is complete.
\bs

\section{Spectral Analysis of the periodic mKdV breather}\label{Sect:6}

\ms

\subsection{Mathematical description} In this section we give further evidence of the stable-unstable character 
of the KKSH breather solution depending on the parameters phase space. First of all, let us notice that, thanks to \eqref{EcBp},
the linearized operator for a KKSH breather is given by the expression
\be\label{L_per}
\begin{aligned}
\mathcal L_\#[z] & :=z_{(4x)} +(5 B^2 - a_1) z_{xx} + 10 B B_x z_x  + \Big(a_2 + 5B_x^2 +  10BB_{xx} + \frac{15}2 B^4 - 3a_1 B^2 \Big)z. 
\end{aligned}
\ee
This operator is defined acting on functions in $H^2(\T)$, $\T=(0,L)$, see \eqref{H2T}, and reduces to the standard aperiodic mKdV breather operator
\eqref{LmKdV} as the length of the interval tends to infinity (or $k\to 0$). The constants $a_1$ and $a_2$ were introduced in \eqref{a1} and \eqref{a2}, 
and we assume the convention \eqref{dependence}.  

\medskip

In the following lines, we analyze the spectral stability of the KKSH breather, namely the understanding of the 
spectrum of $\mathcal L_\#$ in \eqref{L_per}. First, we prove some useful expression for the mass of a breather. 
As a second step, we compute numerically the spectrum of $\mathcal L_\#$, and conclude that it has the desired spectral properties. 
Then, we analyze which property is the main responsible of the KKSH stability.

\subsection{Mass calculations}

The purpose of this paragraph is to compute the mass of the KKSH breather as a function of $k$ and $\beta$. 
This explicit function will be essential for the study of the nonlinear stability of the solution. 

\medskip

First of all, recall the breather profile in \eqref{Bper}. Since the mass $M_\#$ in \eqref{Mass_per} is conserved,
we can simply assume $t=x_1=x_2=0$ in \eqref{Bper}-\eqref{Y1_Y2}.  Consider the functions $F$ and $G$ defined in \eqref{F_G}, 
and the length of the interval $L$ defined in \eqref{Largo}. We have from \eqref{Mass_x},
\be
\frac 12\int_0^L B^2 =  4\al^2\bt^2 \frac{F}{G}(x=L) -4\al^2\bt^2 \frac{F}{G}(x=0)  -2\al^2\frac{k}{\bt^2} \int_0^L (\bt^2\sn_1^2 - \al^2\nd_2^2).
\ee

Now, note that from \eqref{Cond1} and the periodic character of the involved functions, we have
%
\be\label{B23}
\begin{aligned}
 \frac12\int_0^L B^2 & = 4\bt m \frac{ \cd(\frac{\bt}{2}L,m)\sd(\frac{\bt}{2}L,m) }{\nd(\frac{\bt}{2}L,m)} - 4\bt m \frac{\cne(\frac{\bt}{2}L,m) \sne(\frac{\bt}{2}L,m)}{\dne(\frac{\bt}{2}L,m)} \\
 & \quad -8\al K(k) + 4\al E( A(2 K(k),k),k) + 4\bt E\Big(A\Big(\frac{\bt}{2}L,m\Big),m \Big),
\end{aligned}
\ee
where $E$ denotes the {\it complete elliptic integral of the second kind} \cite{By}, defined as
\be\label{Eint}
E(r):=\int_0^{\pi/2}(1-r\sin^2(s))^{1/2}ds = \int_0^1(1-t^2)^{-1/2}(1-rt^2)^{1/2}dt,
\ee
and $A$ is the {\it Jacobi amplitude}, that can be defined by
\be\label{Aint}
A(x,r):=\int_0^{x}dn(s,r)ds.
\ee
Some simple exact values for $E$ are $E(0)=\frac{\pi}{2}$ and $E(1)=1$, and for $A$ are $A(0,r)=0$, $A(K(r),r)=\frac{\pi}{2}$. Note also that  $E(A(2 K(k),k),k) = E(A(0, k) + \pi,k)=2E(k).$
%
Using \eqref{Cond1}, \eqref{B23} simplifies as follows:
\be\label{B24} \frac12 \int_0^L B_0^2 = 4\bt E(m) + 8\al E(k) - 8\al K(k).\ee
\medskip
We are able to go one step further in the simplification of \eqref{B24}, using \eqref{Cond1} again. We have 
\be\label{relper}
\frac k{1-m}=\frac{1}{16}\frac{K(m)^4}{K(k)^4}.
\ee
Hence, with these relations, \eqref{B24} simplifies to the simple expression:
\be\label{B25}
M_\#[B]= \frac{1}{2}\int_0^L B_0^2 = 4\bt\left(E(m) + 4\frac{ K(k)}{ K(m)}( E(k) -  K(k))\right).
\ee
\medskip
When $k\rightarrow0$, we have $m\rightarrow1$, and $ M_\#[B] \rightarrow4\bt$, which is the value of the mass of the aperiodic mKdV breather solution in the
real line (see \cite[p.6, Lemma 2.1]{AM}).

\begin{figure}[!htb]
\centering
\includegraphics[width=11.0cm,height=12.5cm]{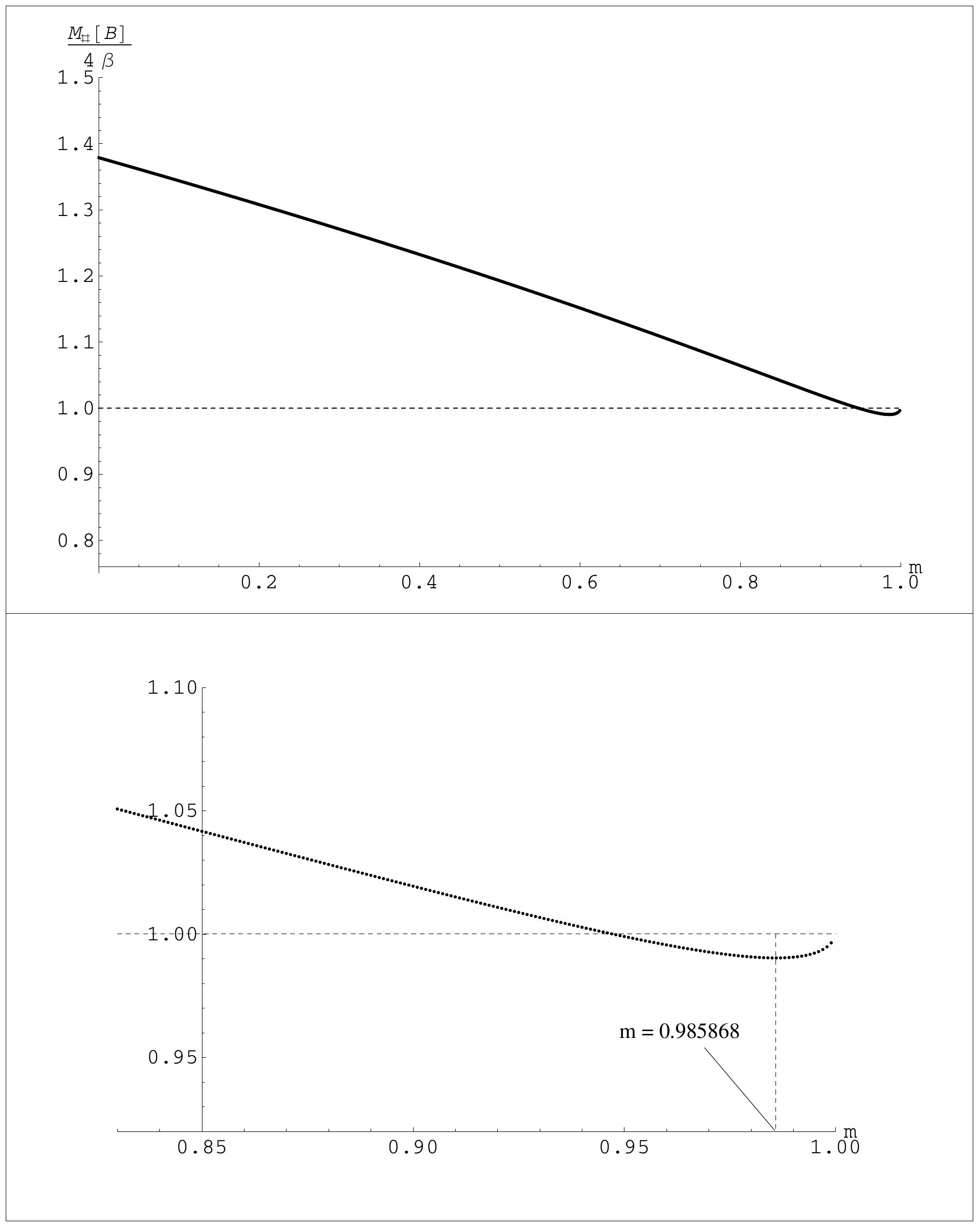}\caption{The mass of the periodic breather $M_\#[B]/4\bt$ as a function of $m$. 
Note that the resulting function is {\bf decreasing} except for $m\gtrsim 0.98$ (see zoom figure below),
corresponding to $k\lesssim 0.025$,  a parameter region very close to the stable, aperiodic mKdV breather.}\label{MassBper}
\end{figure}

\medskip

For $k\neq0,m\neq1$, the dependence of the {periodic mass} \eqref{B25} with respect to the parameter $m$ is computed in the following way:
for each value of $m$, we solve numerically the implicit equation \eqref{relper} in $k$. We then substitute these two pairs 
of values $(m,k)$ verifying \eqref{relper} inside the expression \eqref{B25}.  The resulting plot of \eqref{B25} versus $m$ is given in Fig. \ref{MassBper}.

\subsection{Numerical analysis}
Given the complicated functions that define the KKSH breather and its related linearized operator, a rigorous description of the spectra of $\mathcal L_\#$ has escaped to us. We perform then some numerical simulations to get a good understanding of these spectral properties. We could expect, given the instability results in \cite{KKS2}, that the KKSH breather should give rise to a very different spectrum, with respect to the previous aperiodic breathers. However, as we will show later, this is not the case.

\medskip

In the following lines we explain the main ideas of the numerical method used to compute the eigenvalues of $\mathcal L_\#$. The core of the algorithm is the same as in the previous sections, the main difference being the required test functions, which now must be periodic on $[0,L]$. We have used the classical orthonormal basis of $L^2(0,L)$:
\[
\Bigg\{ \frac 1{\sqrt{L}}, ~\sqrt{\frac{2}L} \cos \Big( \frac{2\pi n x}{L}\Big),~\sqrt{\frac{2}L} \sin \Big( \frac{2\pi n x}{L}\Big) \Bigg\},
\]
where $n \in \{1, \ldots, N\}$. For the standard numerical computations, it is enough to take 
$N=40$, although when approaching the critical values $k\to k^*$ ($m\to 0$) or $k\to 0$ ($m\to 1$), more and more test functions are naturally required.

\begin{figure}[!h]
\begin{center}
\includegraphics[width=10.0cm,height=5cm]{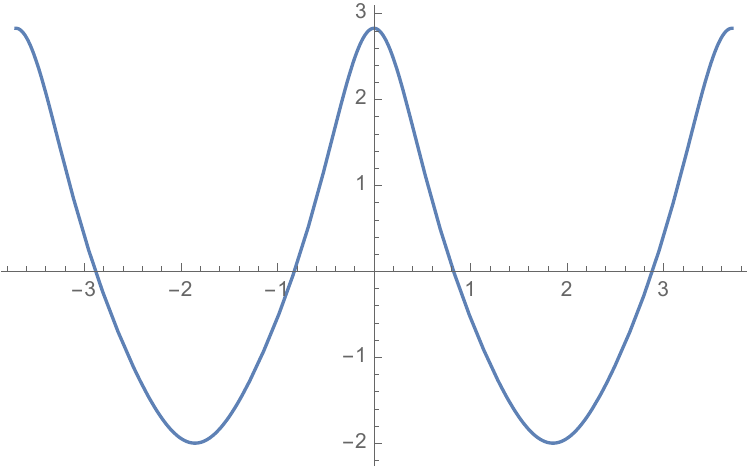}
\caption{Periodic KKSH breather \cite{KKS2} at time $t=0$ for $\beta=1$, $m=0.5$, $k=0.057$, here $L\sim 3.71$. 
Although numerically unstable \cite{KKS2}, this breather leads to a linearized operator $\mathcal L_\#$ that is spectrally stable, that is, it possesses
a ``standard'' one negative eigenvalue and a two dimensional kernel \eqref{results1}.}\label{KKHS_b}
\end{center}
\end{figure}

\ms

First of all, we run a specific computation for the case described in \cite{KKS2}, which the authors describe as numerically unstable. In this case, $\beta=1$, $m=0.5$, $k=0.057$, and $L\sim 3.71$ (see Fig. \ref{KKHS_b}). 
Numerically, we have found that this breather possesses the same spectral structure of all our previous breather solutions. 
To be more precise, running our numerical algorithm with $N= 40$ test functions, 
we have found the following approximations of the first four eigenvalues of $\mathcal L_\#:$
\be\label{results1}
\{-4.86, ~ -1.23\times 10^{-8}, ~ 3.22\times 10^{-10}, ~ 35.35\}.
\ee
Clearly the two components of the kernel are recovered with high precision (recall that a second negative eigenvalue but
very close to zero is unlikely just by continuity arguments on the coefficients of the original breather), 
and a distinctive negative eigenvalue appears, far from the kernel itself. This behavior repeats for all cases we have studied.

\begin{figure}[!h]
\begin{center}
\includegraphics[width=13.0cm, height=7cm]{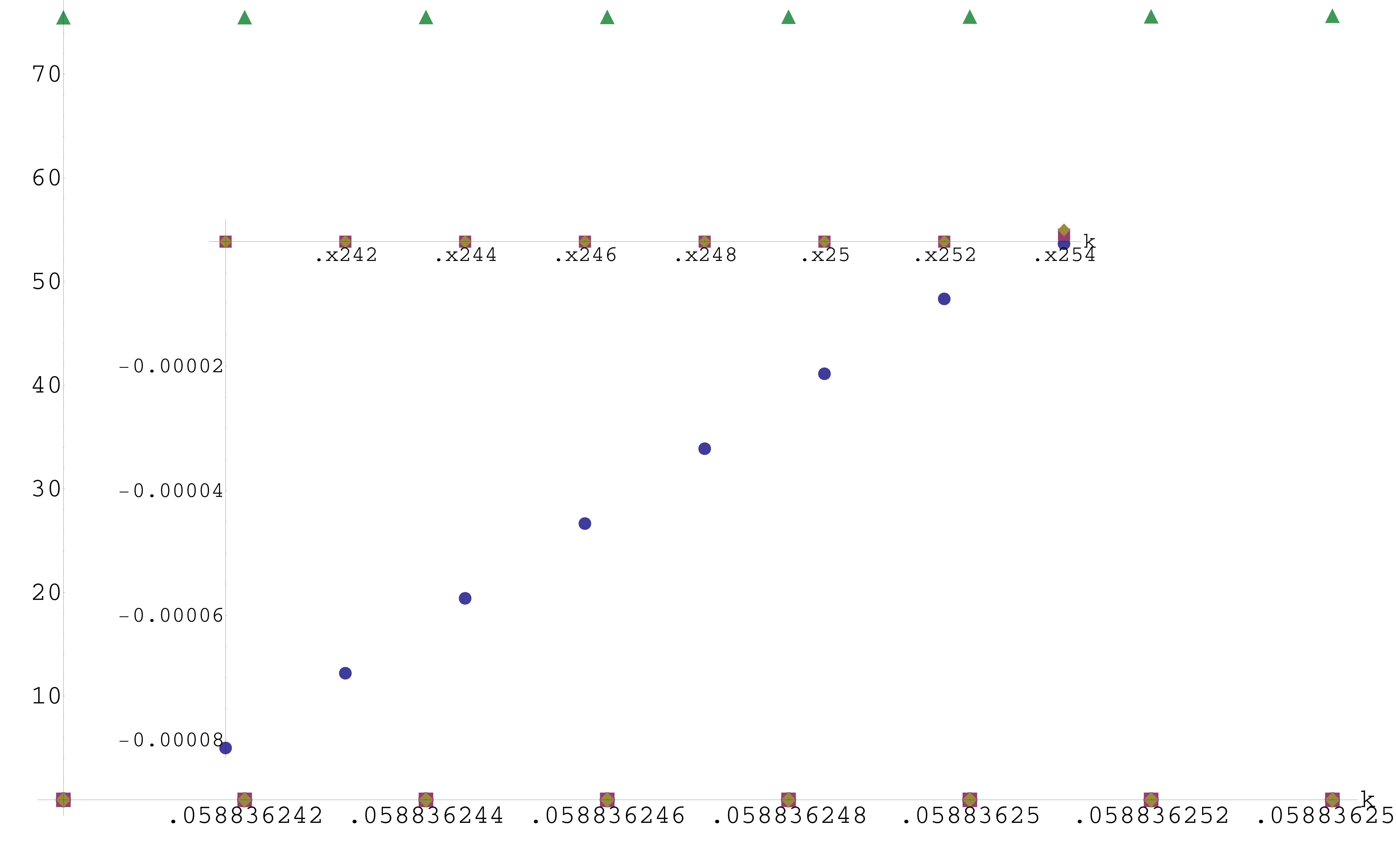}
\caption{The  double zero kernel (purple box and brown diamond) and the first positive eigenvalue (green triangle) of the linearized operator $\mathcal L_\#$ around a periodic KKSH 
breather for $N=40$, $\bt=1$ and $k$ increasing.  Note that the numerical method returns only two eigenvalues very 
close to zero, as expected from the conjectured linear spectral stability. Inside, the representation of the negative (blue circle) and double zero kernel. 
We used the notation $.x242\equiv 0.058836242,~~x=058836$.}\label{v26}
\end{center}
\end{figure}

\ms

As a second test, we perform several eigenvalue computations for the same parameter $\bt=1$, and $k$ moving. 
For the most difficult case,  the one where $k$ approaches the critical value $\sim 0.0588$, we obtain the results described in Fig.   \ref{v26}.
Also, in Fig. \ref{TablePer1}, we exactly describe those eigenvalues, that we obtain for different values of $k$.  
It is important to mention that we always get one negative eigenvalue and a two dimensional kernel, as well as a clearly defined spectral gap.



%
{\small
\begin{figure}[!h]
\begin{center}
 \begin{tabular}{||c c c c c||}
 \hline
 $k$ & 1st. eig & 2nd & 3rd & 4th \\ [0.5ex] 
 \hline\hline
 $.x240$ & -0.00008  & $-7.750\cdot10^{-10}$&  $-1.565\cdot10^{-9}$ & 75.490 \\ 
 \hline
 $.x242$  & -0.00006 & $-6.449\cdot10^{-10}$ &  $-1.561\cdot10^{-9}$ & 75.502 \\
 \hline
 $.x244$  & -0.00005 & $-5.952\cdot10^{-10}$ & $-1.558\cdot10^{-9}$ & 75.514\\
 \hline
 $.x246$  & -0.00004 & $-6.333\cdot10^{-10}$ & $-1.554\cdot10^{-9}$ & 75.528 \\
 \hline
 $.x248$  & -0.00003 & $-6.830\cdot10^{-10}$ & $-1.552\cdot10^{-9}$ & 75.544 \\ 
 \hline
 $.x250$  & -0.00002 & $-6.160\cdot10^{-10}$ & $-1.546\cdot10^{-9}$ &  75.563\\ 
 \hline
 $.x252$  & $-9.189\cdot10^{-6}$ & $-6.774\cdot10^{-10}$ & $-1.541\cdot10^{-9}$ & 75.589 \\ 
 \hline
 $.x254$  & $-3.463\cdot10^{-7}$ & $-1.135\cdot10^{-6}$ & $-1.869\cdot10^{-6}$ & 75.640 \\ 
 \hline
\end{tabular}
\end{center}\caption{The first four eigenvalues of $\mathcal L_\#$ for $\bt=1,~ x_1 = 0.1,~ x_2 = 0$, and 
$k$ varying from $.x240\equiv0.058836240,~~x=058836$  to $.x254$, as in Fig. \ref{v26}. All computations were made with $N=40$ test functions.
The third and fourth columns represent  approximate kernel of $\mathcal L_\#$.}\label{TablePer1} 
\end{figure}
}

\begin{figure}[!h]
\begin{center}
\includegraphics[width=11.0cm, height=5.5cm]{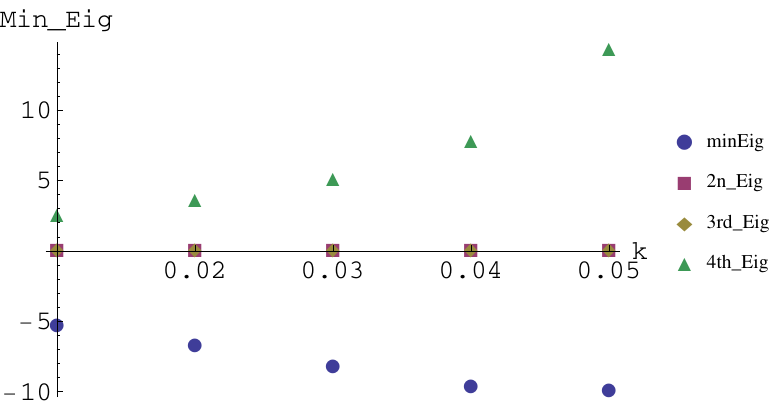}
\caption{For intermediate values of $k$: we plot the negative eigenvalue, the double zero kernel and the fourth eigenvalue of the
linearized operator $\mathcal L_\#$ around a periodic KKSH breather, 
for $\bt=1$ and $k$ increasing from $0.01$ to $0.05$, as expressed before in Table \ref{TablePer2}. 
Note that the numerical method returns only two eigenvalues very close to zero, 
as expected from the conjectured linear spectral stability. Computations were made with $N=50$ test functions.}\label{kinterm}
\end{center}
\end{figure}

In the intermediate case, for  pairs $(k,m)$, with $k=0.01,\dots,0.04$, 
 we obtain the results described in Fig. \ref{kinterm}. Also, in the Table \ref{TablePer2}, we describe the eigenvalues  of $\mathcal L_\#$
 that we obtain for different values of $k$.  Once again, we get one negative eigenvalue and a two dimensional kernel.

\medskip


{\small
\begin{figure}[!h]
\begin{center}
 \begin{tabular}{||c c c c c||}
 \hline
 $k$ & 1st eig. & 2nd & 3rd  & 4th  \\ [0.5ex] 
 \hline\hline
 0.01 & -5.343  & $1.023\cdot10^{-6}$ &  0.0001 & 2.540 \\ 
 \hline
 0.02 & -6.751  & $5.905\cdot10^{-10}$ &  $2.683\cdot10^{-6}$ & 3.561 \\
 \hline
 0.03 & -8.216 & $-2.651\cdot10^{-9}$ & $1.163\cdot10^{-6}$ & 5.067\\
 \hline
 0.04 & -9.623 & $3.997\cdot10^{-9} $ & $4.896\cdot10^{-8}$ & 7.756 \\
 \hline
 0.05 & -9.922 & $-2.173\cdot10^{-7}$ & $-1.143\cdot10^{-8}$ & 14.329 \\ 
 \hline
\end{tabular}
\end{center}\caption{The first four eigenvalues of  $\mathcal L_\#$ for $\bt=1,~ x_1 = 0.1,~ x_2 = 0$, and $k$ varying, as corresponding to Fig. \ref{kinterm}. 
All computations were made with $N=50$ test functions. The third and fourth columns represent the approximate kernel  of $\mathcal L_\#$.}\label{TablePer2} 

\end{figure}
}

\begin{figure}[!h]
\begin{center}
\includegraphics[width=12.0cm, height=6cm]{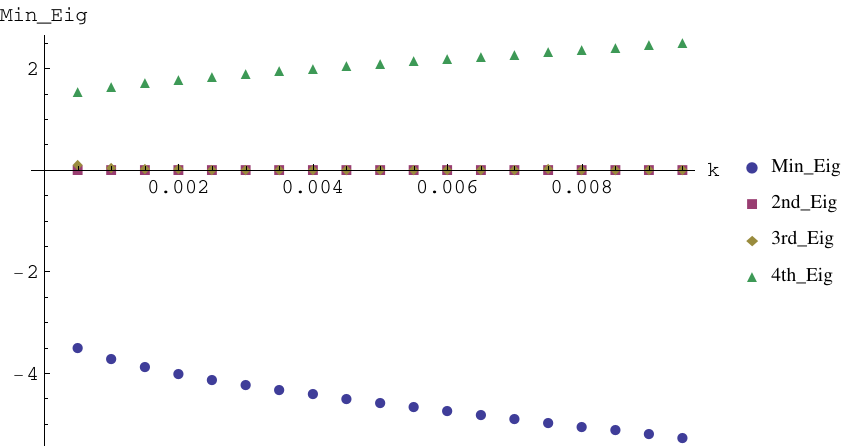}
\caption{For $k$ (\emph{small}): the negative eigenvalue, the double zero kernel and the fourth eigenvalue for the 
linearized operator $\mathcal L_\#$ around
a periodic KKSH breather, for $\bt=1$ and $k$ increasing from $0.0005$ to $0.0095$, as expressed before in Table \ref{TablePer3}. 
Note that the numerical method returns only two eigenvalues very close to zero, 
as expected from the conjectured linear spectral stability. Computations were made with $N=50$ test functions.}\label{kpeq}
\end{center}
\end{figure}

 Finally, for the  case,  with $k$ small, i.e. $k\approx10^{-3}$, 
 we obtain in Fig. \ref{kpeq} the description of the discrete spectra for the first four eigenvalues for the linearized operator $\mathcal L_\#$.
 Also, in the Table \ref{TablePer3}, we show the explicit numerical eigenvalues 
 which correspond to these small values of $k$.  Once again, we get one negative eigenvalue and a two dimensional kernel.

\medskip
{\small
\begin{figure}[!htb]
\begin{center}
\centering
\begin{tabular}{|p{0.073\textwidth}|p{0.073\textwidth}|p{0.076\textwidth}|%
p{0.073\textwidth}|p{0.073\textwidth}|p{0.0001\textwidth}|p{0.073\textwidth}|p{0.073\textwidth}|p{0.076\textwidth}|%
p{0.073\textwidth}|p{0.073\textwidth}|}
 \hline
\centering $k$ & \centering  1st eig. &\centering  2{\small nd$\cdot10^{-6}$} & \centering 3rd&\centering  4th&& \centering $k$ &\centering  1st eig. &\centering  2{\small nd$\cdot10^{-5}$ }&\centering  3rd &  4th \\ [0.1ex] 
 \hline\hline
 0.0095 & -5.271  & 1.265 &  0.0001 & 2.495 && 0.0045 & -4.505  & 1.531 & 0.0017 & 2.040 \\ [0.1ex]
 \hline 
 0.0085 & -5.127  & 1.962 & 0.0002 & 2.405 && 0.0035 & -4.327  & 2.919 & 0.0034 & 1.941\\ [0.1ex]
 \hline
 0.0075 & -4.971 & 1.248 & 0.0142 & 2.316 &&  0.0025 & -4.127  & 6.234 & 0.0073 & 1.833\\  [0.1ex]
 \hline
 0.0065 & -4.828 & 5.092 & 0.0005 & 2.225 && 0.0015 & -3.883 & 16.47 & 0.0198 & 1.708  \\  [0.1ex]
 \hline
 0.0055 & -4.671 & 8.609 & 0.0009 & 2.134 && 0.0005 & -3.497& 79.70 & 0.0983 & 1.530 \\  [0.1ex]
 \hline
\end{tabular}
\end{center}\caption{The first four eigenvalues for the linearized operator $\mathcal L_\#$ around
a periodic KKSH breather for $\bt=1,~ x_1 = 0.1,~ x_2 = 0$, and $k$ varying in a sample of points of Fig. \ref{kpeq}. 
All computations were made with $N=50$ test functions. The third and fourth columns (left) and eigth and ninth columns (right) represent respectively the approximate kernel of $\mathcal L_\#$.}\label{TablePer3} 

\end{figure}
}

Therefore, we can certainly claim that there is strong evidence that, according to numerical simulations, KKSH breathers are spectrally stable, 
in the sense that they have only one negative eigenvalue, and a nondegenerate, two-dimensional kernel. For clarity reasons, we state a rigorous statement of both properties for the case of the periodic KKHS breather.

\ms 

\begin{itemize}\label{A14}
\item[] {\bf Assumption 1:}  (\emph{Nondegeneracy of the kernel})
 For each $k\in (0,k^*)$, $x_1,x_2\in \R$ and $\bt>0$, $\ker \mathcal L_\#$ is spanned by the two elements $\partial_{x_1}B$ and $\partial_{x_2} B$; 
 and there is a uniform gap between the kernel and the bottom of the positive spectrum. 
\end{itemize}

\ms

\begin{itemize}
\item[] {\bf Assumption 2:} (\emph{Unique, simple negative eigenvalue})
For each $k\in (0,k^*)$, $x_1,x_2\in \R$ and $\bt>0$,  the operator  $\mathcal L_\#$ has a unique simple, negative eigenvalue $\la_{1}=\la_{1}(\bt,k,x_1,x_2)<0$ associated to the unit $L^2$-norm  eigenfunction $B_{-1}$. Moreover, there is $\la_{1}^0<0$ depending on $\bt$ and $k$ only, such that  $\la_{1} \leq \la_{1}^0$ for all $x_1,x_2$.
\end{itemize}

\subsection{Duality stability/instability} 
\begin{figure}[!h]
\begin{center}
\includegraphics[width=11.0cm,height=6.8cm]{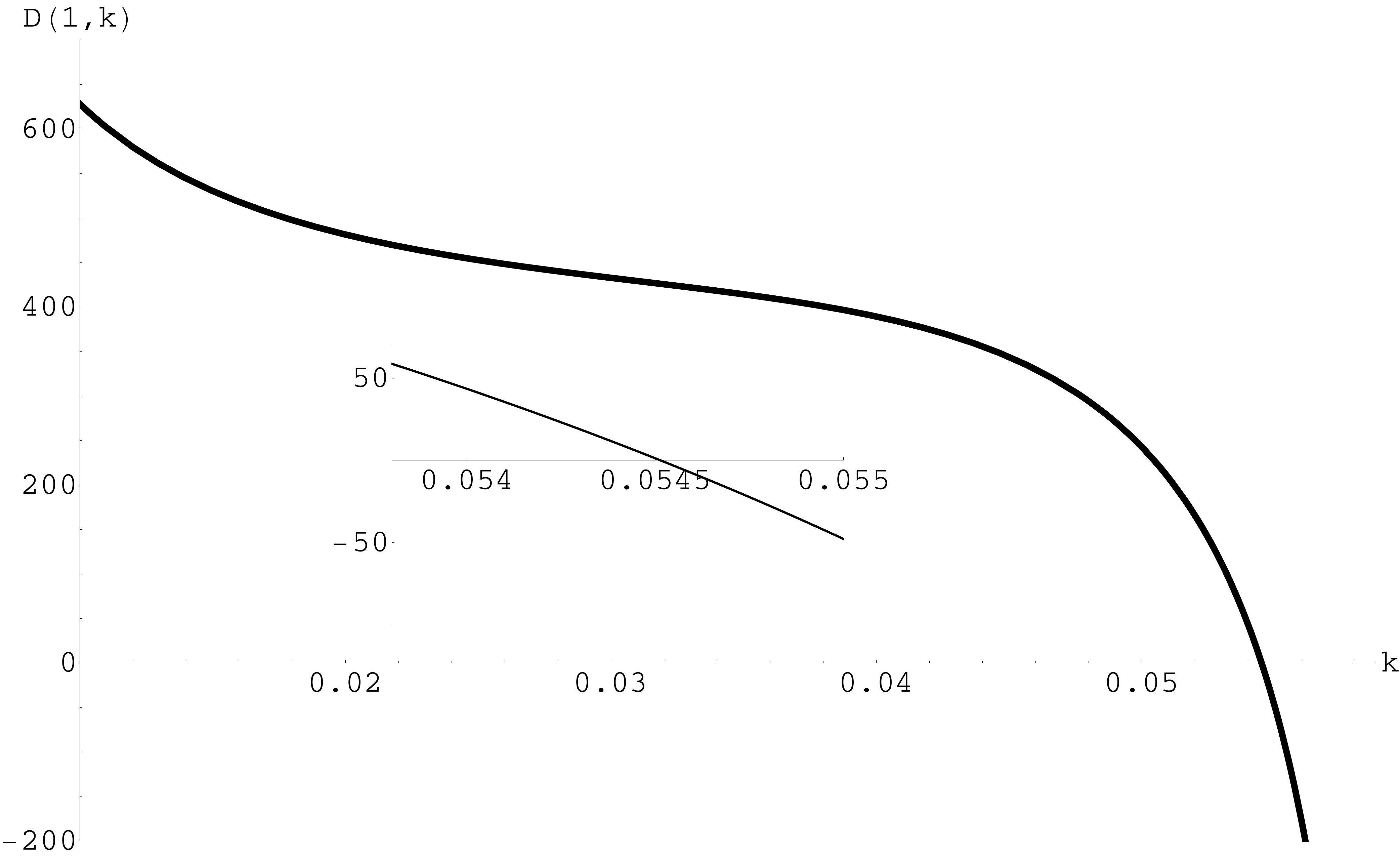}
\caption{The discriminant function $D(1,k)$ in terms of $k$, as described in \eqref{Discr}. Note that $D(1,k)$ changes sign (see zoom figure inside)
for $k \geq k_* \sim 0.0545$. 
When $k$ approaches its limit value $\sim 0.058836$ (see equation \eqref{km}), we recover the aperiodic mKdV breather.}\label{KKHS_c}
\end{center}
\end{figure}

\begin{figure}[!h]
\begin{center}
\includegraphics[width=13cm, height=8.5cm]{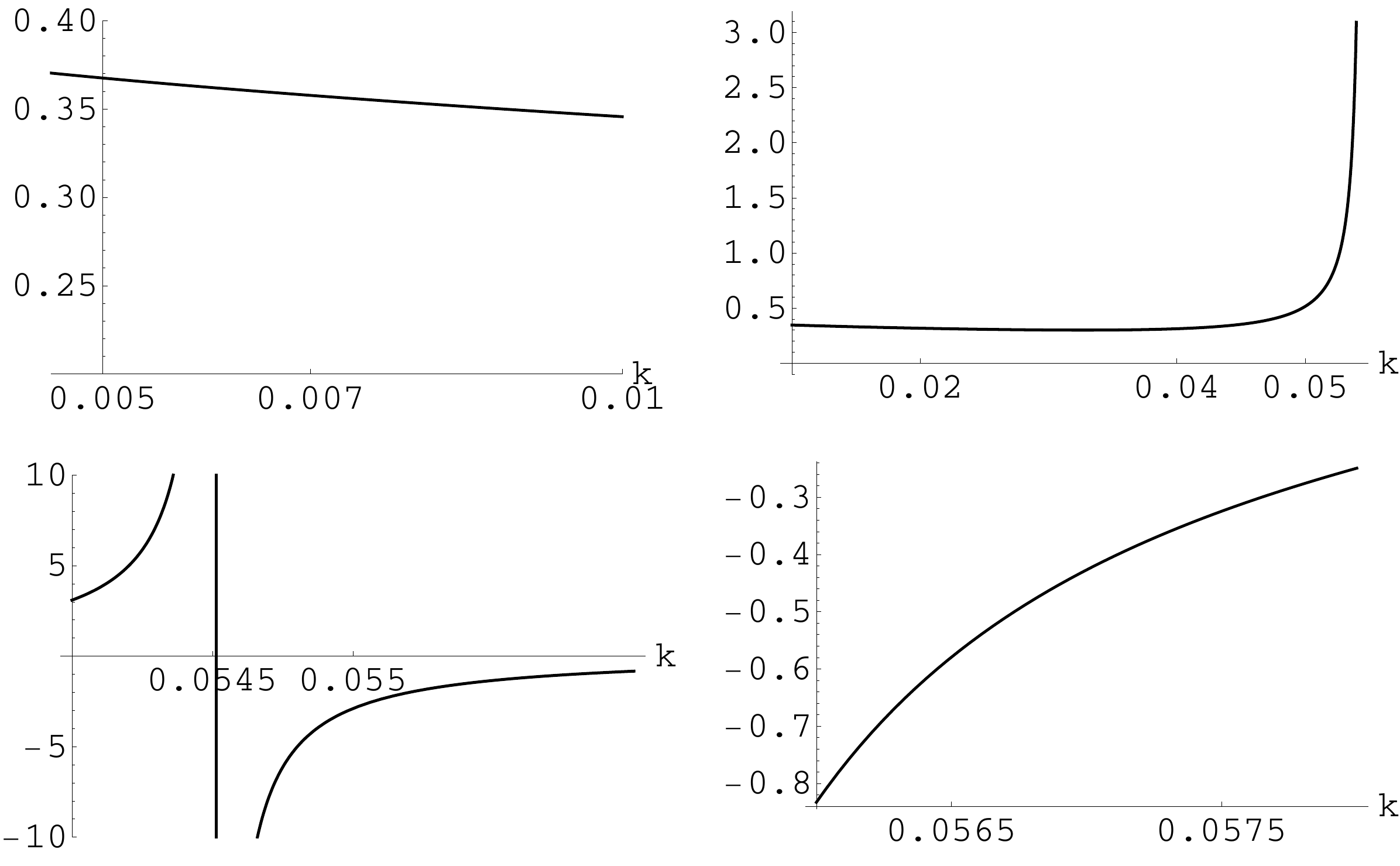}\caption{The Weinstein's condition $HG(\bt,k)$ \eqref{HG}, for the periodic  KKSH
breather, in the case $\bt=1$, for $k\in [0.0045,0.01]$ (above left), for $k\in [0.01,0.054]$ (above right), $k\in [0.054,0.056]$ (below left),
and $k\in [0.056, 0.058]$ (below right). In order to run our argument for a stability proof \cite{AM}, we require $HG(1,k)>0$, 
which is only satisfied for $k< k_*$, where $k_*\sim 0.0545$ is the approximate point where $D(1,k)$ vanishes (see Fig. \ref{KKHS_c}). 
Note also that the case $k\sim 0.057$ assumed in \cite{KKS2}, that leads to instability, is not included in the stability region described in Theorem \ref{Thm000}, but in the region where $HG(1,k)<0$.}\label{KKHS_d}
\end{center}
\end{figure}

Now the main problem is to figure out 
where our nonlinear stability proof does/does not work. For this purpose, the \emph{discriminant}
\be\label{Discr}
D= D(\bt,k):= \partial_{k}a_1 \partial_\bt a_2 -\partial_{k}a_2 \partial_\bt a_1 \qquad (a_1,a_2 \hbox{ from } \eqref{a1}-\eqref{a2}),
\ee
is the key element to look at.   In order to explain why this element is important, let us notice that from \eqref{EcBp} and \eqref{L_per}, we readily have (compare with Corollary \ref{Scaling})
\[
\begin{aligned}
\mathcal L_\# (\partial_k B)&  = \partial_k a_1 (B_{xx}+B^3) -\partial_k a_2 B, \\ \mathcal L_\# (\partial_\bt B)&  = \partial_\bt a_1 (B_{xx}+B^3) -\partial_\bt a_2 B.
\end{aligned}
\]
Therefore, as soon as $D\neq 0$,
\be\label{B_0per}
B_{0,\#} := \frac1D (\partial_k a_1 \partial_\bt B - \partial_\bt a_1 \partial_k B ),
\ee
satisfies the equation
\[
\mathcal L_\# (B_{0,\#}) =-B,
\]
see also \cite[Corollary 4.5]{AM}. Using this fact, we can easily prove, as in Proposition \ref{Coer}, that the eigenfunction associated to the negative eigenvalue of $\mathcal L_\#$ can be replaced by the breather itself, which has better behavior in terms of error controlling, unlike the first eigenfunction. This simple fact allows us to prove the nonlinear stability
result as in the standard approach, without using scaling modulations.
Recall that using the first eigenfunction as orthogonality condition does not guaranty a suitable control on the scaling modulation parameter, because the control given by this direction might be not good enough to close the stability estimates. However, the breather can be used as an alternative direction, and all these previous arguments remain valid, exactly as in \cite{AM}, provided the Weinstein's sign condition 
\be\label{Cond3}
\int_0^L B_{0,\#} B >0  \qquad \hbox{\Big(or equivalently $\displaystyle{\int_0^L B_{0,\#} \mathcal L_\#[B_{0,\#}] <0}$\Big),} 
\ee
do hold.  Using \eqref{B_0per}, we are lead to the understanding of the quantity
\be\label{HG}
\begin{aligned}
\int_0^L B_{0,\#} B & =  \frac1D \int_0^L (\partial_k a_1 \partial_\bt B - \partial_\bt a_1 \partial_k B ) B \\
& = \frac1D (\partial_k a_1 \partial_\bt M_\#[B] -\partial_\bt a_1 \partial_k M_\#[B]) =: HG(\bt,k),
\end{aligned}
\ee
where $ M_\#[B]$ was computed in \eqref{B25}. Recall that $a_1$ and $a_2$ are almost explicit from \eqref{a1}-\eqref{a2}. An exact expression for $HG(\bt,k)$ has escaped to us, however, we can graph this new function in some interesting cases. In particular, for the case considered in \cite{KKS2}, we assume $\bt=1$ 
and we graph $D=D(1,k)$ and $HG(1,k)$, to obtain the results in Fig. \ref{KKHS_c} and Fig. \ref{KKHS_d}. We note that condition \eqref{Cond3} 
holds provided $k$ is small enough.  However, the values for which $HG(\bt,k)>0$ do not coincide with the values for which the standard Weinstein's condition
(positive derivative with respect to the scaling), deduced from Fig. \ref{MassBper}, holds true, and this is totally natural for the case of breathers, 
as it was explained in \cite[Corollary 2.2]{AM}.

\medskip

Now we are ready to fully state the Assumption 3 described in the introduction of this paper.

\medskip

\begin{itemize}
\item[] {\bf Assumption 3:}   {\it (Positive generalized Weinstein's condition)} The following generalized Weinstein's type sign condition is satisfied: if $M_\#[B]$ is the mass of the breather solution  \eqref{B25}  in terms of $\bt$ and $k$, and $a_1, a_2$ are the variational parameters given in \eqref{a1}-\eqref{a2}, then
\be\label{Weinstein23}
 HG(\bt,k) =  \frac{1}{D}(\partial_k a_1 \partial_\bt M_\#[B] -\partial_\bt a_1 \partial_k M_\#[B]) >0.
\ee
\end{itemize}
Now we can state a rigorous statement for Theorem \ref{T1p8}.

\begin{thm}\label{Thm000}
Under Assumptions 1, 2  in page \pageref{A14}, and Assumption 3 above, KKHS breathers are orbitally stable under small, $L$-periodic $H^2$ perturbations. More precisely, there are $\eta_0>0$ and $K_0>0$, only depending on $\bt$ and $k$, such that if $0<\eta<\eta_0$ and if $u_0$ is $L$-periodic with
\[
\| u_0 - B(t=0,\cdot ; 0, 0)    \|_{H^2(\T)} <\eta,
\]
then there are real-valued parameters $x_1(t)$ and $x_2(t)$ for which the global $H^2$, $L$-periodic solution $u(t)$ of \eqref{mKdV} with initial data $u_0$ satisfies
\[
\sup_{t\in \R} \| u(t) - B(t,\cdot ; x_1(t), x_2(t))    \|_{H^2(\T)} <K_0\eta,
\]
with similar estimates for the derivatives of the shift parameters $x_1,~x_2$.
\end{thm}

The proof of this result is completely similar to the proof of Theorem \ref{MT}, after following the same steps. Assumption 3 is used to ensure that an expression like  \eqref{Positivity} has a never zero denominator.

\medskip
Additionally, based on this numerical evidence, we conjecture the following alternative stability theory for KKSH breathers:

\ms 

\begin{center}\label{Conj1}
{\bf Conjecture:}
{\it Assume that $HG(\bt,k)<0$. Then $B$ is  unstable under small $H^2(\T)$ perturbations.}
\end{center}

\medskip

Examples of unstable structures that have a nondegenerate kernel and only one negative eigenvalue are solitons for the nonlinear Klein-Gordon equation in $\R_t \times \R_x^d$:
\[
u_{tt} - \Delta u + u + u^p =0, \quad p>1.
\]
Here, unlike our case, the lack of stability is related to the absence of a scaling symmetry controlling the negative direction (appearing because of the negative eigenvalue).

\bs

\section{Periodic mKdV breathers with nonzero mean}\label{Sect:7} 

\medskip

\subsection{Introduction} Although KKSH breathers are periodic in space, the are still zero-mean solutions. We would like to see if nonzero mean 
periodic\footnote{Non periodic and nonzero mean breather solutions of mKdV were already known, see \cite{Ale0, Ale}.}
breather solutions may exist. By periodic breather we mean the object in Definition \ref{DEF_B_per}, that is, any solution that is periodic in time and space, having two
independent and different space variables (i.e. not being a one profile solution being translated over time, as solitons are), 
and finally, having oscillatory behavior, unlike standard 2-soliton solutions. Numerical evidence of the existence of these solutions
was given by the first author in \cite{Ale}, since these solutions are connected with the (nonzero) curvature of the planar curve evolving
according to a particular law. For more details about this connection, see e.g. \cite{Ale}.

\ms

\begin{figure}[!htb]
\includegraphics[width=13.5cm,height=6.5cm]{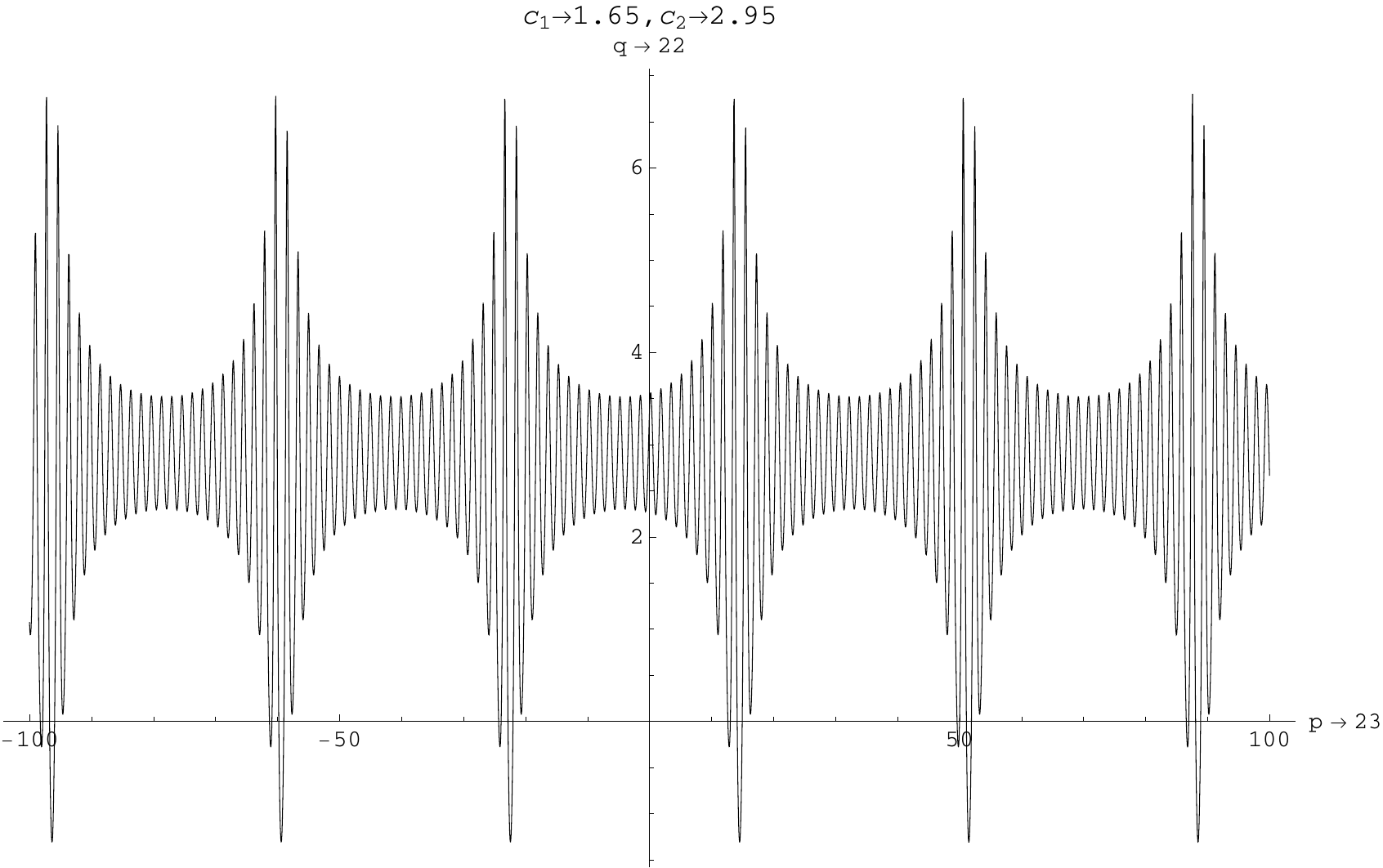}
\caption{Periodic breather of \eqref{mKdV} with non zero mean. The parameters here are $c_1=1.65,c_2=2.95,p=22,q=23,$ and the period $L$ is $\sim 35.7$}\label{fig1}
\end{figure}

\begin{figure}[!htb]
\includegraphics[width=13.5cm,height=6.5cm]{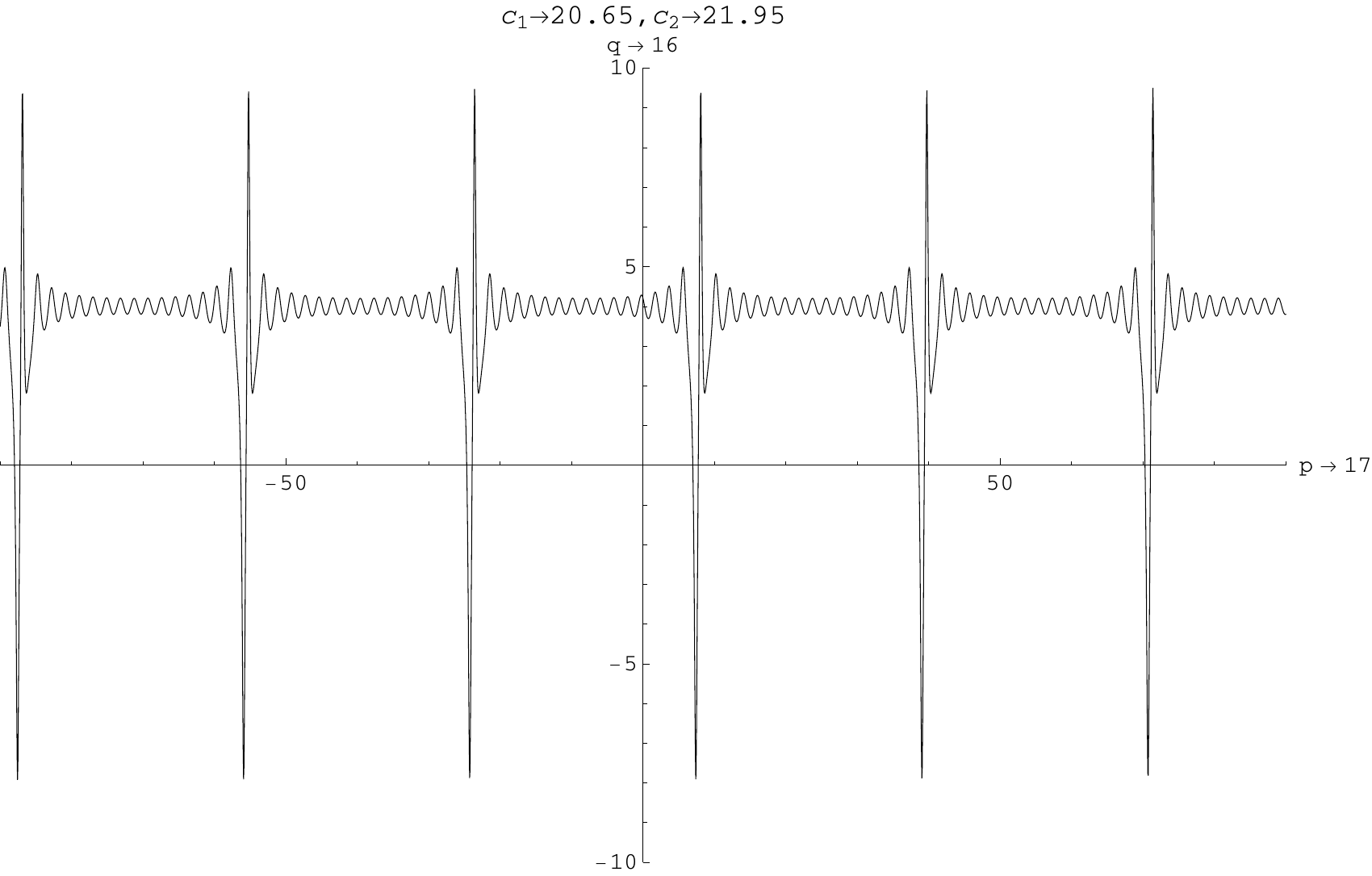}
\caption{Periodic breather of \eqref{mKdV} with non zero mean, for parameters $c_1=20.65, ~c_2=21.95,~p=17,~q=16.$}\label{fig2}
\end{figure}

For the case of the mKdV equation, we have been able to obtain a new set
of periodic breather solutions of \eqref{mKdV} \emph{with nonzero mean}. More explicitly,

\begin{defn}\label{NEW_BREATHER}
Given $c_1,c_2,\mu>0$, $p,q$ nonzero integers, with $p,q$ coprime, such that the following \emph{commensurability} condition is satisfied\footnote{Note that each $c_i$ is less than $2\mu^2$.}
\be\label{Cons2}
 \frac{2\mu^2-c_1}{2\mu^2-c_2}=\frac{p^2}{q^2},
\ee
we define the breather $B=B(t,x;c_1,c_2,\mu,p,q)$ by the formula
\be\label{BP1}
 B:= \mu + 2\sqrt{2}\partial_x\arctan\Big(\frac{f(t,x)}{g(t,x)} \Big),
\ee
where,
\[
\rho:=\frac{\sqrt{c_1}+\sqrt{c_2}}{\sqrt{c_1}-\sqrt{c_2}}, \quad  f(t,x) := -\sqrt{2}\mu \rho\Big(\sqrt{c_1}-\sqrt{c_2} + \sqrt{2\mu^2-c_2}\tan y_2-\sqrt{2\mu^2-c_1}\tan y_1\Big),
\]
and
\[
g(t,x) := 2\mu^2 + \Big( \sqrt{2\mu^2-c_1}\tan y_1 -\sqrt{c_1}  \Big)\Big( \sqrt{2\mu^2-c_2}\tan y_2 -\sqrt{c_2}  \Big).
\]
Here
\be\label{y1y2_fin}
y_1=\frac12\sqrt{2\mu^2-c_1}(x-\delta t),\quad y_2=\frac12\sqrt{2\mu^2-c_2}(x-\gamma t),
\ee
and the speeds are 
\[
 \delta:=\mu^2+ c_1,\quad \gamma:=\mu^2+ c_2.
\]
\end{defn}
Note that condition \eqref{Cons2} is imposed in order to obtain a truly periodic solution, see the formulae for $f$  and $g$. The spatial period of this breather is given by $L = \frac{2\pi q}{\sqrt{2\mu^2 -c_1}}$. See also Fig. \ref{fig1}-\ref{fig2} for some drawings of different breather solutions, depending on the parameters $c_1,c_2,\mu, p$ and $q$.

\medskip

Recall that KKSH breathers cannot have any admissible set of parameters $k$ and $m$; 
they are constrained by conditions in \eqref{Cond1}. Here, the only condition that we need 
to satisfy is \eqref{Cons2}. Therefore, given nonnegative integers $p$ and $q$, with $p$ and $q$
coprime, and given $c_1<2\mu^2$, then there is a unique $c_2$ solution to \eqref{Cons2}. In that sense,
$\mu$ being fixed, the breather in \eqref{BP1} has three different degrees of freedom (in addition to shifts), two of them being discrete. 
Finally, in terms of $p$ and $q$, the larger these integers are, the more oscillatory the breather solution is. Note additionally that this breather has not been constructed by using Jacobi functions, but only standard periodic functions, a fact that simplifies many computations. 

\medskip

It is also relevant to mention that the construction of a breather solution with different patterns as the usually required can be very involved. For example, a different class of periodic breather was discovered by Blank et al. in \cite{BCLS}.  This particular solution is constructed for the so called $\phi^4$ model, by assuming that the equation is no longer autonomous, but it has suitable, well-chosen periodic coefficients.

\subsection{Sketch of proof of Theorem \ref{T1p9}}
We will use the B\"acklund Transformation for mKdV (see \cite{AM1} for a rigorous setting of the computations below) to construct this solution. Given a solution $u_0$ of the form $u_0 =\partial_x \widetilde u_0$ of mKdV and fixed constants $a_1,a_2\in \R$, $a_1\neq a_2$, we can construct a second $u_1 =\partial_x \widetilde u_1$ and third solution $u_2 =\partial_x \widetilde u_2$ by setting
\[
u_1 - u_0 = a_1 \sin\Big(\frac{\widetilde u_1 + \widetilde u_0}{\sqrt{2}} \Big), \quad \hbox{and} \quad 
u_2 - u_0 = a_2 \sin\Big(\frac{\widetilde u_2 + \widetilde u_0}{\sqrt{2}} \Big).
\]
These two solutions can be combined to construct a fourth solution $u$ through the \emph{permutability} condition \cite{AM1}:
\be\label{permu4}
\tan\Big(\frac{\widetilde u -\widetilde u_0}{2\sqrt{2}}\Big) = 
-\Big(\frac{a_1+a_2}{a_1-a_2}\Big) \tan\Big(\frac{\widetilde u_2 - \widetilde u_1}{2\sqrt{2}}\Big).
\ee
Fix $\mu>0$. Starting with the constant solution $u_0 =\mu$,  $\widetilde u_0= \mu x$ and two constants $a_1=\sqrt{2c_1}$, $a_2=\sqrt{2c_2}$, $c_1,c_2>0$, we have
\be\label{bt1}
u_1 - \mu =  \sqrt{2c_1}\sin \Big( \frac{\tilde u_1 + \mu x}{\sqrt{2}} \Big), \qquad u_2 - \mu =  \sqrt{2c_2}\sin \Big( \frac{\tilde u_2 + \mu x}{\sqrt{2}} \Big).
\ee
Looking for a solution of the form $u_i= \partial_x \tilde u_i$, $i=1,2$, we get
\be\label{bt12}
\widetilde u_i(t,x) = -\mu x + 2\sqrt{2}\arctan\Big(\frac{1}{2\mu}\Big(- \sqrt{2c_i}+\sqrt{4\mu^2-2c_i}\tan\Big(\frac{\sqrt{4\mu^2-2c_i}}{2\sqrt{2}}(x-(\mu^2+ c_i)t) \Big) \Big)\Big). 
\ee
The factor $(\mu^2+ c_i)t$ in the above solution appears as a constant of integration, and is chosen in such a form that $u_i$ is actually solution to mKdV. Calling $y_1 =x+\delta t$, $y_2= x+ \ga t$ as in \eqref{y1y2_fin}, and $\rho=\frac{\sqrt{c_1}+\sqrt{c_2}}{\sqrt{c_1}-\sqrt{c_2}},$ the desired breather \eqref{BP1} is obtained by using \eqref{permu4} with $u_0=\mu$, $u_1,u_2$ from \eqref{bt12}:
%
%
\[
\begin{aligned}
B & =\mu + 2\sqrt{2}\partial_x\arctan\Bigg(
\frac{-\sqrt{2}\mu \rho(\sqrt{c_1}-\sqrt{c_2}+\sqrt{2\mu^2-c_2}\tan(y_2)-\sqrt{2\mu^2-c_1}\tan(y_1))}
{2\mu^2+(\sqrt{2\mu^2-c_1}\tan(y_1)-\sqrt{c_1})\cdot(\sqrt{2\mu^2-c_2}\tan(y_2)-\sqrt{c_2})}\Bigg).
\end{aligned}
\]
The fact that this is a solution of mKdV is a tedious, lengthy but straightforward computation.

\subsection{Final comments} 
A detailed study of this new breather solution will be done elsewhere. For the moment, we advance that $B$ \eqref{BP1} satisfies the elliptic ODE
\[
\begin{aligned}
& B_{4x}  -   (c_1+c_2-4\mu^2) (B_{xx} +3\mu (B-\mu)^2 + (B-\mu)^3) +  
(c_1-2\mu^2)(c_2-2\mu^2)(B-\mu) \\
& \quad + 5 (B-\mu)B_{x}^2 + 5(B-\mu)^2 B_{xx}+ \frac 32 (B-\mu)^5  + 5\mu B_{x}^2 \\
& \quad + \frac{15}{2} \mu (B-\mu)^4  + 10\mu  (B-\mu) B_{xx}  + 10\mu^2 (B-\mu)^3=0.
\end{aligned}
\]
In particular, $B$ is a critical point of the functional
\[
\mathcal H_{\mu p}[w](t) := F_{\mu p}[w](t) +  (c_1+c_2-4\mu^2)E_{\mu p}[w](t) +(c_1-2\mu^2)(c_2-2\mu^2) M_{\mu p}[w](t),
\]
where $M_{\mu p}$ and $E_{\mu p}$ are defined in the following way:
\be\label{M1p}
M_{\mu p}[w](t)  :=  \frac 12 \int_\R (w-\mu)^2(t,x)dx = M_{\mu p}[w](0),
\ee 
\be\label{E2p}
E_{\mu p}[w](t)  :=  \frac 12 \int_\R w_x^2 -\mu\int_\R (w-\mu)^3 - \frac 14 \int_\R (w-\mu)^4= E_{\mu p}[w](0),\ee
and
\be\label{F1p}
\begin{aligned}
 F_{\mu p}[w](t)  &  :=   \frac 12 \int_\R w_{xx}^2dx - 5\mu\int_\R (w-\mu)w^2_xdx + \frac{5}{2}\mu^2\int_\R (w-\mu)^4dx \\
& \quad   - \frac{5}{2} \int_\R (w-\mu)^2w_x^2dx+ \frac{3\mu}{2} \int_\R (w-\mu)^5+ \frac{1}{4} \int_\R (w-\mu)^6dx
\end{aligned}
\ee
is a third conserved quantity for the mKdV equation. Finally, it is interesting to note that we can recover the aperiodic breather with non zero mean (see \cite{Ale0,Ale1} choosing in \eqref{BP1} 
complex conjugate scalings $c_1=\bt+i\al,~c_2=\bt-i\al$. For the sake of simplicity, we only show a picture of it:

\begin{figure}[!h]
\begin{center}
\includegraphics[width=10cm, height=6cm]{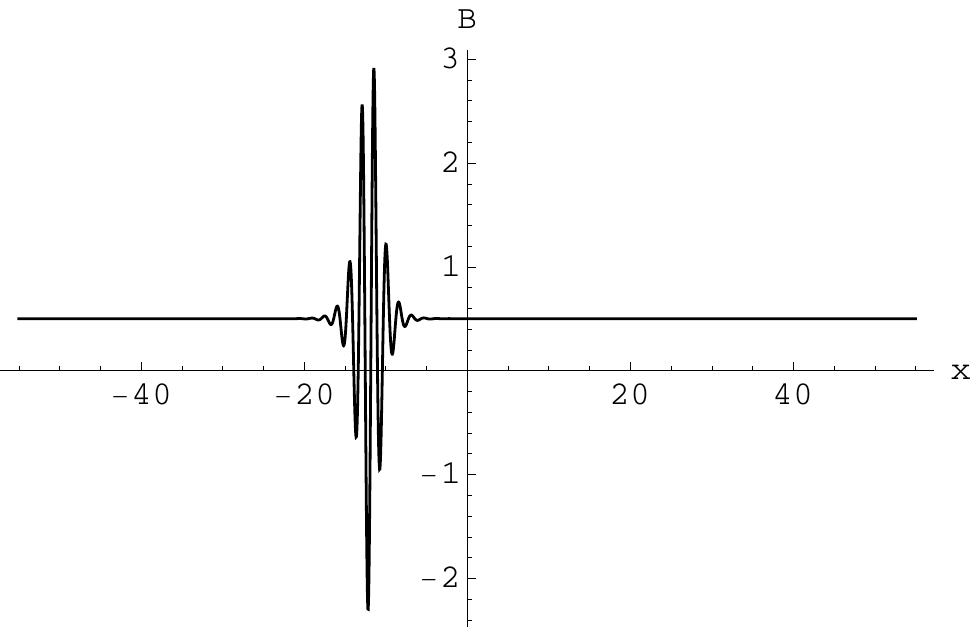}
\caption{Aperiodic mKdV breather with non zero mean at $t=0.2$ with $\bt=1$, $\al=4$, $\mu=0.5$, obtained from periodic breather \eqref{BP1}.}\label{mKdV_per1}
\end{center}
\end{figure}

\appendix

\section{Proof of Corolary \ref{corSG}}\label{B}

\medskip

In this Section we prove that, using the conserved quantity from \eqref{F1}, $ F[u,u_t]$, SG breathers are critical points for the functional $\mathcal{H}[u,u_t]$ defined in \eqref{H1}. It is clear that $\mathcal H[u,u_t]$ represents a real-valued, conserved quantity, well-defined for $H^2\times H^1$-solutions of (\ref{SG}). Moreover, one has the following decomposition property.

\begin{lem}\label{EE0}
Let $(z,w)\in H^2(\R)\times H^1(\R)$ be any pair of functions with sufficiently small $H^2\times H^1$-norm, and $(B, B_t)$ be any breather solution, 
with corresponding parameters $\bt>0$ and $v\in (-1,1)$.  Then, for all $t\in \R$,  one has
\be\label{EE}
\mathcal{H}[B+z, B_t + w](t) - \mathcal{H}[B,B_t] (t) = \frac 12\mathcal Q[z,w] + \mathcal N[z,w],
\ee
with $\mathcal Q$ being the quadratic form defined in (\ref{Q}), and $\mathcal N[z,w]$ a small, nonlinear term satisfying
\[
|\mathcal N[z,w] | \lesssim  p(\|z\|_{H^2(\R)}, \|w\|_{H^1(\R)}),
\]
where $p$ is a positive, third order monomial  on its corresponding variables.
\end{lem}

\begin{proof}
We compute:
\[
\begin{aligned}
 \mathcal{H}[B+z, B_t +w] & = \frac 12 \int_\R [(B+z)_{xx}^2+(B_t+w)_{x}^2]  -\frac 1{32} \int_\R [(B_t+w)^4 + (B+z)_x^4] \\
 & \quad - \frac3{16} \int_\R (B_t+w)^2(B+z)_x^2 + \frac 58\int_\R (B+z)_x^2\cos(B+z)\\
&  \quad  + \frac 18\int_\R[\sin^2 (B+z) + (B_t+w)^2\cos(B+z)]  \\
&  \quad + \frac a2 \int_\R [(B+z)_{x}^2+(B_t+w)^2] + a\int_\R [ 1-\cos(B+z)]   +\frac b2\int_\R (B_t+w)(B+z)_x.
\end{aligned}
\]
Expanding every term above, as is done in \cite{AM}, we get
\bee
& & \mathcal{H}[B+z, B_t +w] =  \mathcal{H}[B,B_t] +\\
& & \qquad+ \int_\R \Big[B_{4x}+\frac 38B_x^2B_{xx} + \frac 38B_t^2B_{xx} +\frac 58B_x^2\sin B  + \frac 34 B_tB_xB_{tx} -\frac54B_{xx}\cos B + \frac14 \sin B \cos B \\
& & \qquad\qquad\qquad\qquad  -\frac 18B_t^2\sin B  -a(B_{xx} -\sin B) -\frac b2 B_{tx}\Big]z\\
& &\qquad -\int_\R \Big[B_{txx}+\frac 18B_t^3+\frac38B_x^2B_t-\frac14B_t\cos B -aB_t-\frac b2B_x\Big]w\\
& &\qquad+\frac 12\int_\R\Big\{ \Big[z_{4x} -[a - \frac 38(B_x^2 + B_t^2)+\frac 54\cos B ]z_{xx}   + [\frac 34B_{xx}B_x + \frac 34 B_tB_{tx} - \frac 54B_x\sin B]z_x\\
& & \qquad \qquad\qquad+[a\cos B +\frac 14(\cos^2 B-\sin^2B) -\frac 58B_x^2\cos B   + \frac 54B_{xx}\sin B-\frac 18B_t^2\cos B]z\Big]z\\
& &\qquad\qquad\qquad + [-w_{xx} +\frac 14(\cos B- \frac 32(B_x^2 + B_t^2))]w  -\frac 12B_t\sin B w z + (b-\frac 32B_tB_x)z_xw\Big\}\\
& & \qquad - \frac 18 \int_\R \Big[B_tw^3 + B_xz_x^3 + \frac 1{4}(w^4+z_x^4) + 3 B_t w z_x^2   + 3 B_x z_x w^2 +\frac 3{2}w^2 z_x^2 + \frac5{2}\cos B z^2 z_x^2\\
& &\qquad \qquad \qquad + 5\sin B z z_x^2-\frac 1{4}\sin^2B z^4 + \sin B\cos B z^3   +\frac 1{2}\cos B w^2 z^2 + B_t w z^2 +\sin B z w^2\Big].
\eee
Therefore, we have the decomposition
\bee
& & \mathcal{H}[B+z, B_t +w]    =  \mathcal{H}[B,B_t] + \int_\R \eqref{EcB} z - \int_\R \eqref{EcB2} w  + \frac 12\mathcal Q[z,w] + \mathcal N[z,w],
\eee
where $\mathcal Q$ is defined in (\ref{Q}). Taking into account the Lemma \ref{GB}, the second and third term in the r.h.s of the above equation vanish.  Finally, the term $N[z,w]$ is given by 
\bee
\mathcal N[z,w]&  := &- \frac 18 \int_\R \Big[B_tw^3 + B_xz_x^3 + \frac 1{4}(w^4+z_x^4) + 3 B_t w z_x^2+ 3 B_x z_x w^2 \\
& &\qquad \quad  +\frac 3{2}w^2 z_x^2 + \frac5{2}\cos B z^2 z_x^2 + 5\sin B z z_x^2-\frac 1{4}\sin^2B z^4  \\
& &\qquad \quad+ \sin B\cos B z^3 +\frac 1{2}\cos B w^2 z^2 + B_t w z^2 +\sin B z w^2\Big].
\eee
Therefore, from direct estimates one has $\mathcal N[z,w] \lesssim  p(\|z\|_{H^2(\R)},\|w\|_{H^1(\R)}),$ where $p$ is a positive, third order polynomial in its variables, and where the constant is independent of the size of $(z,w)$ and the time, provided the former are chosen small.
\end{proof}

\section{Jacobi elliptic functions}\label{JEF}
We present here some basic notions and properties on the Jacobi elliptic functions (JEFs) that we are going to use in 
the study of the periodic breather in Section \ref{Sect:5}. For further reading, see \cite{AS,By}. The JEF $sn(\cdot,r),~r=k~~\text{or}~~m$  
is defined as  the solution of the following ODEs:
\be\label{sn}
\begin{aligned}
 \sn'^{2} = 1-(1-r)\sn^2+r\sn^4, \qquad  \sn'' = -(1+r)\sn+2r\sn^3.
\end{aligned}
\ee
It takes some special values when $r=0$ or $r=1$:
\begin{multicols}{3}
      \begin{enumerate}
      \item[] $\sn(x,0)=\sin(x),$
      \item[] $\con(x,0)=\cos(x),$
      \item[] $\dn(x,0)=1,$
       \item[] $\sn(x,1)=\tanh(x),$
        \item[] $\con(x,1)=\cosh(x),$
        \item[] $\dn(x,1)=\sech(x).$
      \end{enumerate}
 \end{multicols} 
Additionally, it is related with the following two JEFs  $\con(\cdot,r)$ and  $\dn(\cdot,r)$ by the algebraic identities
\be\label{IDsn}
\begin{aligned}
& \sn(x,r)^2 + \con(x,r)^2 = 1,\quad \dn(x,r)^2 + r^2\sn(x,r)^2 = 1,
\end{aligned}
\ee
and by the derivatives with respect to $x$:
\be\label{EDOsn}
\begin{aligned}
& \frac{d}{dx}\sn(x,r)= \con(x,k)\dn(x,r),\qquad \frac{d}{dx}\con(x,r)= -\sn(x,k)\dn(x,r),\\
& \frac{d}{dx}\dn(x,r)= -r^2\sn(x,r)\con(x,r).
\end{aligned}
\ee
The $\nd(\cdot,r),~r=k~~\text{or}~~m$ is defined as the inverse of the $\dn$ JEF , i.e. $\nd(x,r)=\frac{1}{\dn(x,r)}$. It is also
characterized to be the solution of the following ODEs (here $\nd \equiv \nd(x,r)$):
\be\label{nd}
\begin{aligned}
& \nd'^{2} = -1+(2-r)\nd^2+(r-1)\nd^4,\qquad  \nd'' = (2-r)\nd+2(r-1)\nd^3. 
\end{aligned}
\ee
It takes some special values when $r=0$ or $r=1$: $\nd(x,0)=1,$ and $\nd(x,1)=\cosh(x)$.
%
%
and by the derivatives with respect to $x$
\be\label{EDOsn2}
\begin{aligned}
& \frac{d}{dx}\nd(x,r)= r^2\sd(x,r)\cd(x,r),\qquad  \frac{d}{dx}\sd(x,r)= \con(x,r)\nd^2(x,r),\\
& \frac{d}{dx}\cd(x,r)= -r^2\sn(x,r)\nd^2(x,r).
\end{aligned}
\ee

\end{document}